\documentclass{JINST}
\pdfoutput=1
\let\ifpdf\relax
\usepackage{epsfig}
\usepackage{xspace}
\usepackage{longtable}
\usepackage{multirow}
\usepackage{graphicx}
\usepackage{booktabs}

\usepackage{fixltx2e}
\usepackage{amsmath,amssymb}
\usepackage[figuresright]{rotating}

\newcommand{\mr}{\multirow}
\newcommand{\mc}{\multicolumn}

\title{\bf 
The MICE Muon Beam on ISIS and the beam-line instrumentation
  of the Muon Ionization Cooling Experiment
}
\author{The MICE Collaboration}
\author{M. Bogomilov, Y. Karadzhov$^{~1}$, D. Kolev, I. Russinov, 
R.~Tsenov, G.Vankova-Kirilova \\
Department of Atomic Physics, St. Kliment Ohridski University of Sofia, 
Sofia, Bulgaria\\
\llap{$^{~1}$}~{Now at DPNC, Universit\'e de Gen\`eve, Geneva, Switzerland}
}
\author{L. Wang, F.Y. Xu, S.X. Zheng\\
Institute for Cryogenic and Superconductivity Technology, Harbin Institute of 
Technology, Harbin, 
PR China}
\author{R. Bertoni, M. Bonesini~\footnote{Corresponding author.
E-mail:Maurizio.Bonesini@mib.infn.it}, F. Ferri$^{~2}$, G. Lucchini, R. Mazza, 
F. Paleari$^{~3}$, F.~Strati \\
Sezione INFN Milano Bicocca, Dipartimento di Fisica G. Occhialini, 
Milano, Italy \\
\llap{$^{~2}$}~{Present address DSM/IRFU, CEA/Saclay, Gif-sur-Yvette, France} \\
\llap{$^{~3}$}~{Now at Quanta Systems, Solbiate Olona, Varese, Italy}
}
\author{V. Palladino\\
Sezione INFN Napoli and Dipartimento di Fisica, Universit\`{a} Federico II, 
Complesso Universitario di Monte S. Angelo, Napoli, Italy}
\author{G. Cecchet, A. de Bari\\
Sezione INFN Pavia and Dipartimento di Fisica Nucleare e Teorica, Pavia, Italy}
\author{M. Capponi, A. Cirillo, A. Iaciofano, A. Manfredini, 
M. Parisi, D. Orestano, 
F.~Pastore, A. Tonazzo$^{~4}$, L. Tortora\\
Sezione INFN Roma Tre e Dipartimento di Fisica, Roma, Italy \\
\llap{$^{~4}$}~{Present address APC, Universit\`e Paris Diderot, Paris, France} \\
}
\author{Y. Mori\\
Kyoto University Research Reactor Institute, Osaka, Japan}
\author{Y. Kuno, H. Sakamoto, A. Sato, T. Yano, M. Yoshida\\
Osaka University, Graduate School of Science, Department of Physics, Toyonaka, Osaka, Japan}
\author{S. Ishimoto, S. Suzuki, K. Yoshimura\\
High Energy Accelerator Research Organization (KEK), Institute of Particle and Nuclear Studies, Tsukuba, Ibaraki, Japan}

\author{
F. Filthaut$^{~5}$\\
NIKHEF, Amsterdam, The Netherlands\\
\llap{$^{~5}$}~{Also at Radboud University Nijmegen, Nijmegen, The Netherlands}\\
}
\author{R.~Garoby, S.~Gilardoni, P.~Gruber, K.~Hanke, H.~Haseroth, P.~Janot,
A.~Lombardi, S.~Ramberger, M.~Vretenar\\
CERN, Geneva, Switzerland}
\author{P. Bene, A. Blondel, F. Cadoux, J.-S. Graulich, V. Grichine$^{6}$, 
E. Gschwendtner$^{7}$, F.~Masciocchi, R.~Sandstrom, V.~Verguilov, H.~Wisting\\
DPNC, Section de Physique, Universit\'e de Gen\`eve, Geneva, Switzerland \\
\llap{$^{~6}$}~{Also at Lebedev Physical Institute, Moscow, Russia} \\
\llap{$^{~7}$}~{Now at CERN, Geneva, Switzerland}
}
\author{C. Petitjean\\
Paul Scherrer Institut, Villigen, Switzerland}
\author{R. Seviour\\
The Cockcroft Institute, Daresbury Science and Innovation Centre, 
Daresbury, Cheshire, UK}
\author{
  J.~Alexander, G.~Charnley, N.~Collomb, S.~Griffiths,
  B.~Martlew, A.~Moss, I.~Mullacrane, A.~Oates, P.~Owens, C.~White,
  S.~York                                                              \\
  STFC Daresbury Laboratory, Daresbury, Cheshire, UK                                                                          \\
}
\author{
  D.~Adams, R.~Apsimon, P.~Barclay,
  D.E.~Baynham, T.W.~Bradshaw, M.~Courthold,
  P.~Drumm$^{~8}$, R.~Edgecock, T.~Hayler, M.~Hills$^{9}$,
  Y.~Ivaniouchenkov, A.~Jones, A.~Lintern, C.MacWaters,
  C.~Nelson, A.~Nichols, R.~Preece, S.~Ricciardi,
  J.H.~Rochford$^{~10}$, C.~Rogers, W.~Spensley$^{~11}$, J.~Tarrant,
  K.~Tilley, S.~Watson, A.~Wilson                                      \\
  STFC Rutherford Appleton Laboratory, Harwell Oxford, Didcot, UK \\
  \llap{$^{~8}$}~{Now at Space Research Centre, Department of Physics and
                  Astronomy, University of Leicester, Leicester, UK}      \\
  \llap{$^{~9}$}~{Now at Mullard Space Science Laboratory, University
                  College London, Dorking, Surrey, UK}      \\
  \llap{$^{~10}$}~{Now at Global Research Centre, General Electric, Albany, NY, USA}      \\
  \llap{$^{~11}$}~{Now at MANTEC SYSTEM Ltd., Newcastle Upon Tyne, UK}      \\
}
\author{D. Forrest, F.J.P. Soler, K. Walaron$^{~12}$\\
School of Physics and Astronomy, Kelvin Building, The University of Glasgow, Glasgow, UK\\
\llap{$^{~12}$}~{Also at Imperial College London, London, UK}
}
\author{P. Cooke, R. Gamet\\
Department of Physics, University of Liverpool, Liverpool, UK}
\author{
  A.~Alecou, M.~Apollonio$^{~13}$, G.~Barber, R.~Benselinck, 
  D.~Clark, I.~Clark, D.~Colling, A.~Dobbs,
  P.~Dornan, S.~Fayer,
  A.~Fish$^{~14}$, R.~Hare, S.~Greenwood, A.~Jamdagni, V.~Kasey,
  M.~Khaleeq, J.~Leaver, K.~Long, E.~McKigney$^{~15}$,
  T.~Matsushita$^{~16}$, J.~Pasternak, T.~Sashalmi, T.~Savidge,
  M.~Takahashi$^{~17}$                        \\
Department of Physics, Blackett Laboratory, Imperial College London,
London, UK    \\                                     \\
\llap{$^{~13}$}~{Now at Diamond Light Source, Harwell Science and
               Innovation Campus, Didcot, Oxfordshire, UK}                  \\
\llap{$^{~14}$}~{BC asset management Ltd., BC House, Poole,
               Dorset, UK}                                        \\
\llap{$^{~15}$}~{Now at Los Alamos Natl. Lab., Los Alamos,
               USA}                                               \\
\llap{$^{~16}$}~{Now at Kobe University, Faculty of Science,
               1-1 Rokkodai-cho, Nada-ku, Kobe-shi,
               Japan}                                                       \\
\llap{$^{~17}$}~{Now at School of Physics and Astronomy, 
                 University of Manchester,
                 Manchester, UK }                             \\
}

\author{V. Blackmore, T. Carlisle, J.H. Cobb, W. Lau, M. Rayner$^{~18}$, 
C.D. Tunnell, H. Witte$^{~19}$, S. Yang \\
Department of Physics, University of Oxford, Denys Wilkinson Building, Oxford, UK\\
\llap{$^{~18}$}~{Now at DPNC, Universit\'e de Gen\`eve, Switzerland} \\
\llap{$^{~19}$}~{Now at Brookhaven National Laboratory, Upton, NY, USA}
}
\author{C.N. Booth, P. Hodgson, L. Howlett, R. Nicholson, E. Overton, M. Robinson, P.~Smith\\
Department of Physics and Astronomy, University of Sheffield, Sheffield, UK}
\author{D. Adey, J. Back, S. Boyd, P. Harrison\\
Department of Physics, University of Warwick, Coventry, UK}
\author{M. Ellis$^{~20}$, P. Kyberd, M. Littlefield, J.J. Nebrensky\\
Brunel University, Uxbridge, UK\\
\llap{$^{~20}$}~{Now at Westpac Institutional Bank, Sydney, Australia}
}
\author{A.D. Bross, S. Geer, D. Neuffer, A. Moretti, M. Popovic \\
Fermilab, Batavia, IL, USA}
\author{M.A.C. Cummings, T. J. Roberts\\
Muons, Inc., Batavia, IL, USA}
\author{A. DeMello, M.A. Green, D. Li,  S. Virostek, M.S. Zisman\\
Lawrence Berkeley National Laboratory, Berkeley, CA, USA}
\author{B. Freemire, P. Hanlet, D. Huang$^{21}$, G. Kafka, D.M. Kaplan, P. Snopok, Y. Torun\\
Illinois Institute of Technology, Chicago, IL, USA \\
\llap{$^{~21}$}~{Now at Shangai Synchrotron Radiation Facility, Shangai,
PR China}
} 
\author{S.~Blot, Y.K. Kim \\
Enrico Fermi Institute, University of Chicago, Chicago, IL, USA}
\author{U. Bravar\\
University of New Hampshire, Durham, NH, USA}
\author{Y. Onel\\
Department of Physics and Astronomy, University of Iowa, Iowa City, IA, USA}
\author{D. Cline, Y.~Fukui, K. Lee, X. Yang\\
Department of Physics and Astronomy, University of California, Los Angeles, CA,
USA}
\author{R.A. Rimmer\\
Jefferson Lab, Newport News, VA, USA}
\author{L.M. Cremaldi, G.~Gregoire $^{22}$,T.L. Hart, D.A. Sanders, 
D.J. Summers\\
University of Mississippi, Oxford, MS, USA \\
\llap{$^{~22}$}~{Permanent address Institute of Physics, 
Universit\'e Catholique de Louvain, Louvain-la-Neuve, Belgium}
}
\author{L. Coney, R. Fletcher, G.G. Hanson, C. Heidt\\
University of California, Riverside, CA, USA}
\author{J.Gallardo, S. Kahn$^{23}$, H. Kirk, R.B. Palmer\\
Brookhaven National Laboratory, Upton, NY, USA\\
\llap{$^{~23}$}~{Now at Muons, Inc., IL, USA}
}

\newpage
\abstract{
The international Muon Ionization Cooling Experiment (MICE), which is
under construction at the Rutherford Appleton Laboratory (RAL), will
demonstrate the principle of ionization cooling as a technique for the
reduction of the phase-space volume occupied by a muon beam.
Ionization cooling channels are required for the Neutrino Factory and
the Muon Collider.
MICE will evaluate in detail the performance of a single lattice cell
of the Feasibility Study 2 cooling channel.
The MICE Muon Beam has been constructed at the ISIS synchrotron at RAL, and
in MICE Step I, it has been characterized using the
MICE beam-instrumentation system.
In this paper, the MICE Muon Beam and  beam-line instrumentation
are described.
The muon rate is presented as a function of the
beam loss generated by the MICE target dipping into the ISIS proton
beam. For a 1 V signal from the ISIS beam-loss monitors downstream 
of our target we obtain a 30 KHz instantaneous muon rate, with a neglible pion 
contamination in the beam.
}

\begin{document}

% ----------------------------------------------------------------
\section{Introduction}
\label{sec:Introduction}

Muon storage rings have been proposed as sources of intense,
high-energy neutrino beams \cite{Koshkarev:1974,Geer:1998} and as
high energy lepton-antilepton colliders \cite{Tikhonin:2008pw,Budker:1996}.
In each of these facilities, the volume of phase space occupied by the
muon beam must be reduced (cooled) before the beam is accelerated and
stored.
The muon lifetime is so short that unacceptable decay losses will
occur if conventional cooling techniques are employed~\cite{Parkhomchuk2008}. 
Ionization cooling is fast, as the cooling is generated by
ionization energy loss as the muon beam passes through an ``absorber''
material, the lost longitudinal momentum being restored in
accelerating cavities.
Ionization cooling has therefore been adopted in the proposed design for
 both the Neutrino Factory and the Muon Collider \cite{Neuffer:1986dg}.

At the Neutrino Factory \cite{Geer:1998}, intense beams
of high energy neutrinos are produced from the decay of muons
circulating in a storage ring.
Long straight sections in the storage ring direct neutrino beams 
to one or more distant detectors.
The energy spectrum and flavour composition of the beam are known
precisely, as compared to those of conventional neutrino 
beams \cite{ Bonesini:2006ik}.
% since muon decay is described with great precision by the
% Standard Model. 
A number of conceptual designs for the Neutrino Factory have been
proposed
\cite{Ozaki:2001bb,Alsharoa:2002wu,Blondel:2004ae,Bandyopadhyay:2007kx,
Choubey:2011zz}
each of which exploits an ionization-cooling channel to increase
significantly the muon flux delivered to the storage ring.
The baseline design for the facility being developed by the
International Design Study for the Neutrino Factory (IDS-NF)
collaboration delivers $10^{21}$ muon decays per year and has been
shown to have a discovery reach that is significantly better than that of
realistic alternatives.
The cooling channel specified in the IDS-NF baseline delivers an
increase of a factor 2.4 in the stored-muon flux and is essential for
the design performance of the facility to be delivered.

The Muon Collider \cite{Ankenbrandt:1999as} offers an attractive
route to multi-TeV lepton-antilepton collisions.
Since the muon mass is 200 times that of the electron it is possible
to deliver very high-energy beams using circular accelerators that can
be designed to fit within the area of existing particle-physics
laboratory sites.
In addition, the large muon mass implies a reduction in the rate of
beamstrahlung by a factor of $\sim 10^4$ over an $e^+ e^-$ collider of
the same centre-of-mass energy.
As a consequence, the annihilation-energy distribution is much
narrower at a Muon Collider than at an $e^+ e^-$ collider of
the same energy.

The $\mu^+ \mu^-$ annihilation cross section falls rapidly as the
centre-of-mass energy in the collision increases, therefore high
luminosity ($\gtrsim 10^{34}$\,cm$^{-2}$s$^{-1}$) is critical to the
success of the Muon Collider.
To achieve the required luminosity, the muon-beam phase space must be
cooled in all six phase-space dimensions: a reduction in the
six-dimensional emittance by a factor $\sim 10^6$ being required.
By contrast, the cooling requirements of the Neutrino Factory, which
requires cooling of only the four-dimensional transverse phase space,
are relatively modest.

Ionization cooling is accomplished by passing the muon beam 
through a low-$Z$ material (the ``absorber'') in which it loses
energy through ionization.
The energy loss results in a reduction in the longitudinal and the
transverse components of momentum.
The lost energy is restored by accelerating the beam such that the
longitudinal component of momentum is increased while the transverse
components remain unchanged.
The net effect of these two operations is to reduce the divergence of
the beam, thereby reducing the volume of transverse phase space it
occupies.
Beam transport through the absorber and the accelerating structures is
achieved using a solenoid-focusing lattice.

The rate of change of normalised emittance due to ionization cooling 
in a medium of thickness $X$ 
may be described as \cite{Neuffer:1986dg}:
\begin{eqnarray}
 \frac{d\epsilon_N}{dX} \approx -\frac{\epsilon_N}{\beta^2 E_{\mu}}\left\langle
 \frac{dE}{dX} \right\rangle+ \frac{\beta_t (0.014 \ \text{GeV})^2}{2 
   \beta^3 E_{\mu} m_{\mu} X_0}\,;  
 \label{eq:cool}
\end{eqnarray}
where $\epsilon_N$ is the normalised transverse (four-dimensional)
emittance of the beam, $X_0$ is the radiation length of the medium, 
$\beta_t$ is the betatron function,
$E_{\mu}$ and $m_{\mu}$ the energy and mass of the muons and 
$\beta=pc/E$ their velocity.
The first (negative) term on the right hand side of equation
\ref{eq:cool} describes a reduction of emittance per unit length,
i.e. cooling. 
The second (positive) term describes the heating effect of multiple
scattering. 
The two effects reach an equilibrium when the emittance of the beam
is:
\begin{eqnarray}
 \epsilon_{eq} \approx \frac{\beta_t(0.014 \ \text{GeV})^2}{2 \beta m_{\mu} X_0 } 
 \left\langle \frac{dE}{dX} \right\rangle ^{-1} \, .
 \label{eq:emieq}
\end{eqnarray}
The ideal cooling channel will produce the lowest equilibrium
emittance.
This is obtained when $\beta_t$ is minimised and 
$ X_0 \cdot \left\langle \frac{dE}{dX} \right\rangle$ is
maximised.
To minimise $\beta_t$ requires strong focusing at the absorber while
hydrogen offers the largest value of 
$ X_0 \cdot \left\langle \frac{dE}{dX} \right\rangle$.

While the principle of ionization cooling is readily described using
equations \ref{eq:cool} and \ref{eq:emieq}, the construction of an
ionization-cooling cell requires  significant engineering.  
In a practical ionization-cooling cell, the energy loss in the
absorber must be replaced by means of RF cavities:
momentum must be restored so that beam
transport through subsequent cells is unaffected.

The international Muon Ionization Cooling Experiment (MICE)
\cite{Blondel:2003,Gregoire:2003} is being constructed at the
Rutherford Appleton Laboratory (RAL).  
The experiment will consist of the full cell of an ionization-cooling
lattice and the instrumentation necessary to measure the emittance of
a muon beam before it enters and after it leaves the cell. 
In this way, the MICE collaboration will measure the cooling
performance of the lattice cell in a variety of modes of operation and
over a range of momentum and emittance.
The results of the experiment will allow the cooling channels of the
Neutrino Factory and Muon Collider to be optimised.
\section{The MICE concept}
\label{sec:mice}

\begin{figure*}
 \begin{center}
\includegraphics[width=\linewidth]{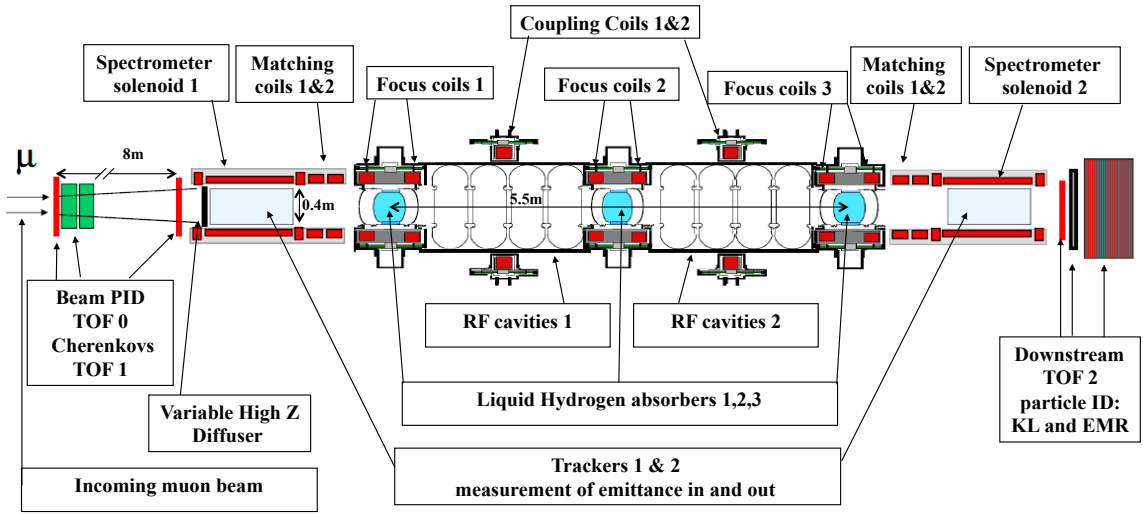}
\includegraphics[width=\linewidth]{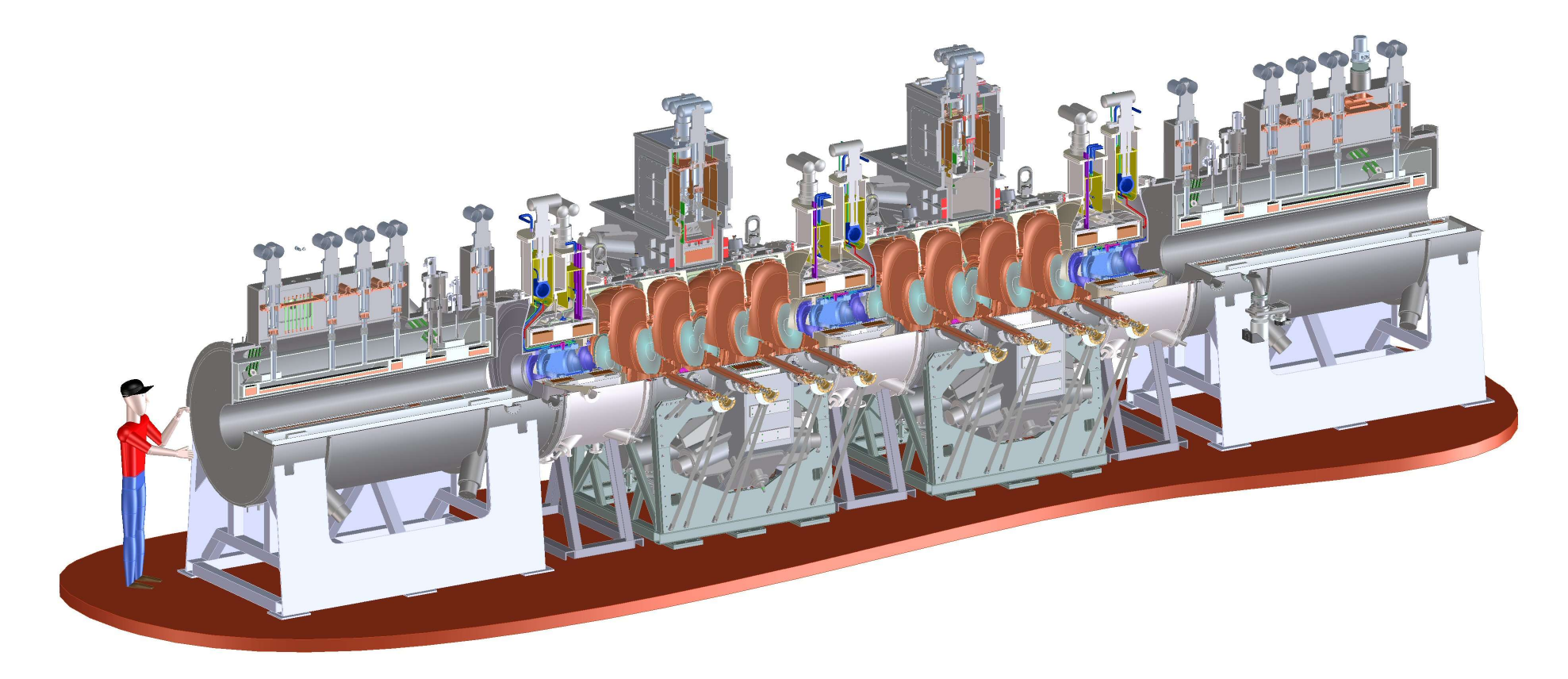}
 \end{center}
\caption{
Top panel:  view of the MICE experimental set-up; the cooling 
channel, with its three liquid hydrogen absorbers and two RF cavity modules,
is sandwiched between two identical trackers. The sequence of
solenoids defining the MICE optics is also visible. The muon beam is
coming from the left. 
Bottom panel: 3D cut-away engineering drawing of MICE, starting at first
  Spectrometer Solenoid. Beam instrumentation is not shown. 
 }
 \label{fig:MICElayout}
\end{figure*}

A schematic diagram of the MICE experiment is shown in figure
\ref{fig:MICElayout}.
The MICE cooling channel, which is based on a single lattice cell of the
cooling channel described in \cite{Ozaki:2001bb}, comprises three
volumes of  $\sim$ 20\,litre of liquid hydrogen and two linac modules each
consisting of four 201\,MHz cavities, with gradients of $\sim 10$ MV/m. Beam 
transport is achieved by means of a series of superconducting
solenoids: the ``focus coils'' focus the beam into the liquid-hydrogen
absorbers, while a ``coupling coil'' surrounds each of the linac
modules.

Detailed Monte Carlo simulations of the passage of muons through the
cooling cell have been carried out to estimate its performance.
The results indicate that a reduction in the normalised emittance of
$10\%$ is expected for a muon beam entering the cell with a momentum
of 200\,MeV/c and an emittance $\epsilon_N = 6.2 \pi $\,mm$\cdot$rad.
The instrumentation upstream and downstream of the cooling cell is
required to measure this change in emittance, $\Delta \epsilon_N$,
with a relative precision $\Delta \epsilon_N / \epsilon_N = 1\%$;
i.e., measurements of $\epsilon_N$ upstream and downstream of the
cooling cell with an absolute precision of 0.1\% are required,
to allow the extrapolation to a full cooling channel.
Conventional emittance measurement techniques based on beam-profile
monitors do not have the required precision.

While the muon-beam intensity in the ionization-cooling channel at the
Neutrino Factory and the Muon Collider is in excess of 
$10^{14}\,\mu^{\pm}$/s, the phase-space density is always low enough
for space-charge forces to be neglected.
This fact has been exploited to devise a
single-particle experiment in which each muon is measured using
state-of-the-art particle detectors and the bunched muon-beam  is
reconstructed offline.
The instrumentation upstream of the MICE cooling cell includes a
particle-identification (PID) system (scintillator time-of-flight
hodoscopes TOF0 and TOF1 and threshold Cherenkov counters Ckova and
Ckovb) that allows a pure muon beam to be selected. 
Downstream of the cooling channel, a final hodoscope (TOF2) and a
calorimeter system allow muon decays to be identified and rejected.
The calorimeter system is composed of a lead-scintillator section
(KL), similar to the KLOE \cite{Ambrosino:2009zza} design but 
with thinner lead
foils, followed by a fully active scintillator detector (the
electron-muon ranger,  EMR) in which the muons are brought to rest.
Charged-particle tracking in MICE is provided by two solenoidal
spectrometers.
The MICE instrumentation must be sufficiently robust to perform
efficiently in the presence of background induced by X-rays produced
in the RF cavities. 
For a full description of the experiment see \cite{Gregoire:2003}. 

The position, momentum and energy of each muon is measured before and
after the cooling cell.  
For each particle $x$, $y$, $t$,
$x'=dx/dz=p_x/p_z$, $y'=dy/dz=p_y/p_z$ and $t'=dt/dz =E/p_z$ 
are measured; $p_i$ is the $i^{\rm th}$ component of momentum, $E$ is
the energy and the MICE coordinate system is defined such that the $z$
axis is parallel to the nominal beam axis, the $y$ axis points
vertically upward and the $x$ axis completes a right-handed coordinate
system. 
The time, $t$, corresponds to the moment the particle crosses a
reference surface within the tracking volume.
\noindent
The input and output normalised  emittances, $\epsilon_{N_i}$ and $\epsilon_{N_o}$, of a beam of $N$ muons
can be determined from the measurements of the phase-space coordinates of 
single muons in the up- and down-stream trackers.  Because they are measured for the same ensemble of muons,
 $\epsilon_{N_i}$ and $\epsilon_{N_o}$ will be correlated with the effect of reducing
the statistical error on the fractional change in emittance,
$f =  (\epsilon_{N_i} - \epsilon_{N_o})/ \epsilon_N$,
below $1/\sqrt{N}$.  The only stochastic process is multiple scattering in the absorbers and it
can be shown that \cite{Cobb}
\begin{equation}
\sigma_f^2 = {1\over{2N}}{{\bar{p}_s^2}\over{\bar{p}_t^2}}
\end{equation}
where $\bar{p}_t^2$ is the mean square transverse momentum of the beam at the absorbers and
$\bar{p}_s^2$ is the mean square transverse momentum kick due to multiple scattering. It has been
demonstrated with Monte Carlo studies \cite{Forrest:2009zz} that for beams with $\epsilon_N > 3 \pi$\,mm $\cdot$ rad
a sample of $3 \times 10^{5}$ muons is sufficient to reduce the statistical error on $f$ to less
than one percent.

The MICE Muon Beam is required to deliver beams with a pion
contamination below 10\% and a central momentum ($p_\mu$) in the range
$140 \leq p_\mu \leq 240$\,MeV/c.
The beam line must also deliver a transverse emittance in the range 
$3 \leq \epsilon_N \leq 10 \ \pi $\,mm $\cdot$ rad.
The full range of emittance is required over the full range of
momentum.
A tungsten or brass ``diffuser'' of variable thickness is placed at the
entrance to the upstream spectrometer solenoid to generate the
divergence necessary for the required range of emittance.

%%For a muon rate of $R_\mu$ muons per second, a sample of $10^5$ muons may
%%be accumulated in $\sim 28/R_\mu$ hours.  
At a  rate of $\sim 30$\,$\mu$/s a sample of $10^5$ muons may be accumulated in
one hour, giving the possibility to record all the needed momentum/emittance
beam settings in the available running time.

\begin{figure}
  \centering
  \includegraphics*[width=\linewidth]{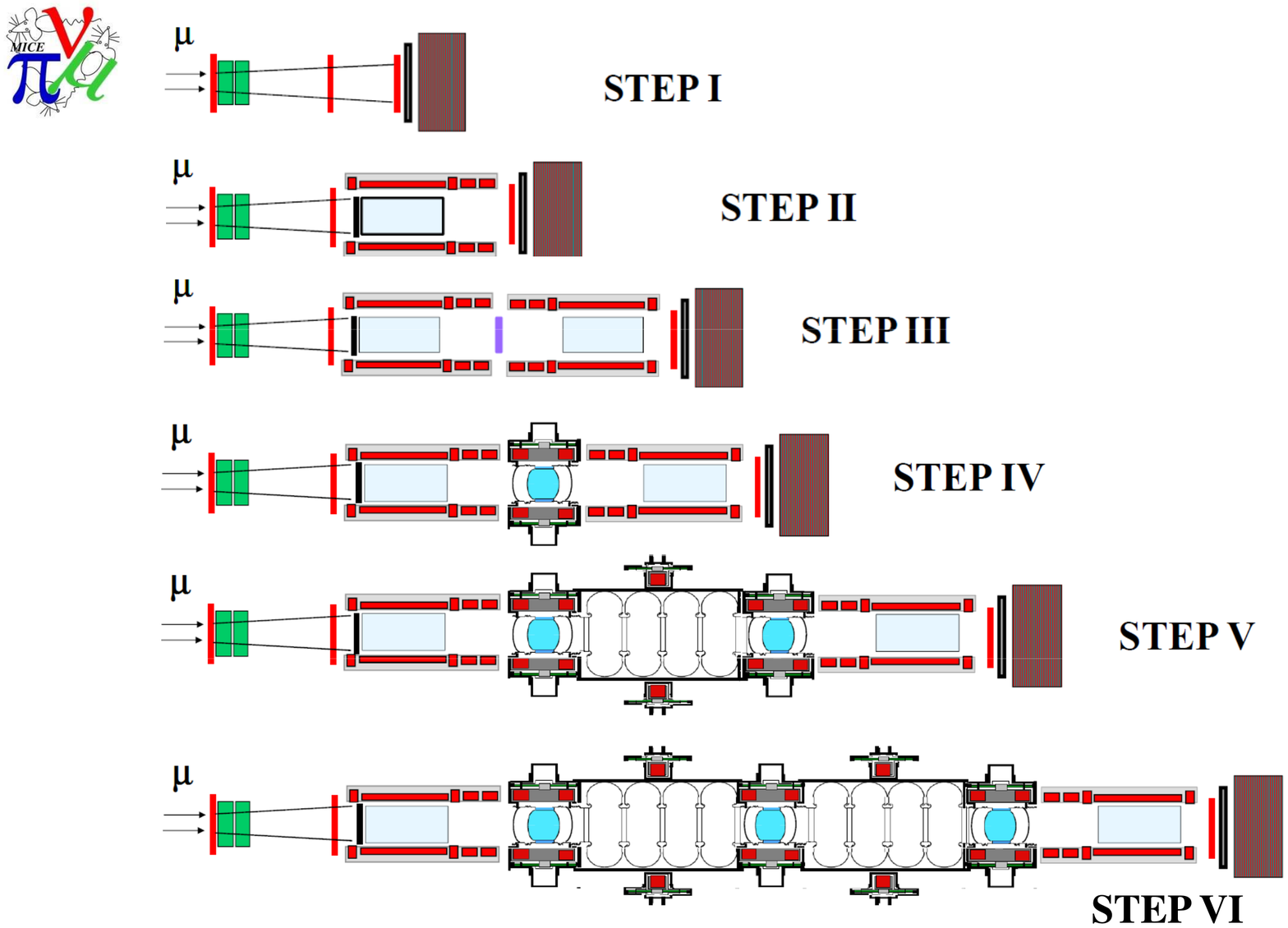}
  \caption{
    Schematic diagrams of the six ``Steps'' in which the MICE
    programme was conceived.
  }
  \label{fig:six_stages}
\end{figure}

The MICE program was conceived to be carried out in six ``Steps'',
summarised in table \ref{tab:MICEstages} and figure
\ref{fig:six_stages}. 
A thorough description of all MICE Steps is beyond the scope of this
paper. 
However, each Step requires that the MICE Muon Beam meets the
specifications outlined above.  
Step I, in which the muon beam was characterized using the
time-of-flight (TOF), Cherenkov and KL detectors, was completed in August
2010.
The EMR calorimeter \cite{Lietti:2009zz}, that will complete 
the downstream calorimetry, will follow soon. 
\begin{table*}
  \caption{
    The six ``Steps'' in which it was conceived to carry out the MICE
    programme. The physics programme of steps II to III will be carried out
    within the STEP IV configuration. } 
  \centering
  \vskip .2cm
  \begin{tabular}{|l|p{9cm}|}
    \hline
    \textbf{Step } & \textbf{Description} \\ 
    \hline
          I    &  \small{Beam characterisation and detector calibration.} \\
          II   &  \small{Upstream spectrometer and diffuser. 
          Incoming beam emittance can be changed and measured precisely.} \\
          III  &  \small{Downstream spectrometer: study of systematics in
          emittance measurement in the two spectrometers. Solid absorber
          with measurement of energy loss and multiple scattering 
          correlations.} \\
          IV  &  \small{Introduction of the first liquid hydrogen
            absorber and focus-coil module.}\\
          V   &  \small{First linac section and second hydrogen
          absorber/focus-coil module. First measurement of cooling
          with momentum recovery.} \\
          VI  &  \small{Second linac module and third liquid
          hydrogen absorber/focus-coil module. Measurement of
          emittance before and after the cooling cell, in various
          optics configurations, with momentum recovery.} \\
    \hline   
  \end{tabular}
  \label{tab:MICEstages}
\end{table*}
\section{MICE Muon Beam}
\label{sec:optics}

\subsection{Overview}

\begin{figure}
  \centering
\includegraphics*[width=0.8\linewidth]{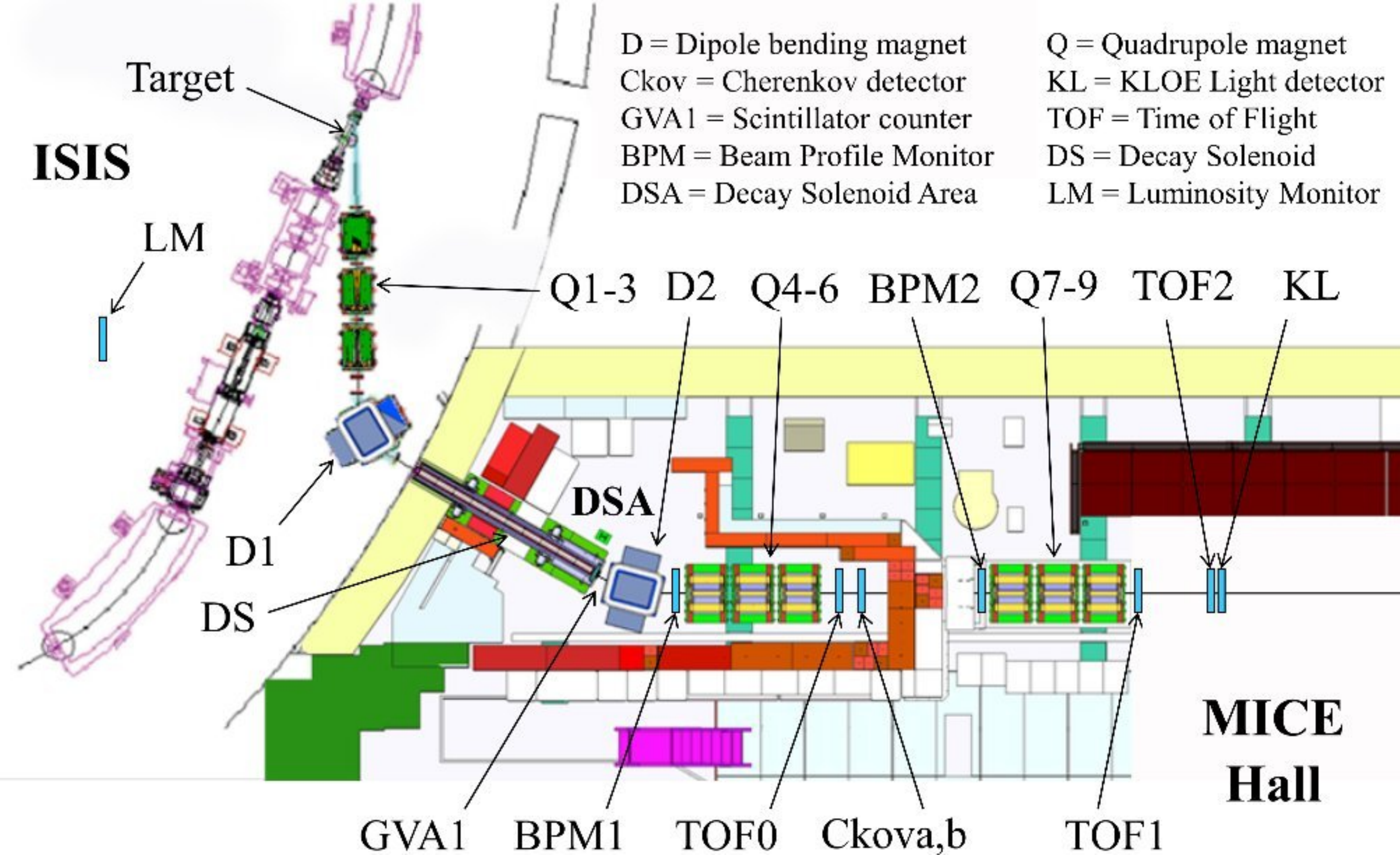}
  \caption{Top view of the MICE beam line with its instrumentation, as
used in Step I.}
 \label{fig:Beamline}
\end{figure}

The design of the MICE Muon Beam is similar to that of the $\mu$E4
beam line at the Paul Scherrer Institute (PSI) in Switzerland.
The MICE Muon Beam may be divided, conceptually, into three sections
(see figure \ref{fig:Beamline}).
In the upstream section, pions produced by the ISIS proton beam
striking a titanium target are captured using a quadrupole triplet
(Q1--3).
The pions are transported to a bending magnet (D1) which directs 
pions of particular momentum into the decay solenoid (DS).
The decay solenoid, the second section of the beam line, causes the
pions to spiral around the nominal beam axis, increasing the path
length. This effect together with the solenoid focusing increases 
the number of muons captured between 
D1 and the second dipole magnet (D2).
D2, the first element of the downstream section, directs muons of a
particular momentum into a quadrupole channel (Q4--6 and Q7--9) that 
transports the muon beam to the MICE experiment.

\begin{figure*}
  \centering
  \begin{tabular}{cc}
    \includegraphics[width=0.51\linewidth]{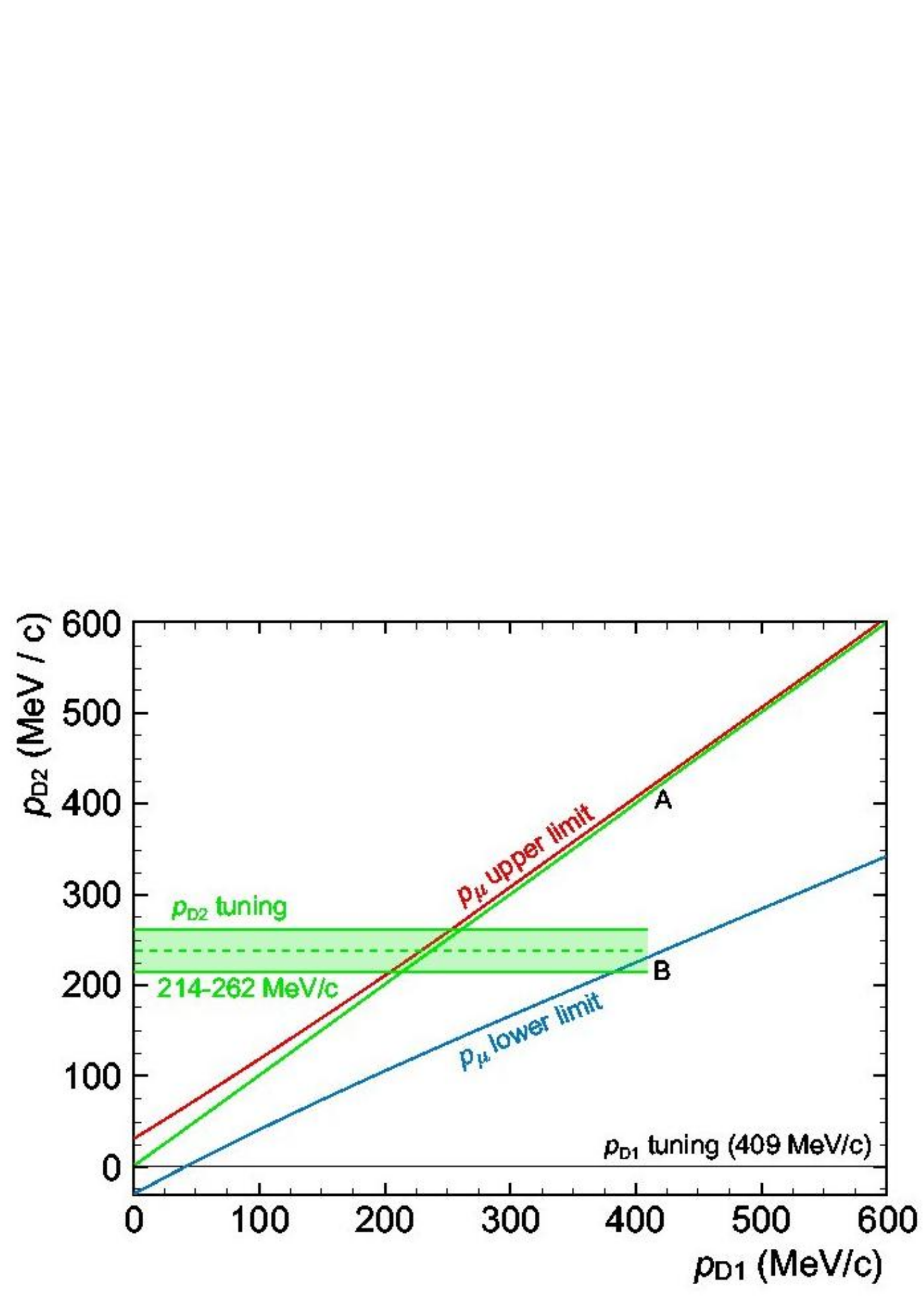}
    \includegraphics[width=0.46\linewidth]{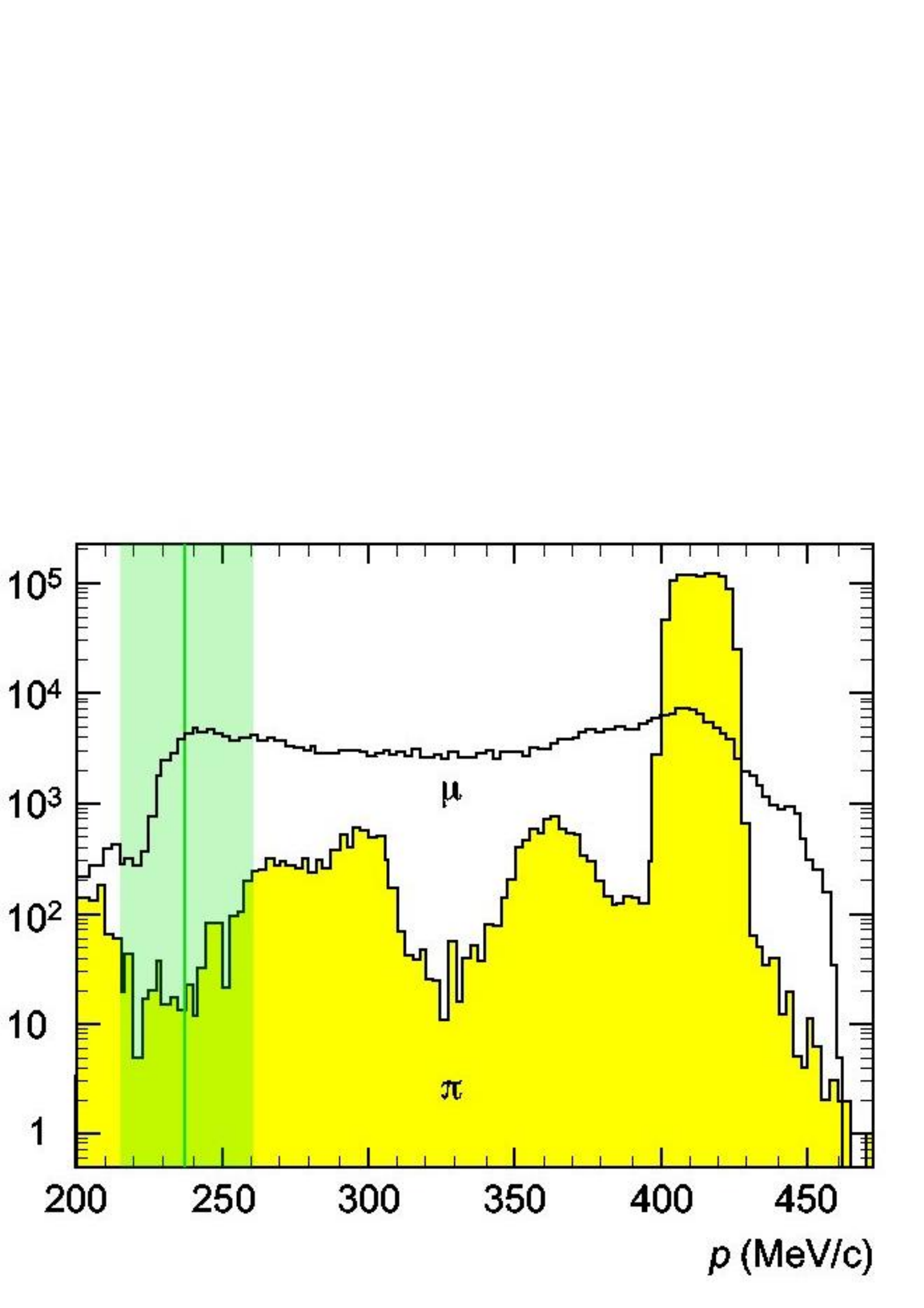}
  \end{tabular}
  \caption{Working principle of the MICE beam line. Left-hand panel: 
  the red and blue lines are the kinematic limits of the spectrum for
  muons produced in pion decays. By tuning D2 to the backward going muon peak an
    almost pion-free sample is produced. Right-hand panel: simulation
    showing pion and muon spectra at the end of the decay
    solenoid. Only high momentum pions survive. The green band
    shows the acceptance of D2, when tuned to the backward-going muon
    peak.}
\label{fig:D1D2_interplay}
\end{figure*}

The composition and momentum spectrum of the beam delivered to MICE
is determined by the interplay between the two bending magnets D1 and
D2.
This is illustrated in figure \ref{fig:D1D2_interplay}.
With D1 set to select a particular pion momentum, $p_{D1}$ (horizontal
axis), the kinematics of pion decay result in muons with momentum in
the range $p_{\mu\,{\rm min}} \le p_\mu \le p_{\mu\,{\rm max}}$.
The kinematic limits are indicated in figure \ref{fig:D1D2_interplay}.
For $p_{D1} \gtrsim 200$\,MeV/c, $p_{\mu\,{\rm max}} \sim p_{D1}$.
When D2 is set to select particles with momentum 
$p_{D2} \sim p_{\mu\,{\rm min}}$, backward-going muons in the pion rest
frame are selected.
The muon-beam purity, $P_{\mu}$, is defined as $P_{\mu}=N_\mu/N_T$, where $N_\mu$
is the number of muons in the beam and $N_T$ is the total number of
particles.
A muon beam of sufficient purity can be selected by setting 
$p_{D2} = p_{\mu\,{\rm min}} \sim p_{D1} /2$.
This can be seen in the right-hand panel of figure
\ref{fig:D1D2_interplay} where the momentum spectra of particles
emerging from the decay solenoid are plotted in a simulation in which
D1 was set to select a central momentum of 400\,MeV/c. 
The figure shows that setting D2 to select a momentum bite of 10\%
around $p_{D2} = 236$\,MeV/c delivers a muon beam with a pion
contamination (defined as $N_\pi/N_T$, where $N_\pi$ is the number of
pions in the beam) less than 2\%.
In practice, a muon beam with a purity in excess of 95\% is obtained
by setting $p_{D1} \simeq 2p_{D2}$. 
This case is referred to as ``$\pi\rightarrow \mu$ mode''.
A momentum-selected beam containing pions, muons, protons and
electrons (or positrons) can be obtained by setting $p_{D1} \simeq p_{D2}$.
Such a setting is referred to as ``calibration mode'' and is used for
the calibration of detectors.

The muon beams delivered to MICE are characterized by their normalised
emittance, $\epsilon_{N}$, and momentum, $p_{z}$.
Different beam settings are identified using the notation 
$(\epsilon_{N},p_{z})$, where $\epsilon_{N}$ is measured in $\pi$ mm $\cdot$ rad and 
$p_{z}$ in MeV/c.
The ``nominal'' values of $\epsilon_{N}$ and $p_{z}$ are defined such
that the nominal value of $\epsilon_{N}$ is evaluated in the upstream
spectrometer solenoid and the nominal value of $p_z$ is evaluated at
the centre of the central liquid-hydrogen absorber in the Step VI
configuration. 
The baseline optics deliver a beam characterized by 
$(\epsilon_{N},p_{z}) = (6, 200)$.
The ($\epsilon_N,p_z$) space is discretised into a matrix of nine
elements (see table \ref{tab:matrix}).
Each cell of this emittance-momentum matrix is associated with a specific downstream
beam-line optics.
Together the cells span the $(\epsilon_{N},p_{z})$ space required to
serve the entire MICE programme.
\subsection{Beam-line elements and  instrumentation}

\subsubsection{The MICE target}

Protons are injected into  ISIS  with a kinetic energy of 70\,MeV and
are accelerated over a period of 10\,ms up to 800\,MeV at a repetition
rate of 50\,Hz~\footnote{Upon reaching 800 Mev energy the protons are
extracted to the ISIS neutron-production targets.}.
The MICE target has been designed to dip into the beam shortly
before extraction at a rate of $\sim 1$\,Hz, i.e. sampling one
in 50 of the ISIS beam pulses.

The MICE target is a hollow titanium cylinder with an outer diameter
of 6\,mm and an inner bore of 4.6\,mm.
The target dips vertically into the beam, 
intercepting it over the last 3\,ms of
the acceleration cycle ($\sim 4\,000$ turns).
Over this period, the target samples proton-beam energies in the range
617--800\,MeV.
The beam envelope shrinks as the beam is accelerated.

\begin{figure}
  \begin{center}
    \includegraphics[width=0.37\linewidth]%
    {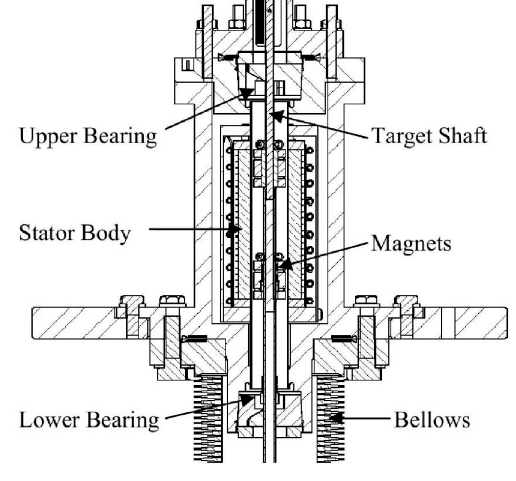}
    \includegraphics[width=0.58\linewidth]%
    {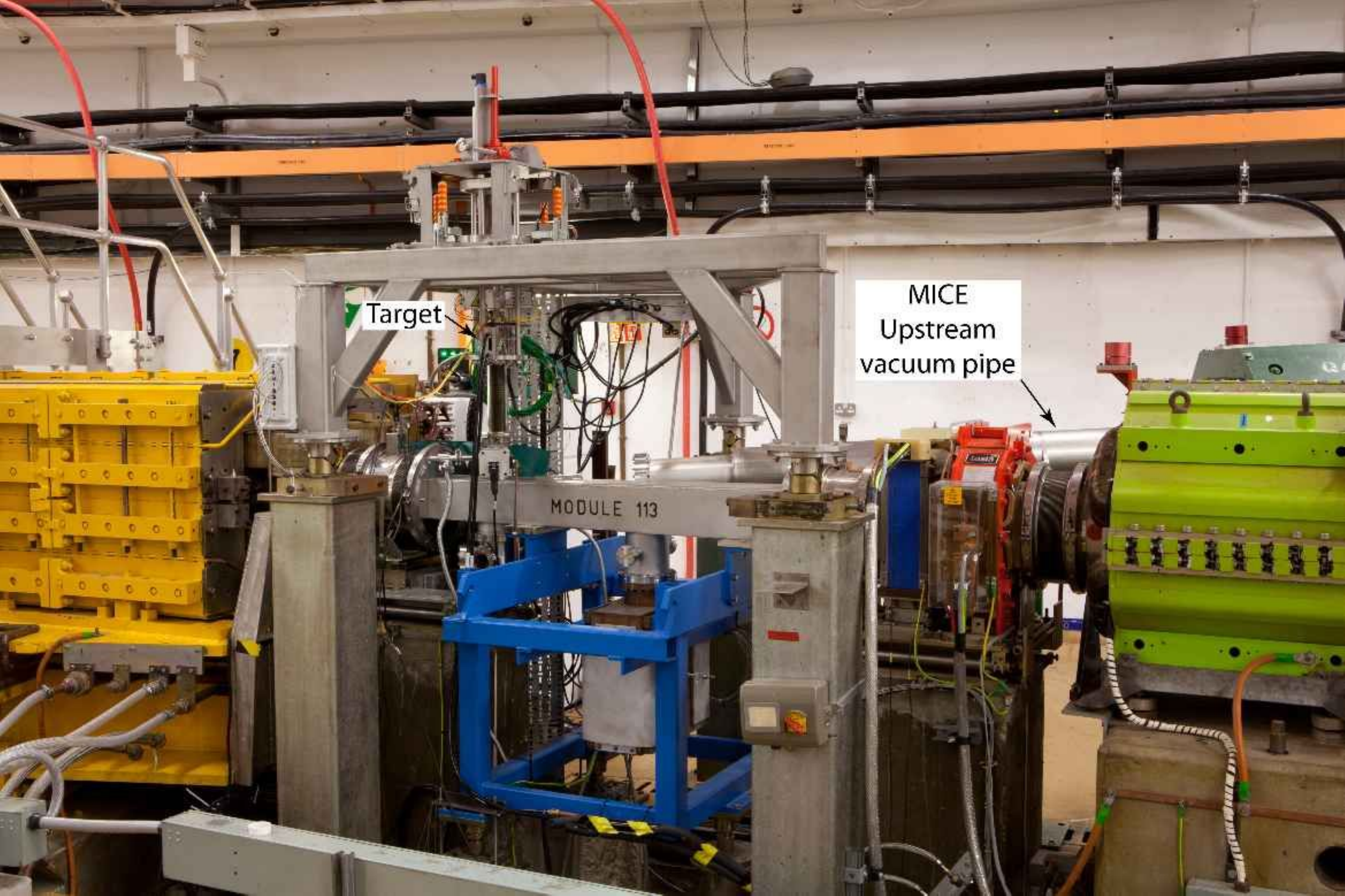}
  \end{center}
  \caption{
    Left panel: MICE target drive schematics. 
    Right panel: The MICE target installed in ISIS.
  }
  \label{target}
\end{figure}

To insert the target into the shrinking beam envelope requires an
acceleration of $\sim 80$ {\it g}.
The target-insertion mechanism consists of a linear electromagnetic
motor in which permanent magnets mounted on the target shaft are
accelerated by a series of 24 coils contained within the stator.
An optical position-sensing system with laser quadrature readout is
used to monitor the position of the shaft and to control the drive
current to the 24 coils (see left-hand panel of figure~\ref{target}).  
A paper describing the design and operation of the MICE target is
in preparation \cite{target}.

\subsubsection{The MICE luminosity monitoring}
\label{sec:lumi}
There are 39 beam loss monitors (BLM) \cite{ISIS_BLM} placed around
the inner radius of ISIS.
These are argon-filled cylindrical ionization chambers, 3 m long, 16
mm in diameter and situated 2--3\,m from the beam axis parallel to the beam.
The BLMs detect particles produced by protons that are lost from the
beam envelope and interact with accelerator components.
The ionization chambers that form the BLMs produce a voltage signal
proportional to the particle rate.
This signal is integrated over some time interval (typically
between 1 and 3 ms) to give an overall measure of beam loss (in units
of V$\cdot$ ms).
The ISIS synchrotron lattice has 10 ``super periods'', numbered from 0
to 9, with the MICE target situated in super period 7 (SP7).
The signal from the four BLMs in SP7 is a measure of beam loss
produced by the MICE target.

\noindent
The MICE Luminosity Monitor (LM) provides a measurement of the
particle rate close to the MICE target independent of the ISIS Beam
Loss Monitors. 
The LM is located inside the ISIS synchrotron ring, 10\,m from the
target (see figures \ref{fig:Beamline} and \ref{UpstreamBeamLine}).
A line drawn from the target to the LM makes an angle of 25$^\circ$
with respect to the direction of the proton beam at the target.
The LM therefore samples particles at an angle similar to that at
which particles are captured by Q1--3.
The rate recorded by LM may therefore be used to
validate the simulation of the upstream beam line and to normalise the
rates measured in the beam line instrumentation to the number of
protons on target (POT).

\begin{figure}
  \begin{center}
    \includegraphics[width=0.49\linewidth]%
       {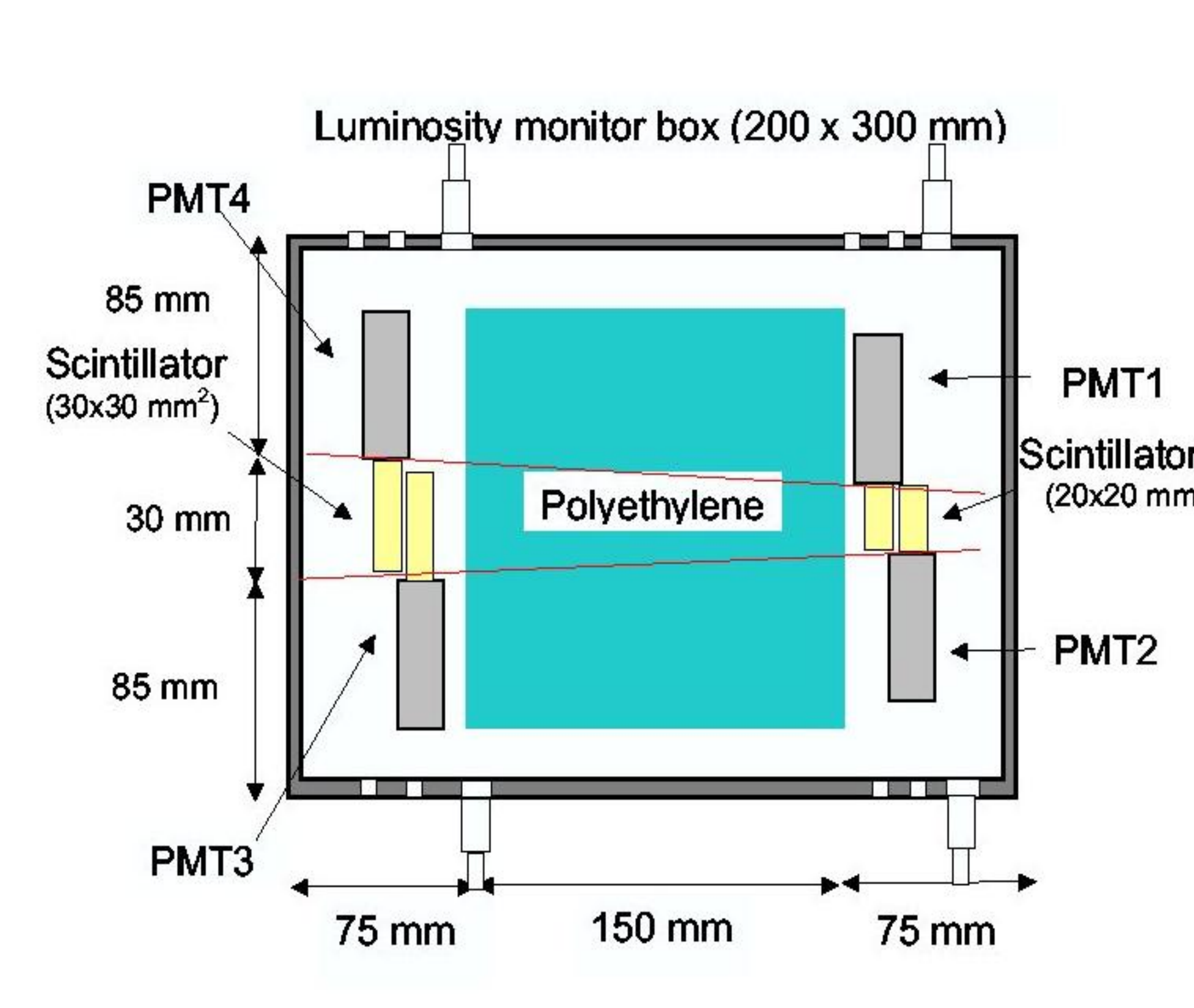}
    \includegraphics[width=0.49\linewidth]%
       {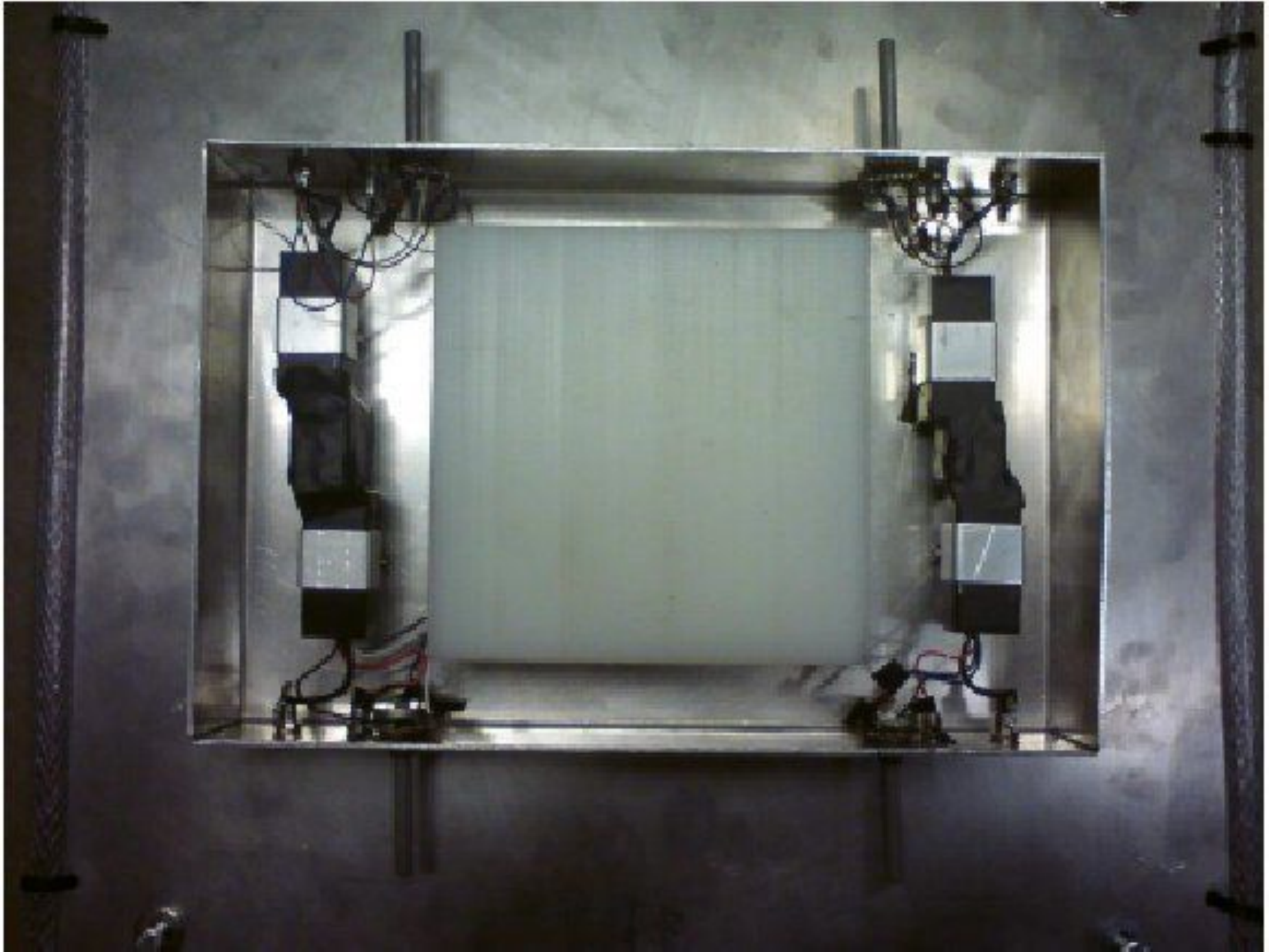}
  \end{center}
  \caption{
    Left panel: Luminosity Monitor design schematic (particles from the beam
    travel from right to left).
    Right panel: photograph of the detector.
  }
  \label{lmdesign}
\end{figure}

The LM consists of two pairs of small scintillators  placed either side of a
polyethylene block that absorbs protons with momentum  below $\sim
500$\,MeV/c and pions with momentum below $\sim 150$\,MeV/c (see
figure \ref{lmdesign}).
The scintillators are read out by Hamamatsu H5783P low noise PMTs which
have a rise time of 0.8\,ns and a gain of $10^5$ \cite{Hamamatsu}.
Upstream of the polyethylene block, PMTs 1 and 2 are bonded to 
scintillator blocks of dimension $20\times 20\times 5$\,mm$^3$;
downstream, PMTs 3 and 4 are attached to $30\times 30\times 5$\,mm$^3$
scintillator blocks.
Neutrons may be detected if scattering inside the polyethylene block
releases a proton.
The PMT signals are fed to LeCroy 623B NIM octal discriminators, the
three outputs of which are used to form three coincidence signals
using a LeCroy 622 NIM quad coincidence unit \cite{LeCroy}.
Coincidences are recorded within the experimental trigger gate
during each spill: PMTs 1 and 2 (LM12), 3 and 4 (LM34) and all four PMTs 
(LM1234)~\footnote{In the following, we use the term
``spill'' for the burst of particles  resulting from one target dip,
even though the primary ISIS proton beam is not extracted in producing
the MICE beam.}.

\begin{figure}
  \begin{center}
    \includegraphics[width=7.5cm,height=6.0cm]%
    {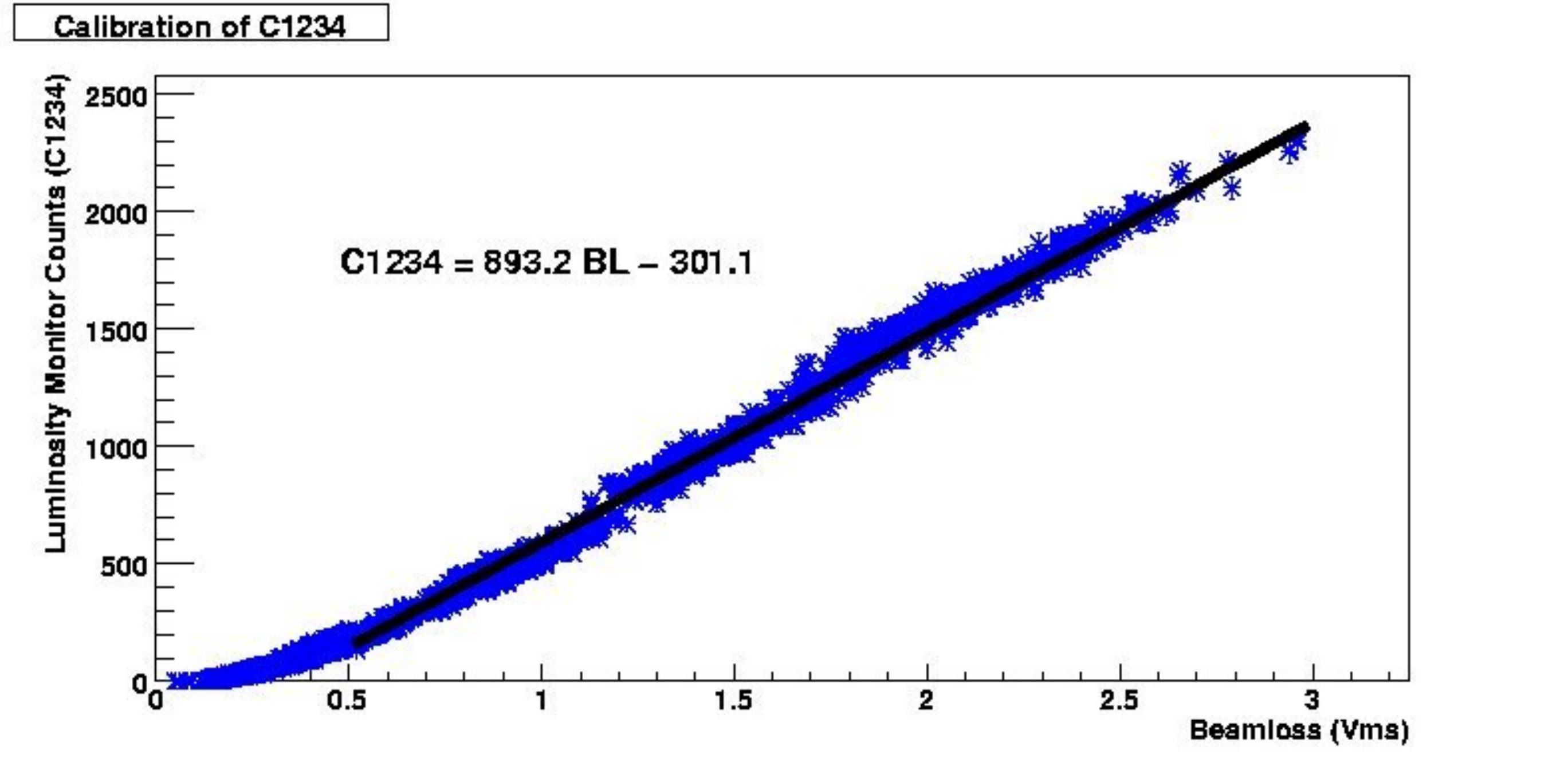}
  \end{center}
  \caption{
    Luminosity monitor data showing a direct relationship between
    luminosity and beam loss on all three scalar output channels.
    Graphs show counts for coincidences of signals for PMTs 1,2,3 and 4.
    The straight lines are a linear fit to the data between $0.5$~V$\cdot$ ms
    and $3$~V$\cdot$ ms.
  }
  \label{fig:firstlmdata}
\end{figure}

The quad LM1234 coincidence rate is plotted as a function of beam loss in figure
\ref{fig:firstlmdata}.
The data show an approximately linear relation between LM counts
and beam loss in SP7 between 0.5\,V$\cdot$ms and 3\,V$\cdot$ms.
Due to beam losses not caused by the MICE target, the relation 
below 0.5\,V$\cdot$ms becomes not linear. Above 3\,V$\cdot$ms
saturation effects become significant.
The rate recorded by the Luminosity Monitor has been used to convert integrated
beam loss to protons on target (POT) \cite{PhDForrest,LM_NUFACT11}.
                                                                               
\subsubsection{The MICE Muon Beam magnets}

The quadrupole magnets that form the upstream pion-capture section of
the upstream beam line are Type-IV quadrupoles recovered from the
NIMROD accelerator which operated at RAL during the 1960s and 1970s
\cite{MICE_NOTE_65}.
Q1--3 have a circular aperture at the pole tips of 203\,mm.
The operating field gradients, shown in table
\ref{tab:magnet_parameters}, are always significantly smaller than the
maximum gradient of 10\,T/m.
The power to each quadrupole is provided by a 200\,A, 30\,V power
supply. 
An aluminium vacuum pipe passes through Q1--3 and extends into a
vacuum box inside D1. 

The quadrupoles that form the two downstream triplets (Q4--6 and
Q7--9) are each Type QC quadrupoles from DESY \cite{MICE_NOTE_65}. 
The downstream quadrupoles have cross-shaped apertures, with a
circular aperture of 352\,mm diameter at the pole tips and a maximum
gradient of 8\,T/m. 
They are each excited by a 400\,A, 70\,V power supply.

The two dipoles (D1, D2) are large, 17\,t, rectangular NIMROD
Type-1 dipoles \cite{MICE_NOTE_65}. 
The nominal apertures for these magnets are rectangular with tapered
pole pieces that provide a 6\,inch vertical aperture and a 20\,inch
horizontal aperture.
D1 is operated at a field close to saturation of the iron yoke.
The power supply for this magnet is limited for operation at a maximum
current of 440\,A at 240\,V.  
D2 is operated at a more modest field and is excited with a power
supply delivering 200\,A at 100\,V.
\begin{table}[ht!]
\centering
\caption{Beam-line magnet physical properties.}
\vspace{.2cm}
\begin{tabular}{|c|c|c|c|}
\hline
Beam line Magnet & Physical length (mm) & Aperture (mm) & Typical field (limit) \\\hline
Quad Type-IV & 1143 & 203 $\phi$& 1.6 (10) T/m \\
Quad Type QC  & 1170 & 352 $\phi$ & 2 (8) T/m \\
Dipole Type-1 6$^{\prime\prime}$ & 1400 & 508 (H)$\times$ 152 (V) & 1.47 (1.5) T \\
PSI Decay Solenoid & 5000 & 115 $\phi$ & 5 (5.7) T\\\hline
\end{tabular}
\label{tab:magnet_parameters}
\end{table}

The large superconducting decay solenoid (DS) was supplied by the Paul
Scherrer Institute (PSI) in Switzerland, where it had been in use at
the $\mu$E4 beam line from 1974 to 2004 \cite{Rohrer_mue4}.
The  coils which make up the DS have an open inner radius of
57.5\,mm (outer radius 65\,mm) and a length of 5\,m.  
The conductor is a Cu/NbTi mixture with proportions Cu:NbTi of 3.5:1.
The magnet has a stored energy of 1.5\,MJ.
The DS can operate at a maximum current of 1000\,A, with a current
density of 220\,A\,mm$^{-2}$.
The nominal operating field in MICE is 5\,T, corresponding to
a current of 870\,A.

\begin{figure}
  \begin{center}
    \includegraphics[width=0.95\linewidth]%
    {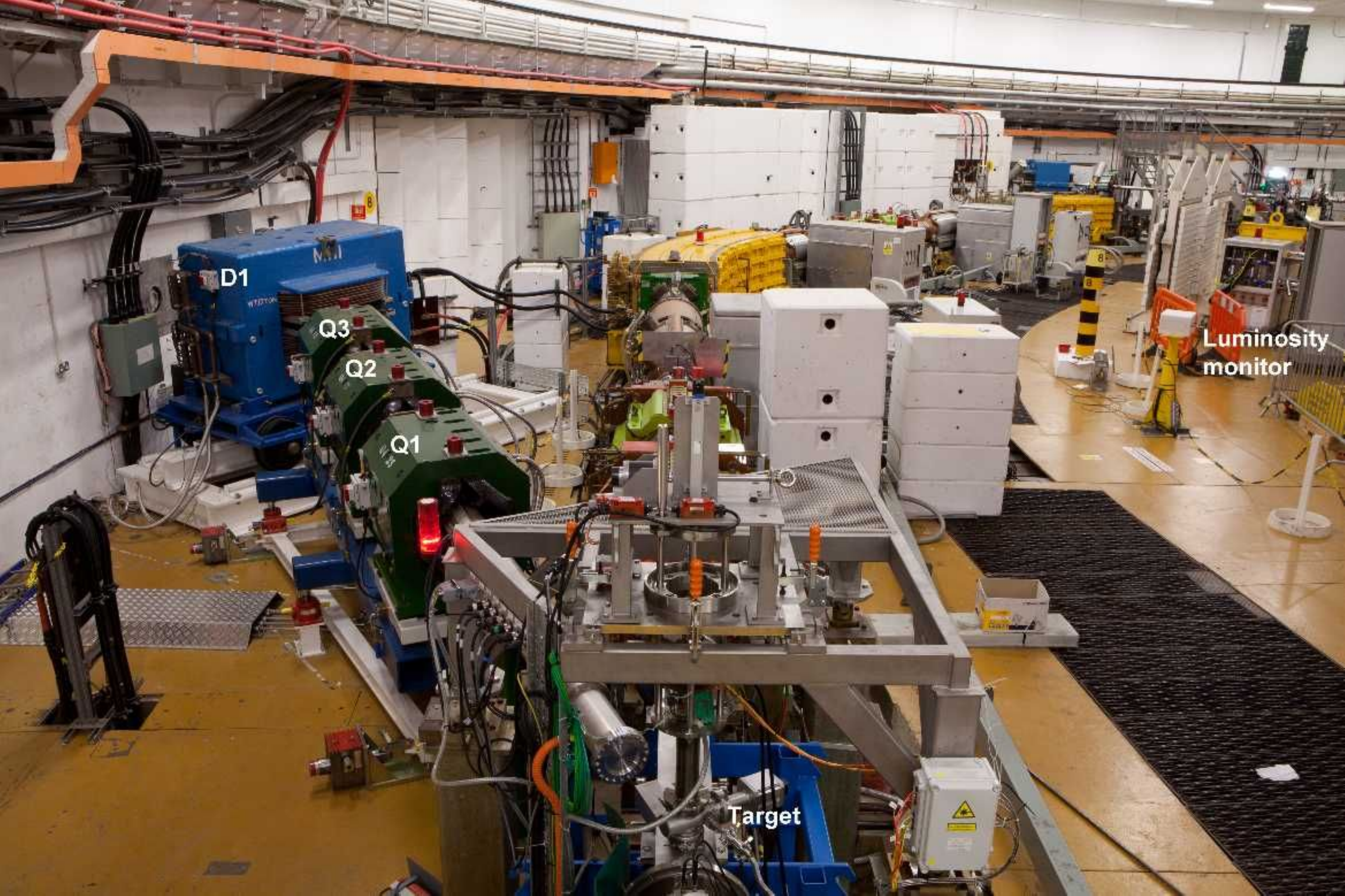}
  \end{center}
  \caption{Upstream beam line.  The target may be seen in the foreground of
  the photograph.  The magnets that make up the upstream beam line 
  (Q1--3 and D1) lie to the left of straight 7 of the ISIS proton synchrotron 
  which can be seen in the centre of the figure.  The luminosity monitor, 
  surrounded by its white, borated-polythene shielding may be seen close to the   right-hand edge of the photograph.}
  \label{UpstreamBeamLine}
\end{figure}

\begin{figure}
  \begin{center}
    \includegraphics[width=0.95\linewidth]%
    {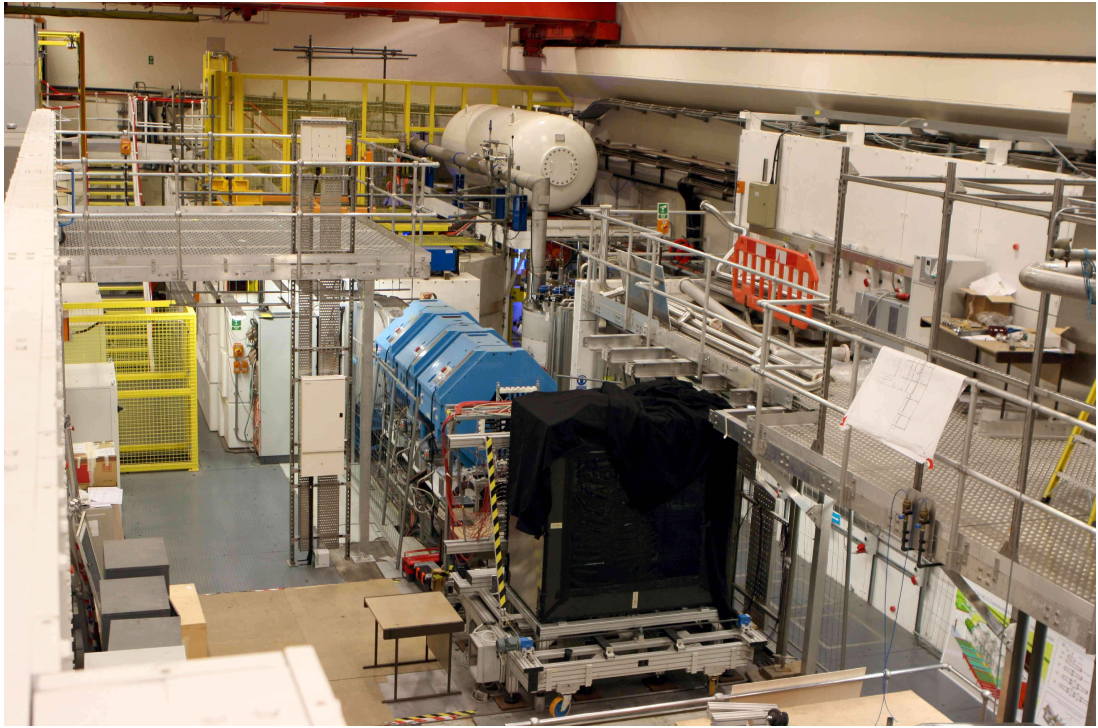}
  \end{center}
  \caption{
     Downstream beam line.  The ISIS synchrotron lies behind the curved
wall in the background.  The downstream quadrupoles (blue, hexagonal
magnets) may be seen just downstream of the white concrete shielding.  The
time-of-flight counter TOF1 is seen immediately downstream of the final
quadrupole.  TOF1 is followed by the KL and the prototype EMR.  
%%In the foreground of the picture, the prototype liquid-hydrogen delivery 
%%system and R\&D cryostat may be seen.
}
  \label{DownstreamBeamLine}
\end{figure}

A summary of the position and dimensions of all the components along the 
MICE beam line is presented in table~\ref{tab:blbrkdown}. 
A photograph of the upstream section of the beam-line is shown in
figure \ref{UpstreamBeamLine} and of the downstream beam
line in figure \ref{DownstreamBeamLine}.
\begin{table*}[hbt] \small
\begin{center}
\caption{The MICE beam-line elements and detectors for Step I.}
\vspace{.2cm}
\begin{tabular}{|c|c|c|c|c|c|c|c|}
\hline
 Element  & Distance from target                 & $L_{e\!f\!f}$ & \mc{2}{c|}{Max
 field/gradient} & Aperture & \mc{2}{c|} {$1/2$-aperture}  \\
   & \scriptsize{[along nominal } & & \mc{2}{c|}{} & Radius & \mc{2}{c|}{
\scriptsize{[H/V]}} \\
   & \scriptsize{beam axis]} & & \mc{2}{c|}{} &\scriptsize{(Pole tip)} &  \mc{2}{c|}{} \\
\hline
\mc{1}{|c|}{} & (mm) & (mm) & (T) & (T/m) & (mm) & mm  & mm\\
\cline{2-8}
\hline
 Q1  & 3000.0 & 888  & - & 1.6 & 101.5 & & \\
 Q2  & 4400.0 & 888  & - & 1.6 & 101.5 & & \\
 Q3  & 5800.0 & 888  & - & 1.6 & 101.5 & & \\
\cline{7-8}
 D1  & 7979.1  & 1038 & 1.6 & - &  & 254.0 & 76.0 \\
\cline{7-8}  \hline
\mc{1}{|c|}{Decay Solenoid}  & 12210.7 & 5000 & 5.7 & - & 57.5 &  & \\ \hline
\mc{1}{|c|}{Proton absorber} & 14880 & \mc{6}{|c|}{ Plastic sheets 15, 29, 49, 54 mm} \\ \hline
\mc{1}{|c|} {GVA1} & 15050.0 &
\mc{6}{|c|}  { Scintillation counter (0.02 X$_0$)} \\ \hline
 D2  & 15808.1 & 1038 & 0.85 & - &  & 254.0 & 76.0   \\ \hline
\mc{1}{|c|} {BPM1} & 16992.0 &
\mc{6}{|c|}  { Scintillating fibres (0.005 X$_0$)} \\ \hline
 Q4  & 17661.6 & 660 & - & 2.3 & 176.0 &  & \\
 Q5  & 18821.6 & 660 & - & 2.3 & 176.0 &  & \\
 Q6  & 19981.6 & 660 & - & 2.3 & 176.0 &  & \\ \hline
% TOF0 } &  & - & - &  - \\ \hline
\mc{1}{|c|} {TOF0} & \mc{1}{c|}{21088.0} &
\mc{6}{c|}  {Segmented scintillator (0.12 X$_0$)} \\ \hline
\mc{1}{|c|} {Ckova} & 21251.5 &
\mc{6}{c|}  { Aerogel threshold Cherenkov (0.019 X$_0$)} \\ 
\mc{1}{|c|} {Ckovb} & 21910.9 &
\mc{6}{c|}  { Aerogel threshold Cherenkov (0.031 X$_0$)} \\ \hline
\mc{1}{|c|} {BPM2} & 24293.7 &
\mc{6}{|c|}  { Scintillating fibres (0.005 X$_0$)} \\ \hline
 Q7  & 25293.7 & 660 & - &  2.3 & 176.0 &   & \\
 Q8  & 26453.7 & 660 & - &  2.3 & 176.0 &   & \\
 Q9  & 27613.7 & 660 & - &  2.3 & 176.0 &   & \\ \hline
% TOF1} &  & - & - &  - \\ \hline
% KL }     &  & - & - &  - \\ \hline
\mc{1}{|c|} {TOF1} & \mc{1}{c|}{28793.1} &
\mc{6}{c|}  {Segmented scintillator (0.12 X$_0$)} \\ \hline
\mc{1}{|c|} {TOF2} & \mc{1}{c|}{31198.1} &
\mc{6}{c|}  {Segmented scintillator (0.12 X$_0$)} \\ 
\mc{1}{|c|} {KL} & \mc{1}{c|}{31323.1} &
\mc{6}{c|}  {Lead + scintillator (2.5 X$_0$)} \\
\mc{1}{|c|} {Tag counters} & \mc{1}{c|}{31423.1} &
\mc{6}{c|}  {Scintillation bars  (0.06 X$_0$)} \\ 
\hline 
\end{tabular}
\label{tab:blbrkdown}
\end{center}
\end{table*}

The MICE beam line works with both positively and negatively
charged muons. The change\-over is achieved by physically disconnecting 
and swapping over the magnet power-supply
cables. A simple system for monitoring and recording the operating 
polarity of the
two bending magnets (and by inference the other beam-line elements) has
been deployed \cite{nebre} consisting of a pair of Honeywell
LOHET-II Hall-effect
sensors that are operating past saturation when placed in the dipole bores,
even at minimal field strengths, and thus return one of the two well-defined
voltages corresponding to the two possible polarities of each magnet.

\subsubsection{The proton absorber}

With the beam line set to transport positive particles, a large flux
of protons is observed to exit the decay solenoid.
The range of proton momenta selected by D1 is such that a large
fraction of the protons reaching D2 will be accepted by Q4--6.
As a result, unless action is taken to remove protons from the beam as
it enters D2, an unacceptably large flux of protons would be
transported to TOF0.
In the momentum range of interest, the energy loss per unit length of
material traversed is larger for protons than for pions. 
Borated-polyethylene sheets placed in the beam
as it emerges from the decay solenoid are used to remove the protons.
The boron loading of the polyethylene serves to increase the absorption
rate of neutrons.
The sheets have thicknesses of 15\,mm, 29\,mm, 49\,mm and 54\,mm and
are raised and lowered into the beam using a cable-pull mechanism. 
The absorbers can be operated together or independently. 
When all four absorbers are placed in the beam, the 147\,mm of material
is enough to stop protons with momentum up to $\sim 500$\,MeV/c.

The choice of absorber thickness depends on the required muon-beam
momentum.
In a beam line set to deliver a nominal muon momentum of 140\,MeV/c,
D1 is set to select 360\,MeV/c particles. 
In this case, a proton-absorber thickness of 44\,mm is required to
reduce the proton contamination in the muon beam delivered to MICE to
less than 1\%.
The beam-line tune that delivers the  nominal muon momentum of
240\,MeV/c requires that D1 be set to select particles with a
momentum of 507\,MeV/c, and the full 147\,mm proton-absorber thickness
is then required to reduce the proton contamination to the few-\% level.
Pure positive muon beams of intermediate momentum are delivered using
appropriate combinations of the four sheets.

\subsubsection{Diffuser}

\begin{figure}[ht!]

  \begin{center}
    \includegraphics[width=0.70\linewidth]%
    {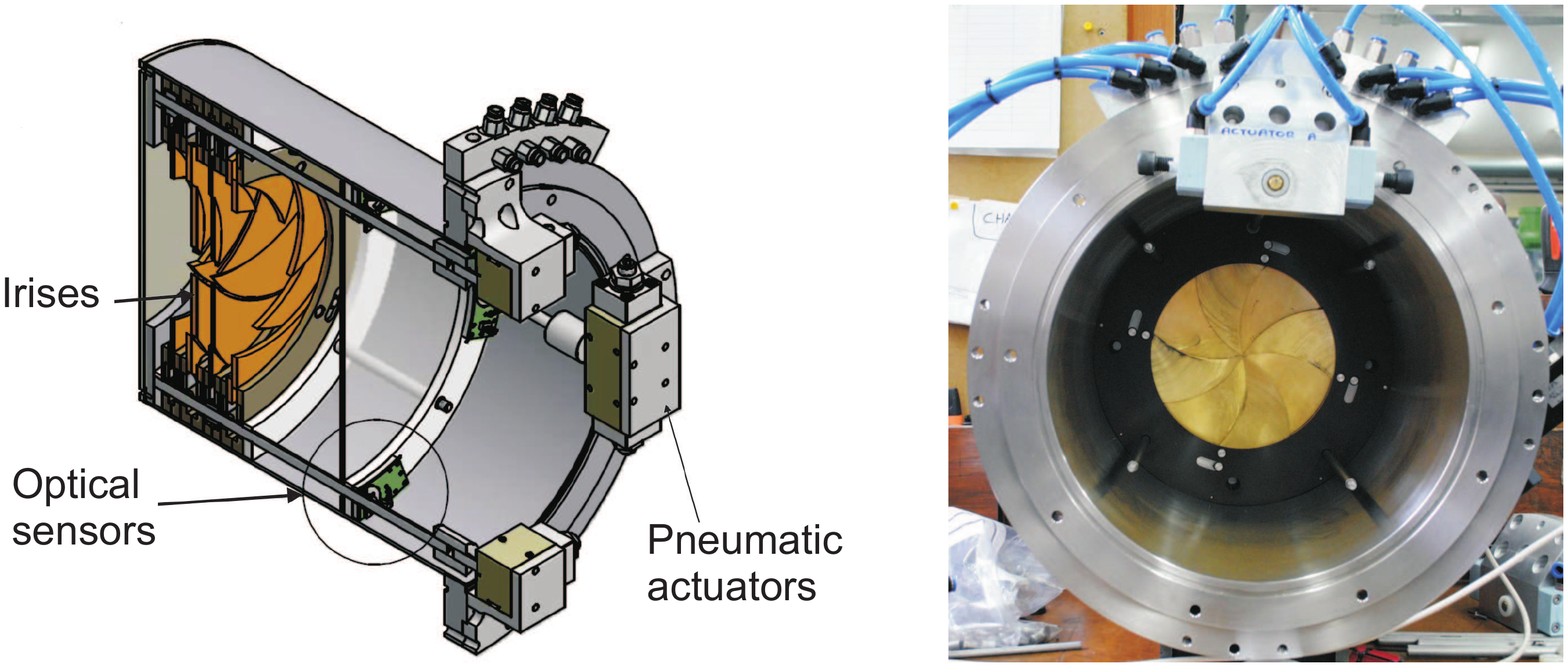}
  \end{center}
  \caption{
    Left panel: schematic of the MICE diffuser showing the irises. 
    Right panel: one of the brass irises being tested.
  }
  \label{diffuser}
\end{figure}

The final element of the downstream section of the beam line is the
diffuser.
By introduction of material with a large radiation length into the beam,
multiple Coulomb scattering increases the divergence of the beam,
enlarging the emittance and providing an approximately matched beam in
the upstream spectrometer solenoid.
Since the diffuser must operate in the field of the 4\,T spectrometer,
electric motors and other magnetic components cannot be used.
The diffuser design consists of a stainless steel drum, which is
inserted into the upstream section of the first spectrometer solenoid
\cite{PAC11_diffuser}. 
The drum contains four irises, two of brass (2.97\,mm,
0.2\,X$_0$, and 5.94\,mm, 0.4\,X$_0$ thick) and two  of tungsten
(2.80\,mm, 0.8\,X$_0$, and 5.60\,mm, 1.6\,X$_0$). 
By selecting different combinations of irises, one can achieve a total
of 3\,X$_0$ in steps of 0.2\,X$_0$. 
Each iris consists of two planes of four segments of material,
supported by a Tufnol~\cite{tufnol} ring of radius 100\,mm
(see figure~\ref{diffuser}) and is operated by a non-magnetic,
air-driven actuator.
The position of each iris is monitored by a set of optical sensors. 
When closed, the iris inserts solid petals into the beam; when it
is open, the petals are stowed away within the Tufnol ring.   

\subsubsection{Beam-line diagnostics}
\label{sec:diagnostic}

The downstream beam line is instrumented with a scintillation counter
(GVA1), placed just after the proton absorber,  upstream of D2.
GVA1 consists of a single slab of scintillator with 
$18 \times 18$\,cm$^2$ active area and 1\,cm thickness, read out by a
2\,inch EMI photomultiplier.  
Two Beam Profile Monitors mounted on the upstream faces of Q4 (BPM1)
and Q7 (BPM2) were used to measure particle rate and beam position.
The BPMs each comprise two planes of Kuraray 1\,mm scintillating
fibres.
BPM1 has an area of $20 \times 20$\,cm$^{2}$, while BPM2 has an area 
of $45 \times 45$\,cm$^{2}$.
The layers of optical fibres are supported by a sheet of
Rohacell~\cite{rohacell} foam 2.55\,mm thick. 
An aluminium box encases the BPMs, except in the acceptance region of
the fibres where four layers of Tedlar~\cite{tedlar} keep the
BPM light-tight. 
The light from the fibres is read out by two 64-channel multi-anode
Burle~\cite{burle} 85011-501 photomultiplier tubes (one per plane).
Further details on both GVA1 and the BPMs are found in
table \ref{tab:blbrkdown}.   
\subsection{Optimisation of the beam line}

\begin{figure}
  \centering
  \begin{tabular}{cc}
    \includegraphics[width=0.49\linewidth]%
    {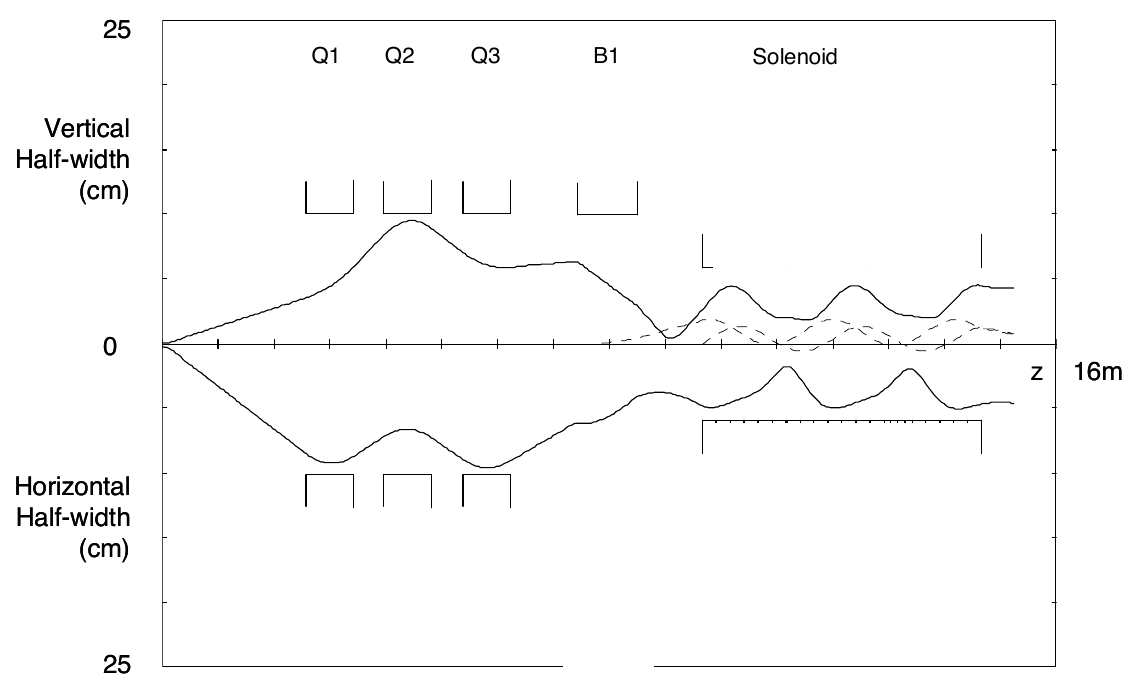} &
    \includegraphics[width=0.490\linewidth]%
    {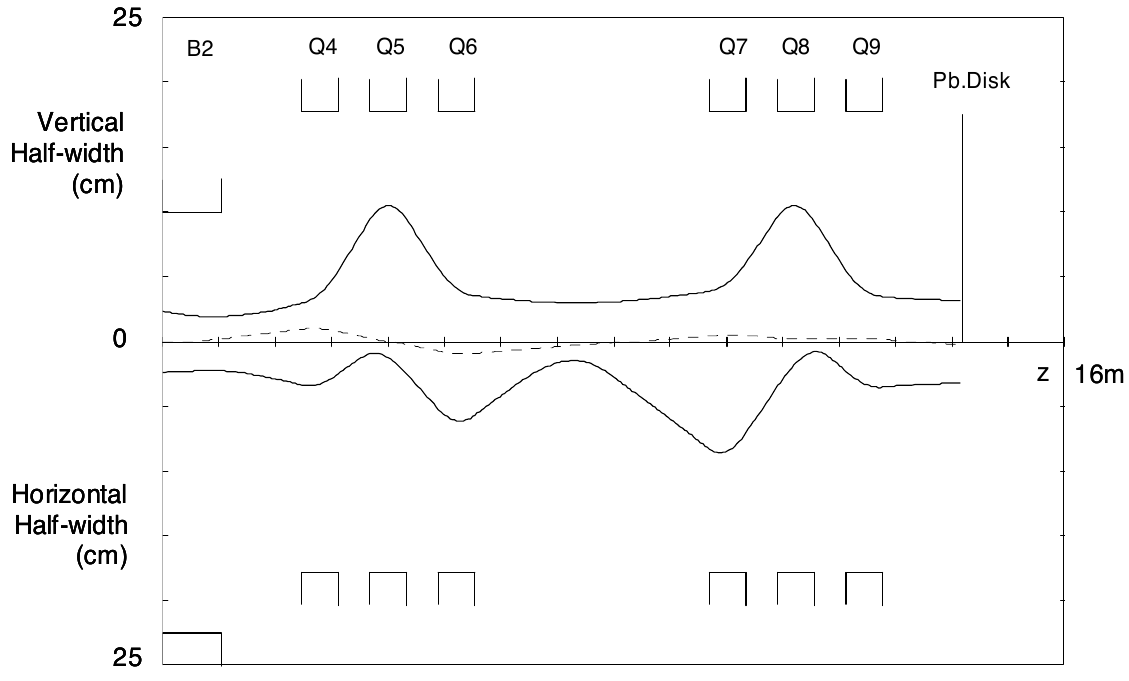}
  \end{tabular}
  \caption{MICE beam-line envelope for the baseline case with 
  $\epsilon_{N} = 6 \pi$ mm $\cdot$ rad and $p_{z}=200$ MeV/c. Left panel:
  pion optics for the upstream section. Right panel: muon
  optics for the downstream section. The lead scatterer (``diffuser'')
  used in this simulation for practical reasons has been realized using 
  brass and tungsten. 
  For both panels the top half of the plot shows the vertical envelope 
  and the bottom half shows the horizontal envelope.} 
  \label{fig:BL_optics}
\end{figure}

The upstream and downstream sections of the beam line are loosely
coupled, allowing each section to be optimised independently.
In the upstream section, where pions travel through an evacuated beam
pipe, the matrix-evolution code TRANSPORT \cite{TRANSPORT,Rohrer} was
therefore used to perform the optical design.
In the decay solenoid, pion decay must be simulated and in the
downstream beam line, where there is no beam pipe, particle
interactions in air and in the detectors must be taken into account.
Therefore, the optics of the decay solenoid and the downstream beam
line were developed using Decay-TURTLE \cite{Rohrer,TURTLE}.
The result of these simulations are illustrated in figure
\ref{fig:BL_optics} \cite{Tilley:2006zz}.
The left-hand panel shows the 1\,$\sigma$ beam envelope obtained using
TRANSPORT for a 454\,MeV/c pion beam generated at the target. 
The upstream quadrupoles (Q1--3) are configured in a
focusing-defocusing-focusing (FDF) configuration in the horizontal plane 
(DFD in the vertical plane).
The particle rate is limited by the aperture of Q2 in the vertical
plane and by the apertures of Q1 and Q3 in the horizontal plane. 
The right-hand panel of figure \ref{fig:BL_optics} shows the
1\,$\sigma$ envelope obtained using TURTLE to track a sample of muons
($p_{\mu} = 238$\,MeV/c at D2) towards the diffuser.
This muon beam was optimised to deliver muons that would have a
nominal momentum of 200\,MeV/c and a nominal emittance of
6\, $\pi$ mm $\cdot$ rad. 
The optical designs were then further refined using two GEANT4-based
\cite{Agostinelli:2002hh,GEANT4} simulation codes: G4beamline
\cite{Roberts:2008zzc,G4Beamline} and G4MICE
\cite{Rogers:2006zz,Ellis:2007zz}.

\noindent
Pion transmission in the upstream beam line was maximized by increasing
the Q1-3 excitation of $\sim 10 \%$ with respect to that obtained from
a TRANSPORT simulation.

\subsubsection{Downstream optimisation}
\label{sec:dsoptim}

%The beam emerging from the decay solenoid enters  the
%14\,m-long downstream section.
The beam optics are tuned for the experiment set up in the Step VI
configuration.
The downstream beam line is designed to provide an optically matched
beam of desired nominal emittance $\epsilon_{N}$ and nominal momentum
$p_z$.
The parameters available for tuning are:
\begin{enumerate}
  \item The thickness of the diffuser, which is used to tune the beam
        divergence in the upstream spectrometer solenoid;
  \item The quadrupole currents, which are used to tune the beam size and
        divergence, i.e., the incoming Twiss parameters 
        ($\beta^{\rm in}_x, \beta^{\rm in}_y$ and 
        $\alpha^{\rm in}_x, \alpha^{\rm in}_y$) at the upstream face
        of the diffuser;
  \item The nominal momentum,  which can be adjusted with the excitation of
        D2.
\end{enumerate}
Ideally, the conditions to be fulfilled simultaneously are:
\begin{enumerate}
  \item The beam is matched in the upstream spectrometer solenoid,
        downstream of the diffuser.
        The matching conditions are: 
        $\beta_x = \beta_y= \beta = 2 p_z{\rm [GeV/c]}/(0.3 B{\rm [T]})$ and
        $\alpha_x = \alpha_y= \alpha = 0$;
  \item The beam size at the entrance of the diffuser (essentially the
        same as the beam size at the exit of the diffuser) is such
        that the rms width of the beam in $x$ and $y$ satisfies 
        $\sigma^2_{x} = \sigma^2_{y} = (\epsilon_N \beta) / \gamma$,
        with $\gamma = E/m_\mu$;
  \item The beam angular divergence at the upstream surface of the
        diffuser ($\sigma^2_{x'}$ and $\sigma^2_{y'}$), increased by
        multiple Coulomb scattering in the diffuser
        ($\sigma^2_{\theta,{\rm MCS}}$), is equal to the desired
        angular divergence in the spectrometer solenoid, i.e.:
        $\sigma^2_{x'} + \sigma^2_{\theta,{\rm MCS}} = 
         \sigma^2_{y'} + \sigma^2_{\theta,{\rm MCS}} = 
         \epsilon_N / (\beta \gamma)$;
  \item The central momentum selected by D2 matches the nominal
        momentum $p_z$, once corrected for the calculated energy loss
        in all material encountered by the beam as it travels from D2
        to the central absorber.
\end{enumerate}
The horizontal beam dispersion is neglected in the calculations.  
It is not possible to fulfil all conditions simultaneously because
the emittance of the beam coming out of D2, and transported in the
quadrupole channel, is different in the horizontal and vertical
planes.  
Nevertheless, an average match can be obtained by using the 4D
covariance matrix, the resulting 4D emittance and the Twiss 
parameters of the beam at the upstream face of the diffuser, and
transporting the beam through the diffuser.
That the beam is only approximately matched results in a somewhat
elliptical beam in the spectrometer solenoid; however this affects the
cooling performance only marginally.  

Beam settings for various combinations of ($\epsilon_N$,$p_z$) are
calculated for a matrix of values $\epsilon_N = (3,6,10) \pi$\,mm $\cdot$ rad and
$p_z = (140, 200, 240)$\,MeV/c.
The calculation procedure \cite{Apollonio:2009zz} starts by fixing the 
emittance, the Twiss
parameters and $p_z$ inside the spectrometer solenoid at the downstream face of
the diffuser.
The Twiss parameters and the momentum are back-extrapolated through
the diffuser.
The diffuser thickness, $d$, and the currents in quadrupoles Q4--6 and
Q7--9 are varied to match the input beam, using G4beamline to simulate
the MICE downstream beam line.
The procedure is first applied to the reference point (6, 200).
This beam line is scaled with momentum to provide a first
approximation, M0, of each of the other points in the emittance-momentum 
matrix.
Then, each point is re-optimised to produce the final set of optimised
beam lines, M1.
The values of the Twiss parameters, diffuser thickness and momentum at
the upstream face of the diffuser for the nine points of the matrix are
given in table \ref{tab:matrix}.
\begin{table}
\centering
\caption{($\epsilon_N$,$p_z$) matrix for the MICE 
programme \cite{Apollonio:2010zza}. 
  The Twiss parameters ($\alpha,\beta$) are those required at the 
upstream face of the diffuser of thickness $d$ for a matched beam.}
\label{tab:matrix}
\vspace{2mm}
\begin{tabular}{cc|c|c|c|c}
\cline{3-5}
& & \multicolumn{3}{c|}{$p_z$ (MeV/c)} \\ \cline{3-5}
& & 140 & 200 & 240  \\ \cline{1-5}
\multicolumn{1}{|c|}{\mr{12}{*}{\begin{sideways} $\epsilon_N$ ($\pi$ mm $\cdot$ rad) \end{sideways}}} &
\multicolumn{1}{c|}{\mr{4}{*}{3}} &
\small{$d=0.0$}&\small{$d=0.0$} & \small{$d=0.0$} \\
\multicolumn{1}{|c|}{}&  & \small{$p_{dif\!f}$=151 MeV/c} & \small{$p_{dif\!f}$=207 MeV/c} &  \small{$p_{dif\!f}$=245 MeV/c} \\ 
\multicolumn{1}{|c|}{}& &\small{$\alpha$=0.2} & \small{$\alpha$=0.1} & \small{$\alpha$=0.1}\\ 
\multicolumn{1}{|c|}{}& &\small{$\beta$=56 cm} &\small{$\beta$=36 cm} & \small{$\beta$=42 cm}  \\ \cline{2-5}
\multicolumn{1}{|c|}{}&\multicolumn{1}{c|}{\mr{4}{*}{6}} &
 \small{$d=0.9X_0$}&\small{$d=1.3X_0$} &\small{$d=1.3X_0$} \\
\multicolumn{1}{|c|}{}& &\small{$P_{dif\!f}$=156 MeV/c} &\small{$P_{dif\!f}$=215 MeV/c } &\small{$P_{dif\!f}$=256 MeV/c}      \\
\multicolumn{1}{|c|}{}& &  \small{$\alpha$=0.3} &  \small{$\alpha$=0.2} & \small{$\alpha$=0.2}\\ 
\multicolumn{1}{|c|}{}& &  \small{$\beta$=113 cm}
&  \small{$\beta$=78 cm} &  \small{$\beta$=80 cm}  \\ \cline{2-5}
\multicolumn{1}{|c|}{}&\multicolumn{1}{c|}{\mr{4}{*}{10}} &
 \small{$d=1.8X_0$}&\small{$d=2.8X_0$} &\small{$d=2.8X_0$} \\
\multicolumn{1}{|c|}{}& &\small{$P_{dif\!f}$=164 MeV/c} & $P_{dif\!f}$=229 MeV/c & $P_{dif\!f}$=267 MeV/c     \\
\multicolumn{1}{|c|}{}& & \small{$\alpha$=0.6} &  \small{$\alpha$=0.4} &  \small{$\alpha$=0.3} \\ 
\multicolumn{1}{|c|}{}& & \small{$\beta$=198 cm}  &  \small{$\beta$=131 cm} & \small{$\beta$=129 cm}  \\ \cline{1-5}
\end{tabular}
\end{table}
\section{The particle identification system in MICE Step I}
\label{sec:instrumentation}

With the exception of the beam-line diagnostics  and the luminosity
monitor described in sections
\ref{sec:diagnostic} and \ref{sec:lumi}, the detectors installed along the 
beam line form
part of the particle identification (PID) system of the MICE
experiment.
The PID system upstream of the first spectrometer solenoid is composed
of two time-of-flight (TOF) stations (TOF0 and TOF1)
\cite{Bertoni:2010by} and two threshold Cherenkov counters (Ckova and
Ckovb) \cite{Cremaldi:2009zj}.
Together the two Cherenkov counters will provide $\pi/\mu$ separation
up to 365\,MeV/c.
TOF0, Ckova and Ckovb are inside the Decay Solenoid Area (DSA).~\footnote{The DSA is a closed
area within the MICE Hall, just outside the ISIS vault area,
that contains the DS and the first particle identification detectors.}

The TOF system is required to reject pions in the incoming muon beam
with an efficiency in excess of 99\%.
In addition, the precision of the TOF time measurement must be
sufficient to allow the phase at which the muon enters the RF cavities
to be determined to 5$^\circ$.
To satisfy these requirements, the resolution of each TOF station must
be $\sim 50$\,ps.
The two Cherenkov detectors have been designed to guarantee
muon-identification purities better than $99.7$\% in the momentum
range 210\,MeV/c to 365\,MeV/c \cite{Sanders:2009vn}.
At lower momenta, $\pi/\mu$ separation is obtained using the TOF
measurement, both Cherenkov detectors being blind to both particle
types.

\begin{figure}
  \begin{center}
    \includegraphics[width=0.54\linewidth]%
      {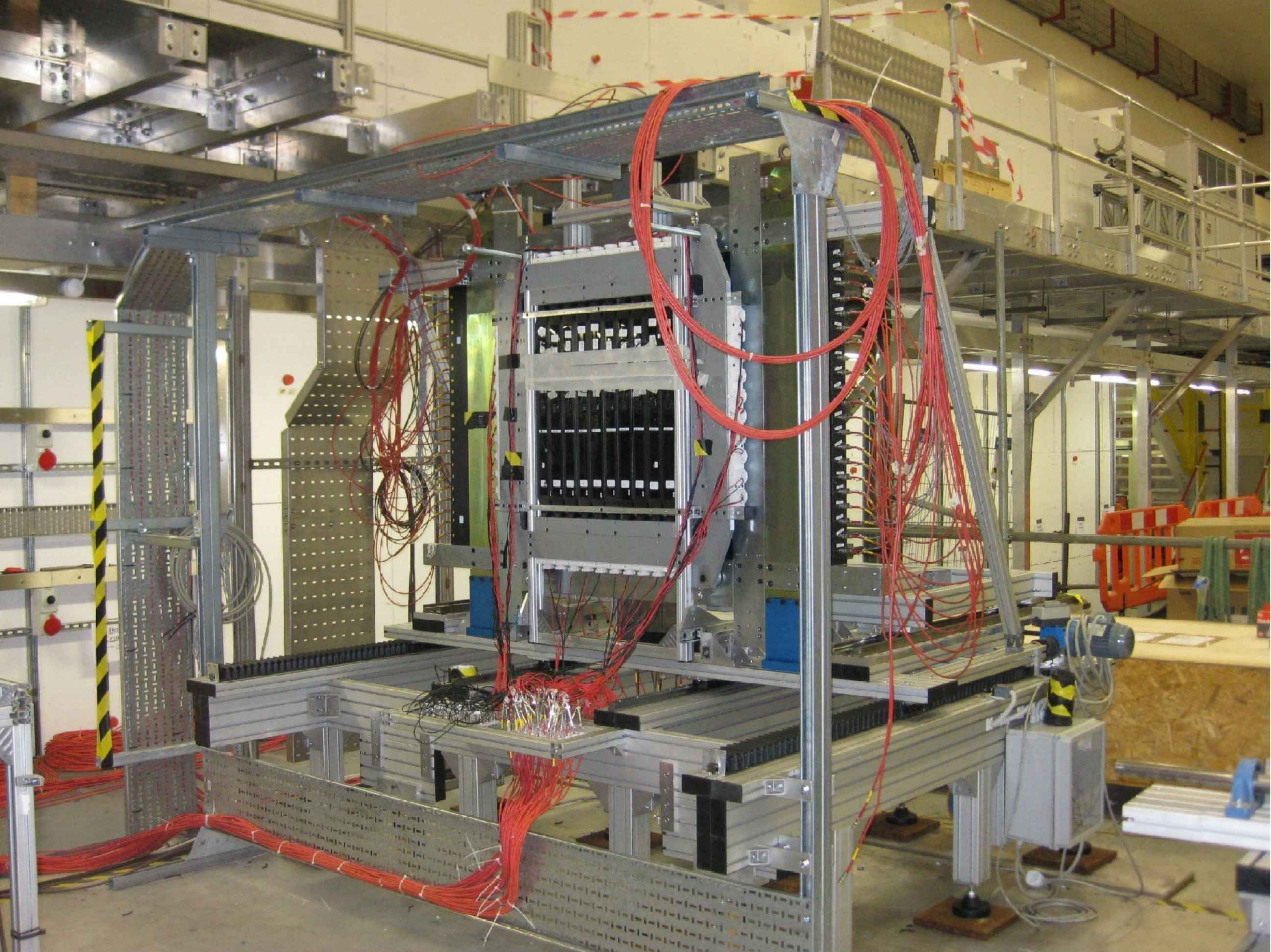}
    \includegraphics[width=0.40\linewidth]%
      {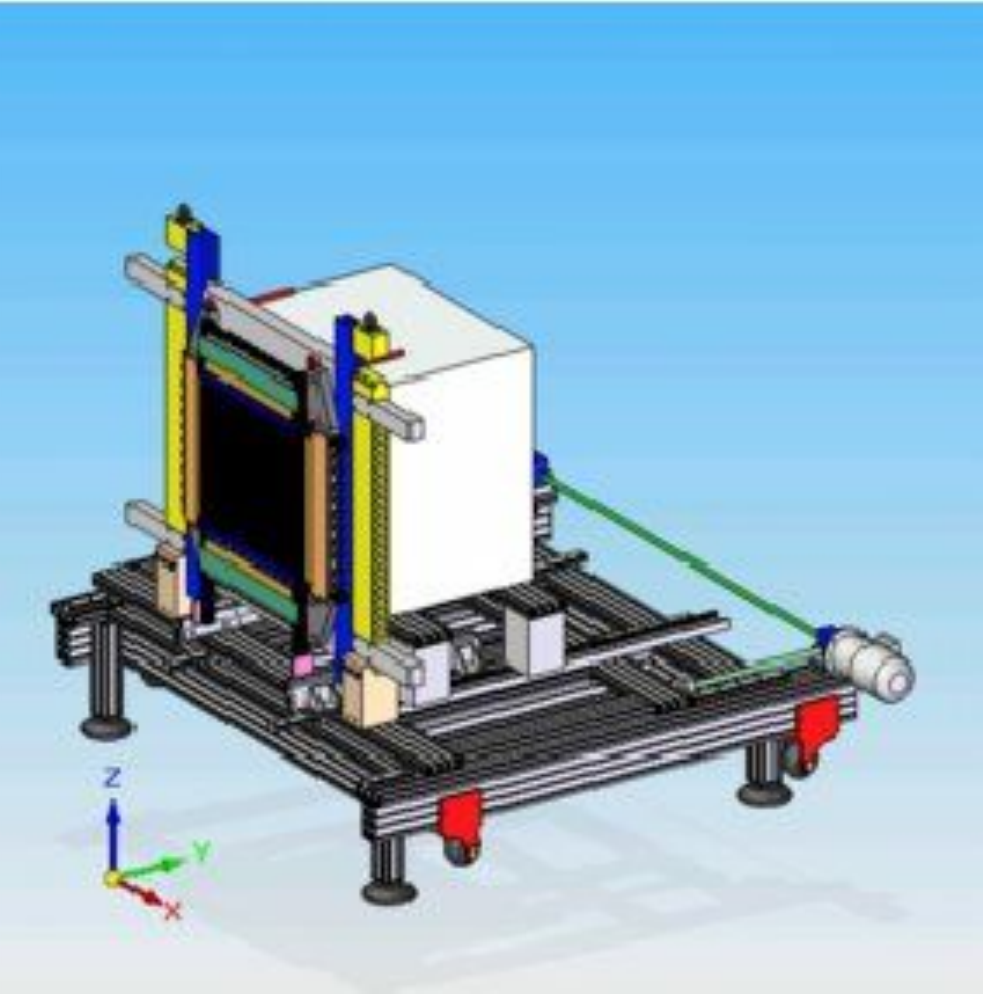}
  \end{center}
  \caption{
    TOF2 is shown in front of  KL on the final downstream platform
    (left panel) and the downstream PID section including TOF2, KL and
    EMR on their platform (right panel).
  } 
  \label{fig:down}
\end{figure}

The identification of particles downstream of the second spectrometer
solenoid is provided by a third TOF station (TOF2) \cite{tof2} 
and an electromagnetic calorimeter (EMC) by which muons may be
distinguished from electrons (or positrons) produced in muon decays
downstream of TOF1.
The electromagnetic calorimeter consists of two parts:  ``KLOE
Light'' (KL), a lead-scintillator device based on the KLOE
calorimeter design \cite{Ambrosino:2009zza}; and the ``Electron-Muon
Ranger'' (EMR), a fully active, scintillator calorimeter
\cite{Lietti:2009zz}).
The left-hand panel of figure \ref{fig:down} shows the positions of
TOF2 and KL in the MICE muon beam line during Step I data-taking.
The right-hand panel shows a CAD drawing of the downstream PID
detectors (TOF2, KL, EMR) on their platform.
The Ckov, TOF0, TOF1, TOF2 and KL detectors were installed in stages
in the MICE Hall.
The EMR was not available for Step I data-taking.

\subsection{The Cherenkov counters}
\label{sec:Cherenkov}

\begin{figure}
  \begin{center}
    \includegraphics*[width=.65\textwidth]%
      {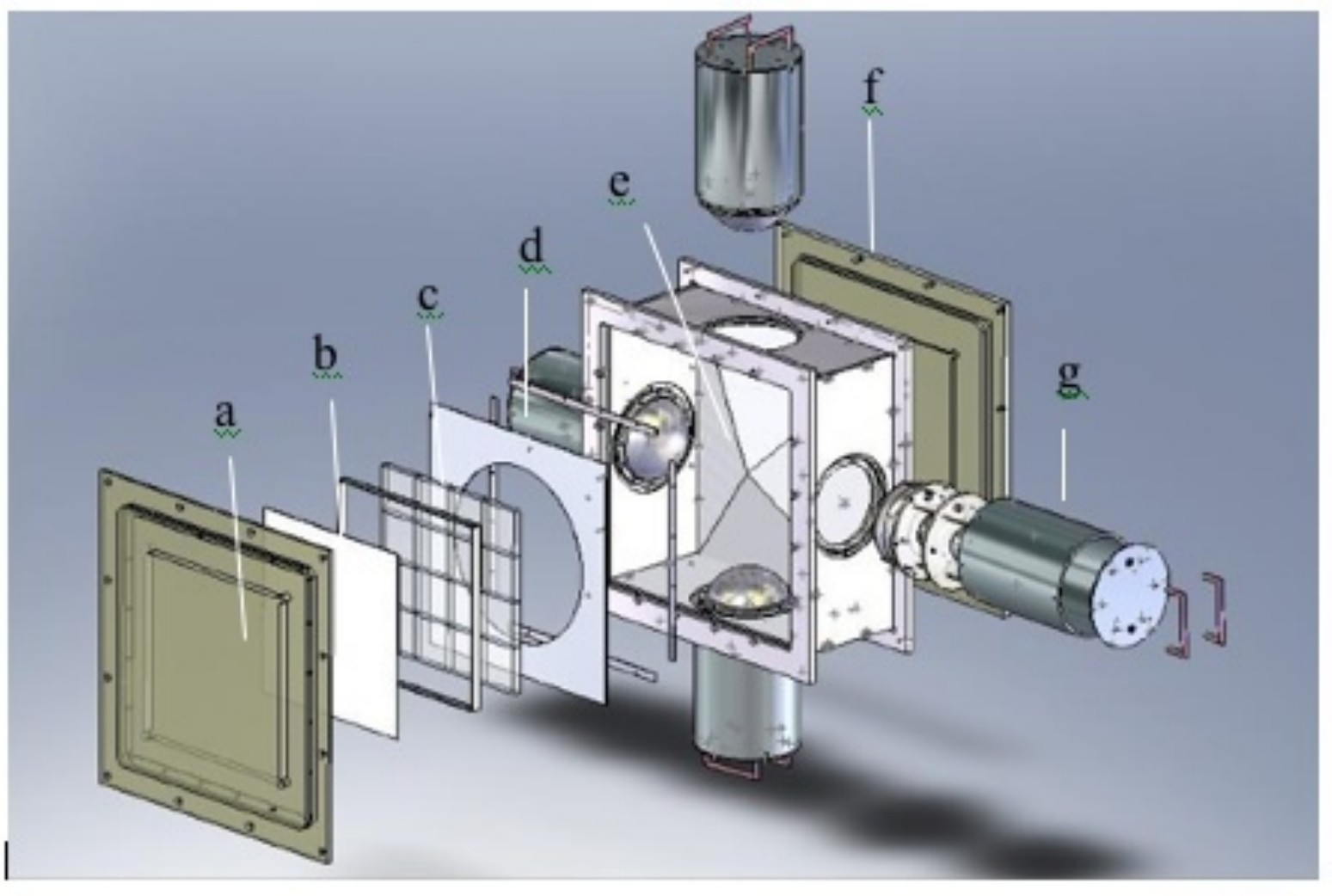}
  \end{center}
  \caption{
    Aerogel Cherenkov counter blowup: a) entrance window, b) mirror,
    c) aerogel mosaic, d) acetate window, e) GORE reflector panel, 
    f) exit window and g) 8 inch PMT in iron shield.  
  }
  \label{ckv_blowup}
\end{figure}

Two aerogel Cherenkov counters are employed
\cite{Cremaldi:2009zj}.
The muon threshold in the upstream counter, Ckova, is set at 
$p_{\mu,a}^{th} = 278$\,MeV/c while the muon threshold for Ckovb is set
at $p_{\mu,b}^{th} = 210$\,MeV/c.
Pions thresholds are $p_{\pi,a}^{th} = 367$\,MeV/c and 
$p_{\pi,b}^{th} = 277$\,MeV/c for Ckova and Ckovb respectively.
The densities of the aerogels are $\rho_{a} = 0.225$\,g\,cm$^{-3}$ and 
$\rho_{b} = 0.370$\,g\,cm$^{-3}$, with indices of refraction 
$n_a = 1.07$ and $n_b=1.12$. 
In figure \ref{ckv_blowup} an exploded view of one counter is shown.
The aerogel tiles are two layers thick (2.3~cm in total) and cover an area of
$46 \times 46$\,cm$^2$.
Four 8\,inch EMI 9356KB PMTs collect the Cherenkov light in each
counter.
Due to the high particle rate, the
digitisation of the pulse profile is performed using a very high
frequency sampling digitiser,  CAEN V1731~\cite{CAEN} (1\,GS/s maximum sampling
rate).
The sampling digitiser is connected directly to the PMTs through a
coaxial cable.

For the 140\,MeV/c beams, both pions and muons are below threshold for
both of the Cherenkov counters. 
For 200\,MeV/c beams, pions are below threshold for both Ckova and
Ckovb, while muons are above threshold only for Ckovb. 
For 240\,MeV/c beams, pions are above threshold for Ckovb while muons
are above threshold for both Ckova and Ckovb. 
Algorithms are being written to use the information from both counters
to provide  $\pi,\mu, e$ likelihoods.
The likelihood information will be combined with the TOF measurement.
Only the TOF system was used for PID in the Step I data analysis presented in
the present paper.

\subsection{The TOF detectors}

\begin{figure}
  \begin{center}
    \includegraphics[width=.49\textwidth]%
      {FigRepository/TOF1_View_2.pdf}
    \includegraphics[width=.49\textwidth]%
      {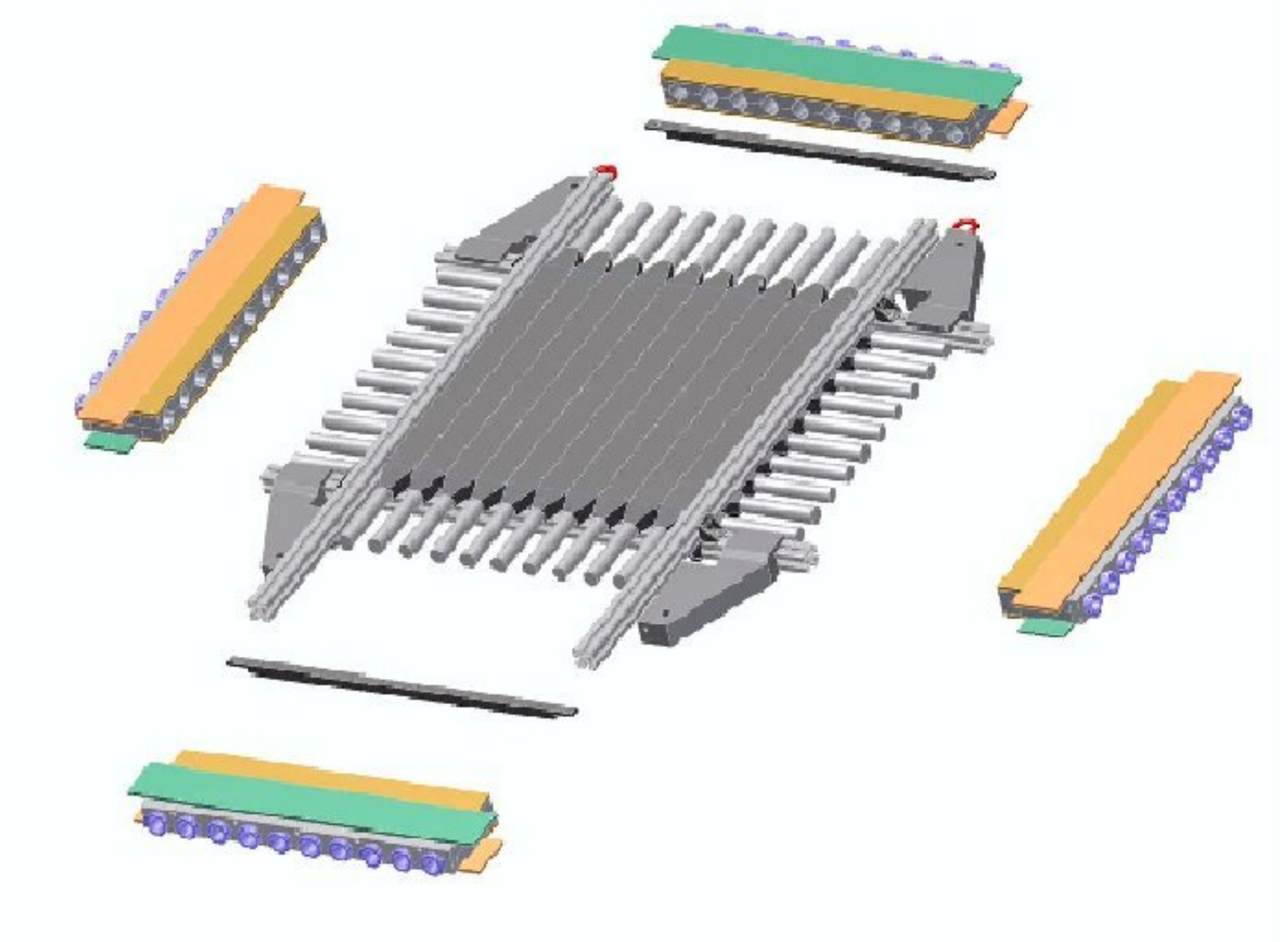}
  \end{center}
  \caption{ 
    Shielding used for TOF PMTs. 
    Left panel: TOF1 global cage showing, A) extraction brackets,
    B/E) shielding plates, C) TOF1 detector in working position, 
    D) rails to move TOF1 outside the shielding cage.
    Right panel: TOF2 local shielding with massive soft iron
    boxes.
  }
  \label{fig:shield}
\end{figure}

All TOF stations share a common design based on 1\,inch thick slabs of
fast scintillator material.
Bicron BC-420 scintillator \cite{Bicron}  was used for TOF0 and BC-404 was used for
TOF1 and TOF2.
The slabs in each station form two planes with {\it x} and {\it y}
orientations to increase measurement redundancy.
Each slab is read out at each end by a Hamamatsu R4998 fast
photomultiplier (rise time $\leq 1$\,ns). 
The active areas of TOF0 ,TOF1 and TOF2  are $40 \times 40$\,cm$^2$, $42 \times 42$\,cm$^2$ and  
$60 \times 60$\,cm$^2$, respectively.
The strip width is 4\,cm for TOF0 and 6\,cm for the other two
stations.
All downstream PID detectors and the TOF1 station will be shielded against
stray magnetic fields of up to 1300\,G (with a $\leq 400$\,G longitudinal
component) due to the presence of the spectrometer solenoids.
TOF1 will be shielded by a double-sided iron cage which fully contains
the detector.
The iron cage has a hole for the beam. 
The TOF2 and KL PMTs are shielded by soft iron boxes
\cite{Bonesini:2011}, as shown in Figure \ref{fig:shield}. 
In MICE Step I, only the local PMT shields for TOF2 and KL were
installed.

The TOF stations must sustain a high instantaneous particle rate (up
to 1.5\,MHz for TOF0).  
The rate capabilities of the R4998 PMT were tested in the laboratory
with a dedicated setup based on a fast laser \cite{Bertoni:2010by}.
The PMT rate capability was increased through the use of an active
divider base. 
These signals, after a splitter, are sent to a leading-edge LeCroy
4415 discriminator followed by a CAEN V1290 TDC for time
measurement.
The signal is also sent, after an RC shaper, 
 to a sampling digitiser CAEN V1724 FADC
(100\,MS/s maximum sampling rate) in order to measure the pulse height for
time-walk correction. 

The passive splitter is designed to match the impedance of the
50\,$\Omega$ coaxial cable, coming from the PMTs, with the
120\,$\Omega$ discriminator and shaper inputs.  
The shaping circuit is used to  extend the duration of the short
PMT pulse, so that it can be finely sampled by the digitiser. 
Software processing of the digitised pulse is needed for the charge
measurement and time-walk correction.

The signal arrival time, $t_{i} \ (i=1,2)$, at the photo-cathode of the
left/right PMT of a scintillator slab  of a TOF detector plane  is
given by:
\begin{eqnarray}
  t_{i}=t_0 + \frac{L/2 \pm x}{v_{e\!f\!f}} + \delta_{i} \, ;
  \label{eq:deltat}
\end{eqnarray}
where $t_0$ is the particle arrival time, $x$ its distance from the
counter centre, {\em L} the scintillator length, $v_{e\!f\!f}$ the
effective velocity of light in the scintillator slab and
$\delta_{i}$ includes all time delays (in cables, PMT transit
time, etc.). 
The transverse impact position, $u$, of a particle on a TOF station
 may be reconstructed from the difference between the time
measurements from the two PMTs $i,j$ at the ends of a counter  as:
\begin{eqnarray}
  u = \frac{v_{ef\!f}}{2} \times ((t_{i} - \delta_{i})
      - (t_{j} - \delta_{j})) \, .
  \label{eq:deltax}
\end{eqnarray}
Transformation to the MICE coordinate system is straightforward: for
vertical planes $x=u$ and for horizontal planes $y=u$.
The measured weighted average for $v_{ef\!f}$ is 
$13.52 \pm 0.30$\,cm/ns giving a spatial resolution 
$\sigma_x = \sigma_y = \sqrt{2} \times v_{eff} \times \sigma_{t} \approx 1$\,cm, 
with $\sigma_t$ the time resolution for the TOF station in question. 

The calculation of the delays $\delta_{i}$ (the ``time
calibration'') is performed using Step I data as described in detail
in \cite{Bertoni:2010by} and \cite{calib}.

The use of leading-edge discriminators in the time-measurement
electronics causes the threshold-crossing time to depend on the
quantity of charge collected, an effect referred to as ``time walk''. 
The difference between the time measured by each TDC and a reference
time was determined as a function of the maximum of the FADC signal
and used to correct for time-walk.
Pre-equalisation of  the amplitude response of the scintillation 
counters in each TOF plane was performed so that the time-walk
corrections for the two PMTs on a particular scintillator counter were
similar.
The pre-equalisation was performed using a YAP pulser unit from
SCIONIX \cite {SCIONIX}  yielding $\sim 20$ counts/s from a YAP:CE
scintillation crystal spot-activated with $^{241}$Am.
Pulse-height spectra were recorded in the laboratory and used to
derive appropriate high-voltage settings for the PMTs.

\begin{figure}
  \begin{center}
    \includegraphics[width=.49\linewidth]%
      {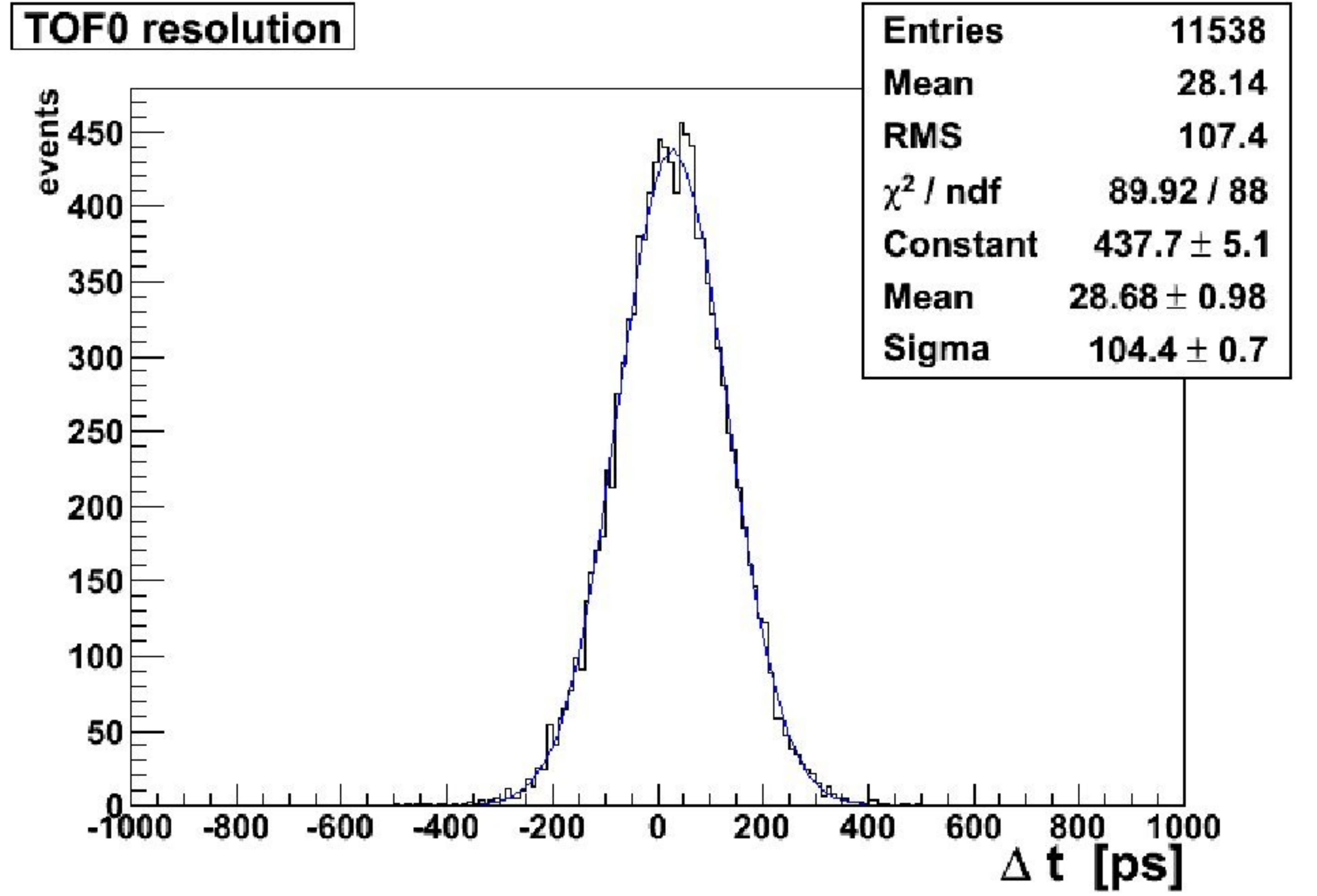}
    \includegraphics[width=.49\linewidth]%
      {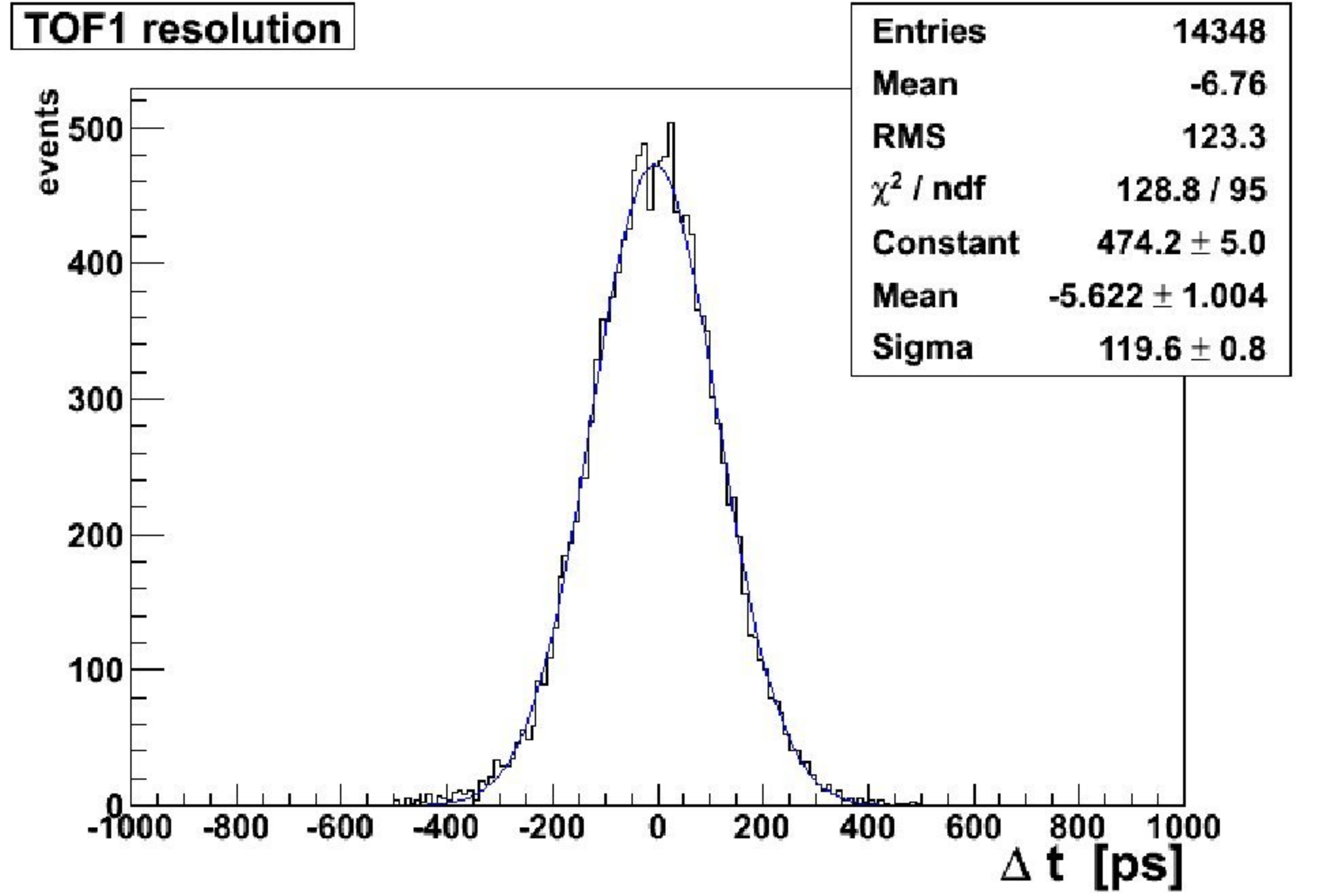}
    \includegraphics[width=.49\linewidth]%
      {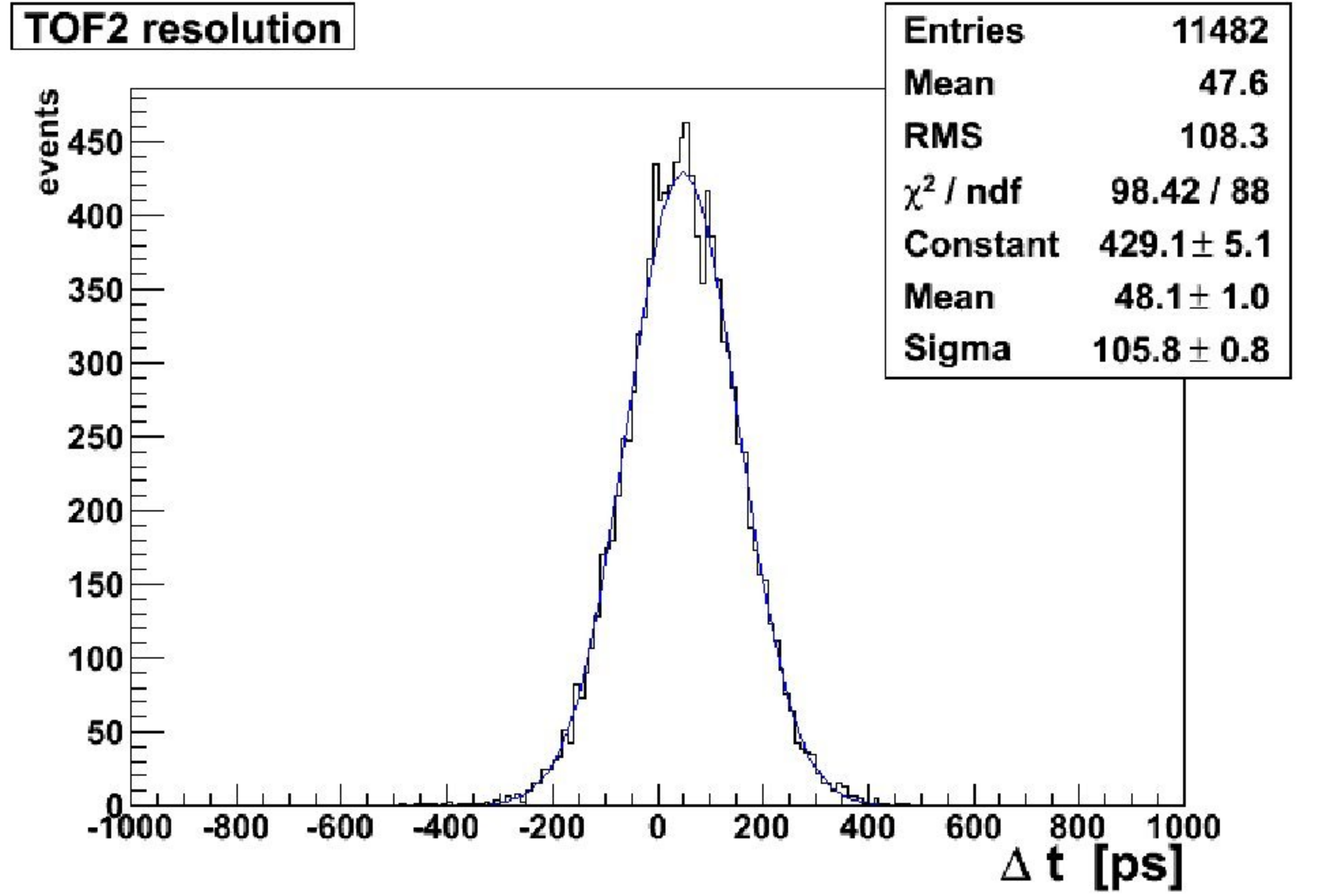}
  \end{center}
  \caption{
    Time difference $\Delta t_{xy}$ between vertical and horizontal 
    slabs in TOF0, TOF1 and TOF2. The trigger is on TOF1. 
  }
  \label{fig:TOF2}
\end{figure}

The performance of the TOF stations was determined by measuring the
difference between the time recorded by the horizontal ($x$) and
vertical ($y$) planes, $\Delta t_{xy}$, of each station.
If it is assumed that the resolution of each plane is the same, the
width of the distribution of $\Delta t_{xy}$ can be used to determine
the time resolution of a single plane, $\sigma_{t_{xy}}$.
Figure \ref{fig:TOF2} shows the $\Delta t_{xy}$ distribution for each
TOF station. They all have remarkably good Gaussian shapes.
The widths of the distributions are $\sim 100$\,ps, corresponding to TOF 
detector resolutions $\sigma_t \sim $ 50 ps.

\begin{figure}
  \begin{center}
    \includegraphics[width=0.49\linewidth]%
      {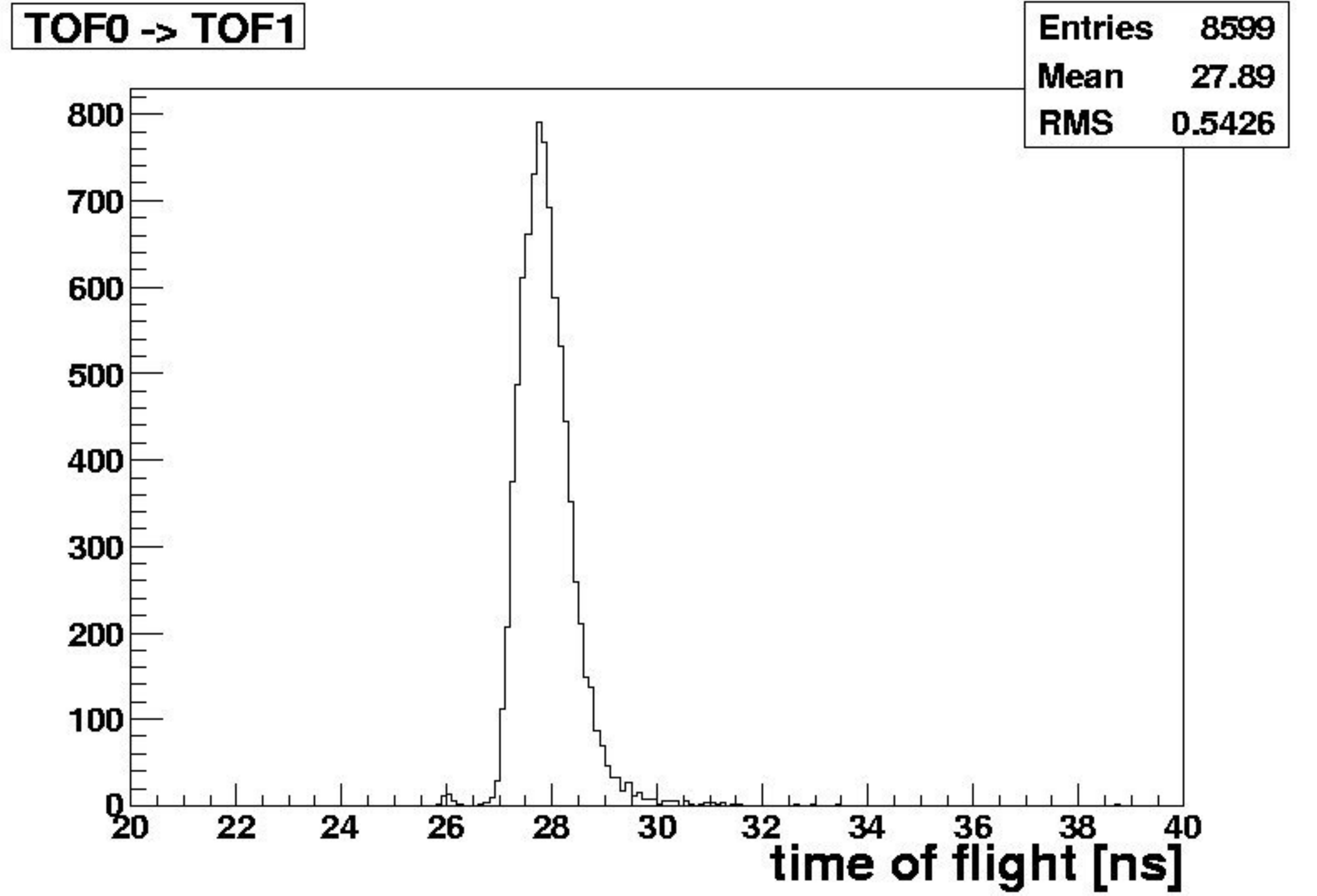}
    \includegraphics[width=0.49\linewidth]%
      {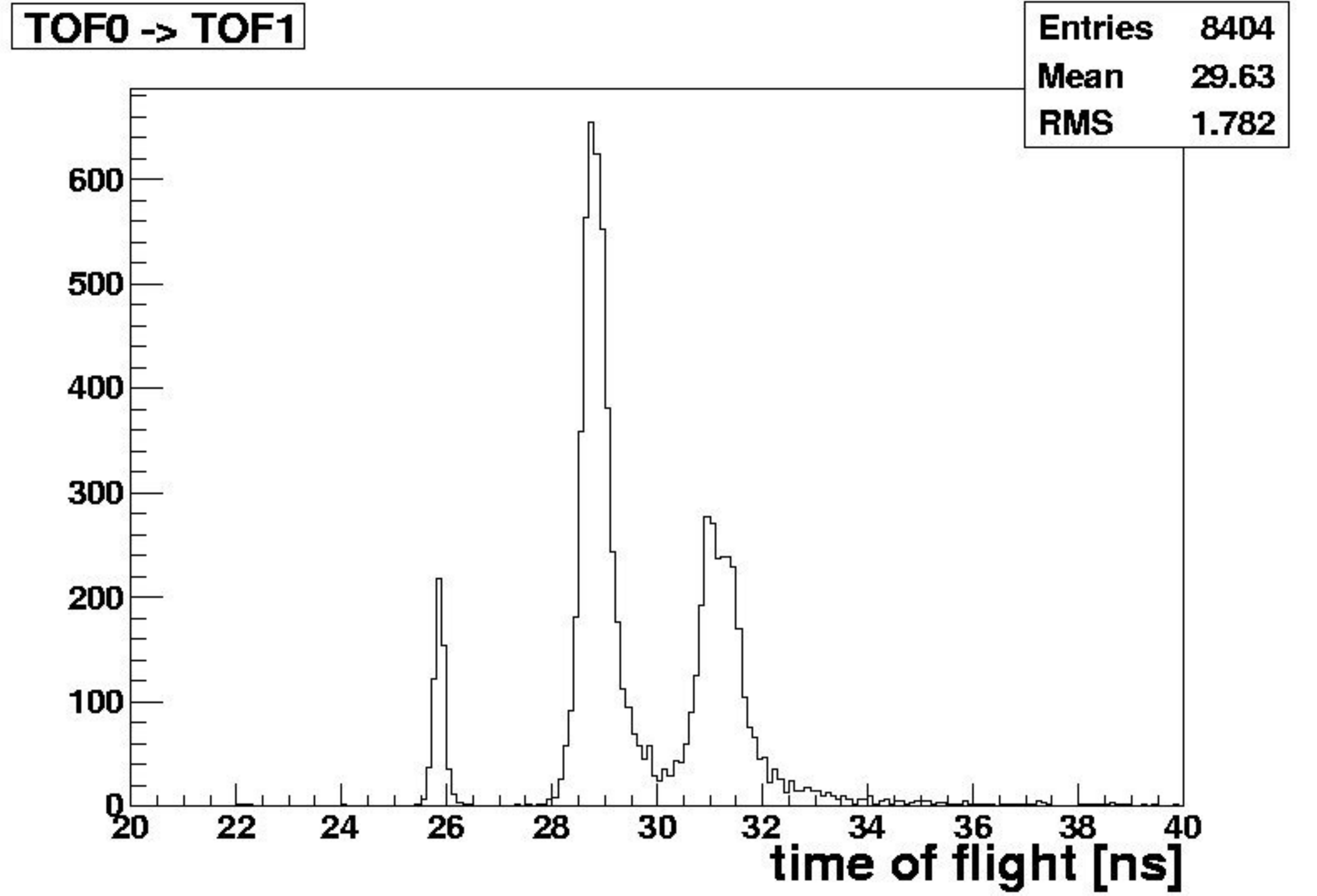}
  \end{center}
  \caption{
    Time of flight between TOF0 and TOF1 for  a muon 
    beam (left)  and a  ``calibration'' beam (right).
  }
  \label{tof}
\end{figure}

\begin{figure}
  \begin{center}
    \includegraphics[width=0.49\linewidth]%
      {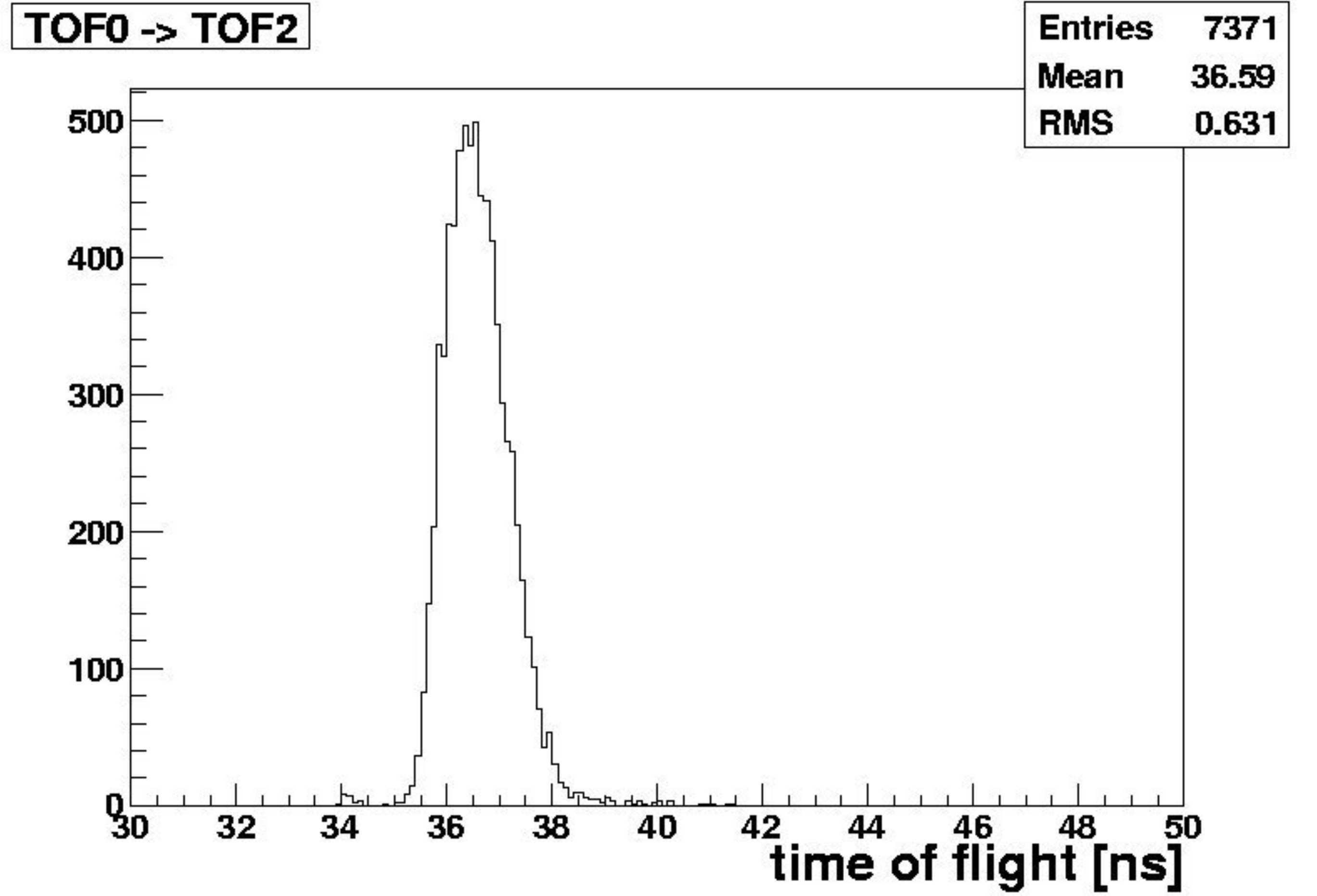}
    \includegraphics[width=0.49\linewidth]%
      {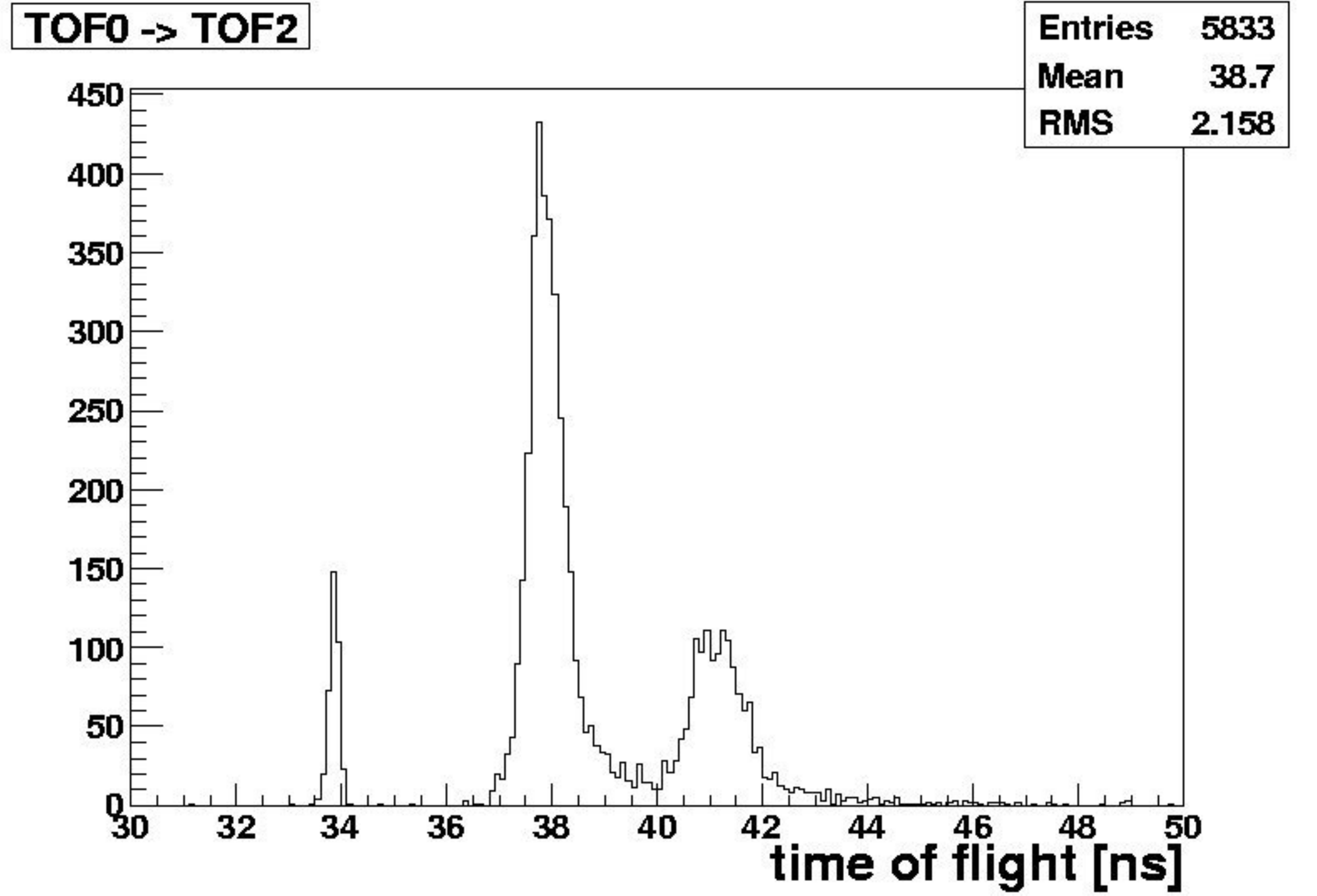}
  \end{center}
  \caption{
    Time of flight between TOF0 and TOF2 for a  muon 
    beam (left) and   a ``calibration beam'' (right).
  }
  \label{tof02}
\end{figure}

Figure \ref{tof} shows distributions of the time-of-flight between
TOF0 and TOF1.
The left panel represents data taken with a  
$\pi\rightarrow\mu$ beam. 
It has a small contamination of electrons and pions. 
Similar beams will be used to demonstrate ionization cooling. 
The right panel shows data taken with a  calibration
beam. 
In this beam configuration electrons, muons and pions fall into three
well-defined peaks. 
Similar plots for the time-of-flight between TOF0 and TOF2 are shown
in figure \ref{tof02}.

\begin{figure}
  \begin{center}
    \includegraphics[width=0.6\linewidth]%
      {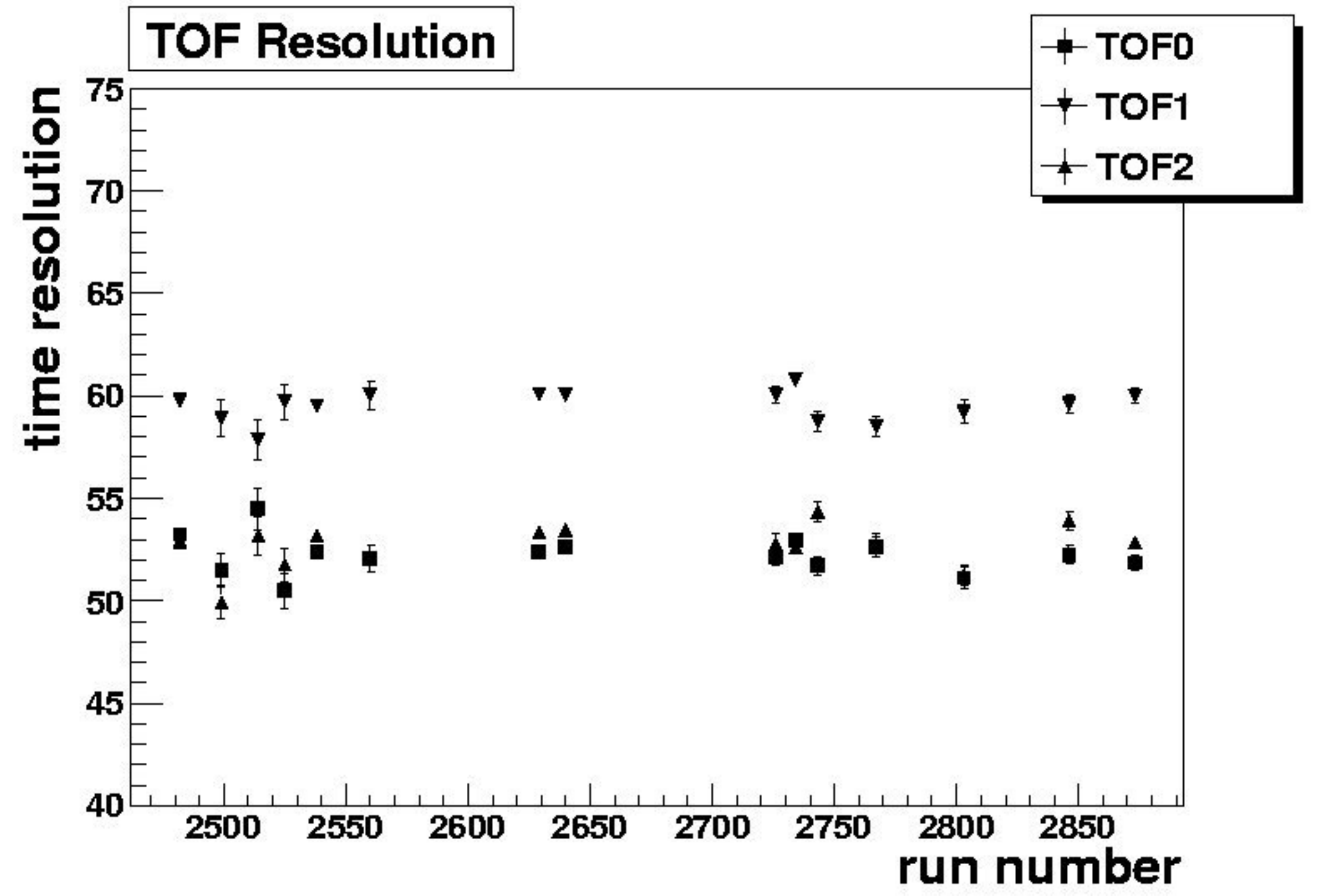}
  \end{center}
  \caption{
    Stability of the time resolution (in ps) of the TOF stations versus
    running time. 
    The covered period is about one month of data-taking.
  }
  \label{fig:stab1}
\end{figure}

\begin{figure}
  \begin{center}
    \includegraphics[width=0.95\linewidth]%
      {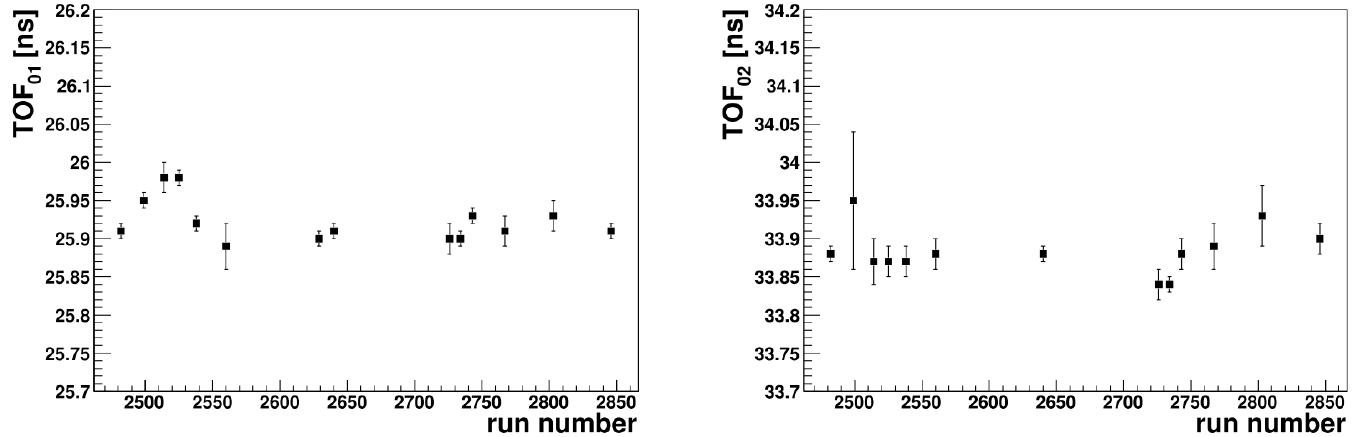}
  \end{center}
  \caption{ 
    Stability of the time-of-flight of electrons between TOF0 and TOF1
    (left) and TOF0 and TOF2 (right) versus run number. 
    The covered period is about one month of data-taking.
  }
  \label{fig:stab2}
\end{figure}

The reference muon beam is a standard setup of the beam-line magnets
that has been used to monitor the stability of the data-acquisition
system and the beam line.
Each data-taking shift begins with a run that reproduces this
reference setup. 
The reference muon beam runs can also be used to validate the
consistency of the calibration procedure over the running period. 
Figure \ref{fig:stab1} shows the variation of the TOF detector
resolutions, as computed from the $\Delta t_{xy}$ distribution, for
each reference run.  
The average time resolutions of TOF0, TOF1 and TOF2 are 
$52.2 \pm 0.9$\,ps, $59.5 \pm 0.7$\,ps and $52.7 \pm 1.1$\,ps
respectively \cite{datax}. 
The slightly worse resolution of TOF1 arises from some of
the PMTs used on TOF1 being of slightly poorer quality than the other
tubes used in the TOF system.~\footnote{ This feature was corrected
later by refurbishing all PMTs of TOF1, obtaining a TOF1 detector resolution
comparable to those of TOF0 and TOF2 \cite{Bonesini:2012}.}
The resolution of the TOF0 station (4\,cm wide slabs) and that of the
TOF2 station (6\,cm wide slabs) are similar, showing that
light path-length fluctuations are negligible.
The stability of the TOF measurement for both upstream
(TOF1--TOF0) and downstream PID (TOF2--TOF0) is of the order
of $\sim \pm 30$\,ps (see figure \ref{fig:stab2} for details). 
This matches well the required resolution for PID ($\sim 100$ ps).

\subsection{KL detector}

\begin{figure}
  \begin{center}
    \includegraphics[width=.70\textwidth]%
      {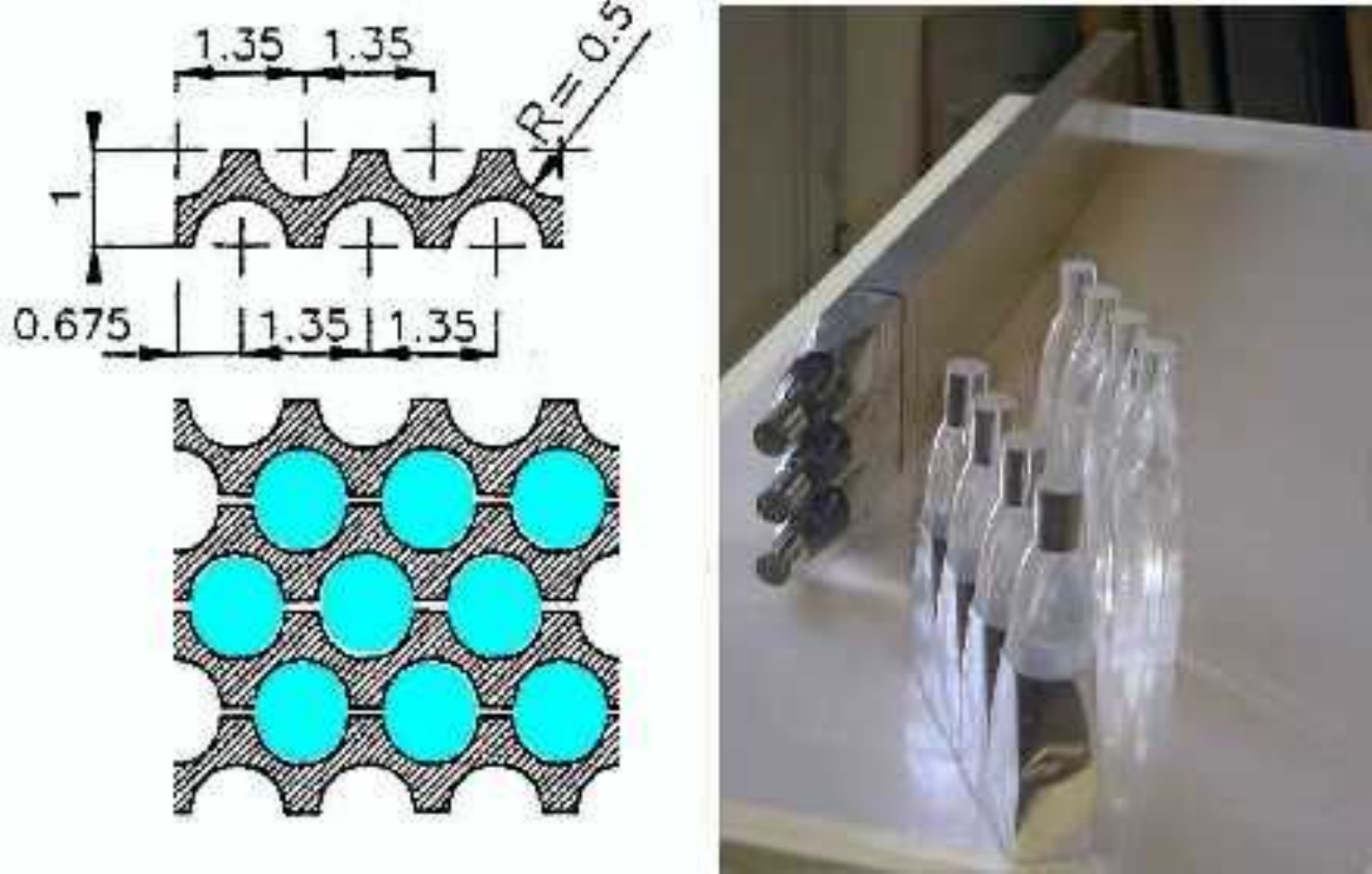}
  \end{center}
  \caption{
    Left panel: schematic layout of KL extruded lead layer and composite
    with fibres; 
    Right panel: a photograph of a three-cell module of KL with Winston
    cone light-guides.
  } 
  \label{fib}
\end{figure}

\begin{figure}
  \begin{center}
    \includegraphics[width=0.70\linewidth]%
      {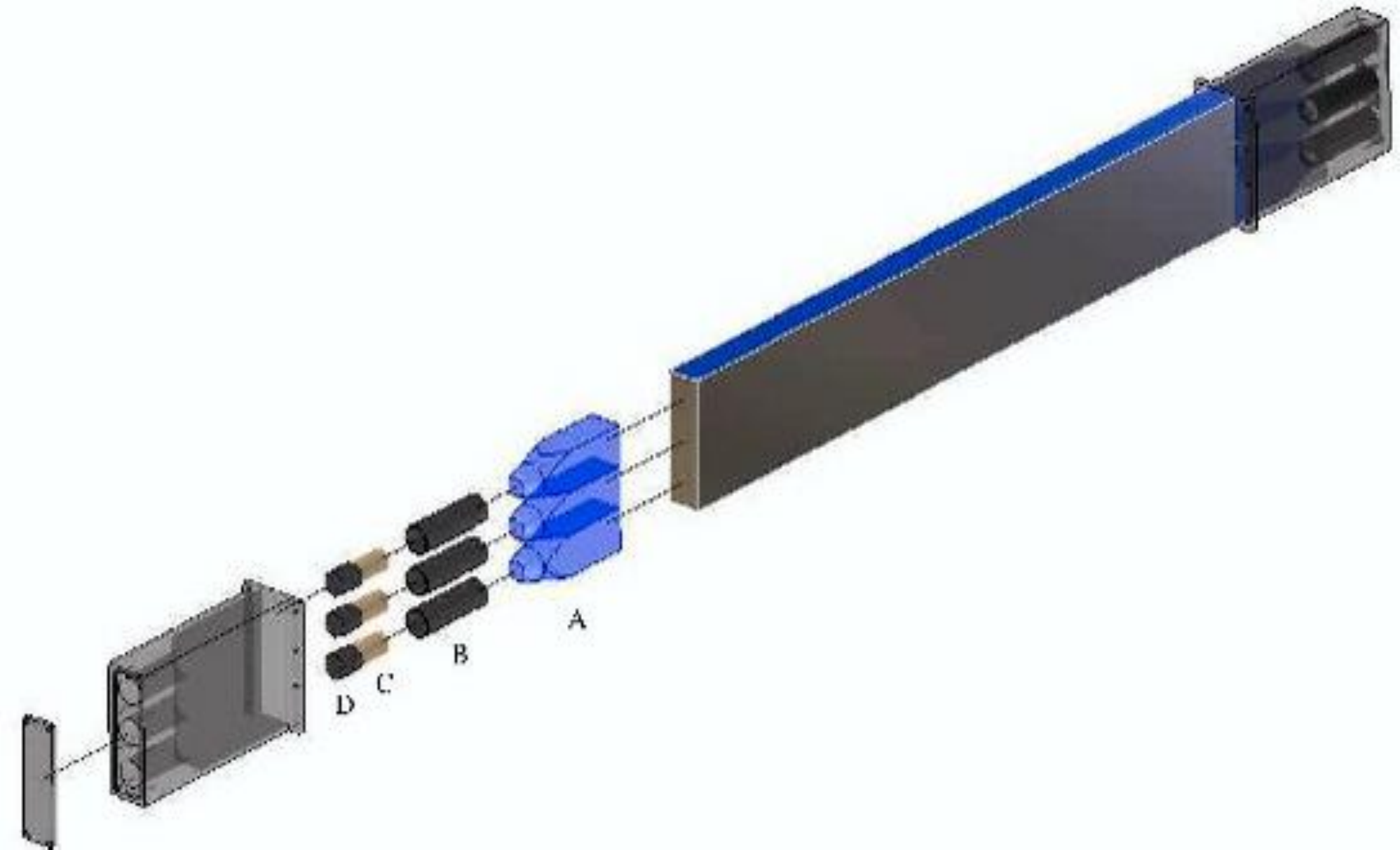}
  \end{center}
   \caption{
     Exploded view of one KL module, showing the assembly of
     the fibre/lead module, light guides (A), mu-metal shielding (B),
     PMTs (C) and voltage dividers (D).
  }
  \label{kl1}
\end{figure}

\begin{figure}
  \begin{center}
    \includegraphics[width=.4\textwidth]%
      {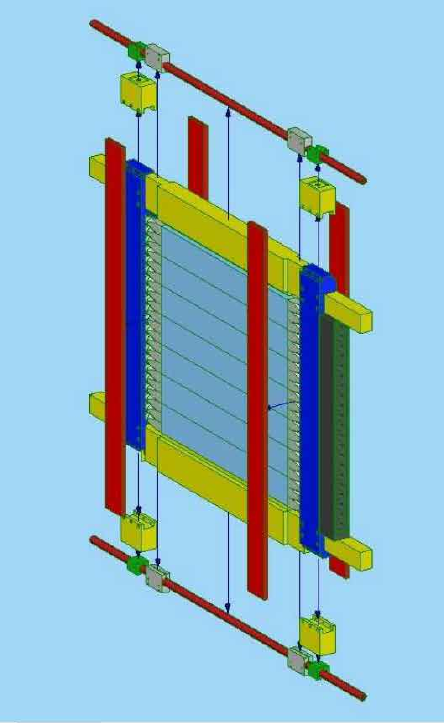}
  \end{center}
  \caption{
    Global layout of KL assembly. The exploded view shows the various
    custom made components for support and magnetic shielding: in yellow
    and dark blue the mechanical support and the PMT soft iron magnetic 
    shields, in green the iron bars housing the PMT voltage dividers and 
    in red the additional iron bars covering the Winston cones light-guides.
  }
  \label{kl2}
\end{figure}

KL is a KLOE-type sampling calorimeter \cite{Ambrosino:2009zza},
composed of extruded Pb foils in which scintillating fibres are placed
in a volume ratio ${\rm Scintillator}:{\rm Pb} \sim 2:1$ (see figure
\ref{fib}).
Since the particle energy is lower in MICE than in KLOE, 
the ratio of lead to fibre has been reduced from that adopted in
\cite{Ambrosino:2009zza} ($\sim 1:1$), hence the name ``KLOE Light''.
The fibres chosen are Bicron BCF-12 with 1\,mm diameter, scintillating
in the blue.
The distance between two neighbouring fibres in the same layer is
1.35\,mm.
The distance between two layers is 0.98\,mm, one layer being shifted
by half the fibre pitch with respect to the next.
The overall detector dimensions, including magnetic shielding and
housing of photomultiplier tubes and voltage dividers, is
approximately $120 \times 4 \times 160$\,cm$^3$. 
In figures \ref{kl1} and \ref{kl2} a schematic view of one exploded KL module 
and the global layout of the KL assembly are  shown.
The active volume of $93 \times 4 \times 93$\,cm$^3$ is divided 
vertically into seven modules, which are supported by an iron frame. 
The iron frame shields the PMTs from magnetic fields.
KL has a thickness of 2.5\,$X_{0}$ and $\sim
0.15$\,$\lambda_{int}$.
From tests performed at the $e^{+}/e^{-}$ test beam facility BTF
\cite{Ghigo:2003gy} of INFN LNF, a time resolution of 
$\Delta t \approx 70$\,ps/$\sqrt{E}$ and an electron-energy
resolution, fully dominated by sampling fluctuations,  of 
$\Delta E/E \approx 7\%/\sqrt{E}$ are obtained for electron energies
between 75\,MeV and 350\,MeV. 

KL has 21 cells and 42 readout channels (one at each end of each
cell).
The light is collected by Hamamatsu R1355 PMTs with E2624-11 
voltage dividers, providing differential output pulses on twisted pair cables
with 120\,$\Omega$ impedance at 50\,MHz. 
The signal from the PMTs is sent to a shaper module, which shapes and
stretches the signal in time in order to match the sampling rate of
the flash ADCs. 
The flash ADC modules are the same 14 bit CAEN V1724 used for the TOF
stations.

In MICE StepI setup, KL is followed by three, 1\,inch
thick, $10 \times 100$\,cm$^{2}$ scintillator bars, placed vertically
side by side behind the centre of the detector, to tag any particles
that pass through KL.
These ``Tag counters'' are used to measure the number of particles that pass
through KL.
They will be removed when the EMR calorimeter is installed.

The TOF system was used to study the response of KL to different
particle types.
From the distance between the TOF1 and TOF2 stations and the measured
time-of-flight, the average momentum between TOF1 and TOF2 of the
impinging particle was determined.
For particles with momentum larger than 300\,MeV/c, TOF2--TOF1 time-of-flight
has insufficient resolution to determine the particle type,
hence the TOF1--TOF0 time-of-flight is used.
To estimate the momentum at the front of KL, the energy lost in
the material of  TOF2 is taken into account.

The procedure described above may be used for muons and pions. 
Since electrons and positrons are relativistic, it is not possible to
use the time-of-flight method. 
The momentum of electrons and positrons is estimated using a model
that takes into account all materials traversed by the electrons and
the magnetic fields traversed.

\begin{figure}
  \begin{center}
 \includegraphics[width=0.49\linewidth]%
      {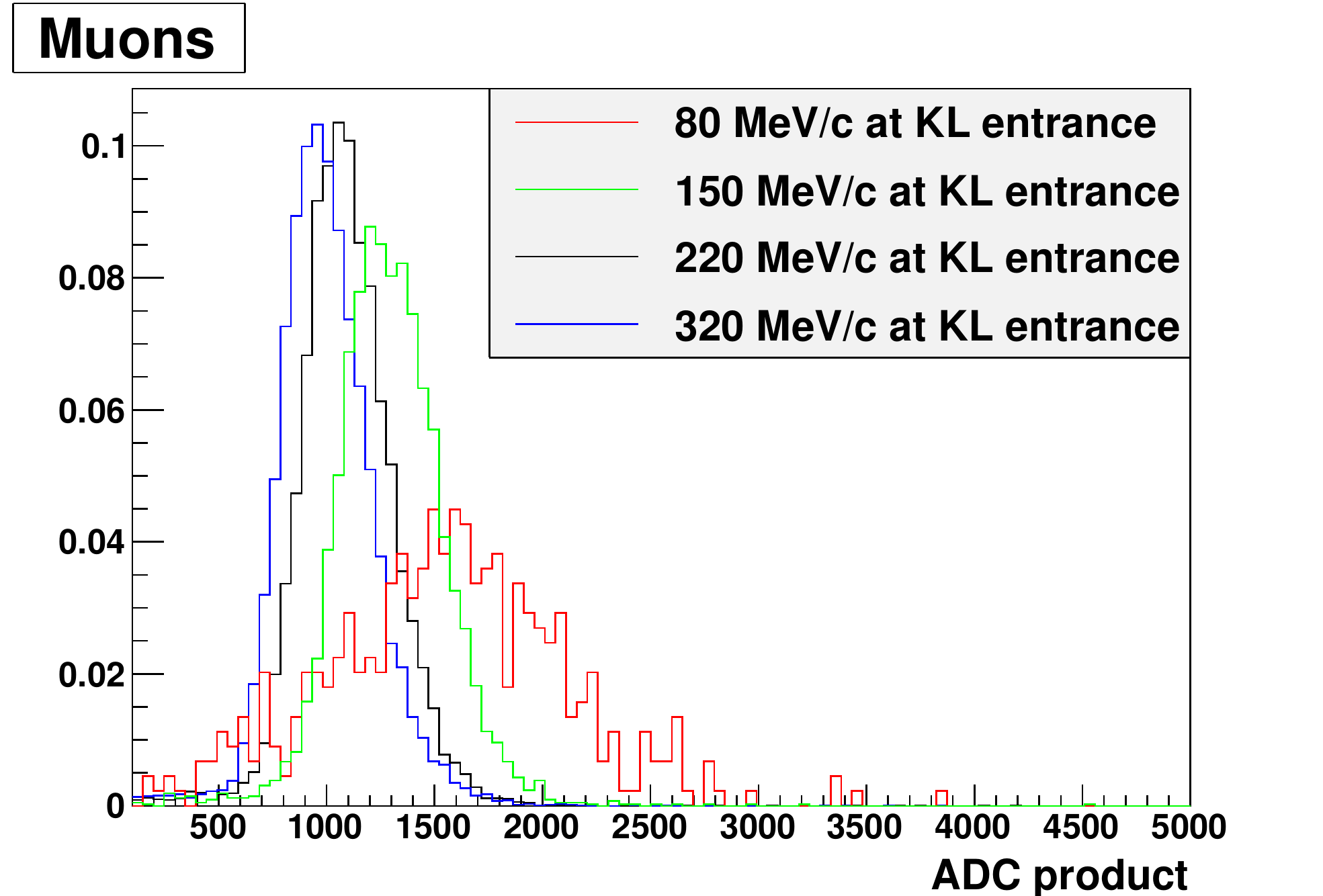}
    \includegraphics[width=0.49\linewidth]%
      {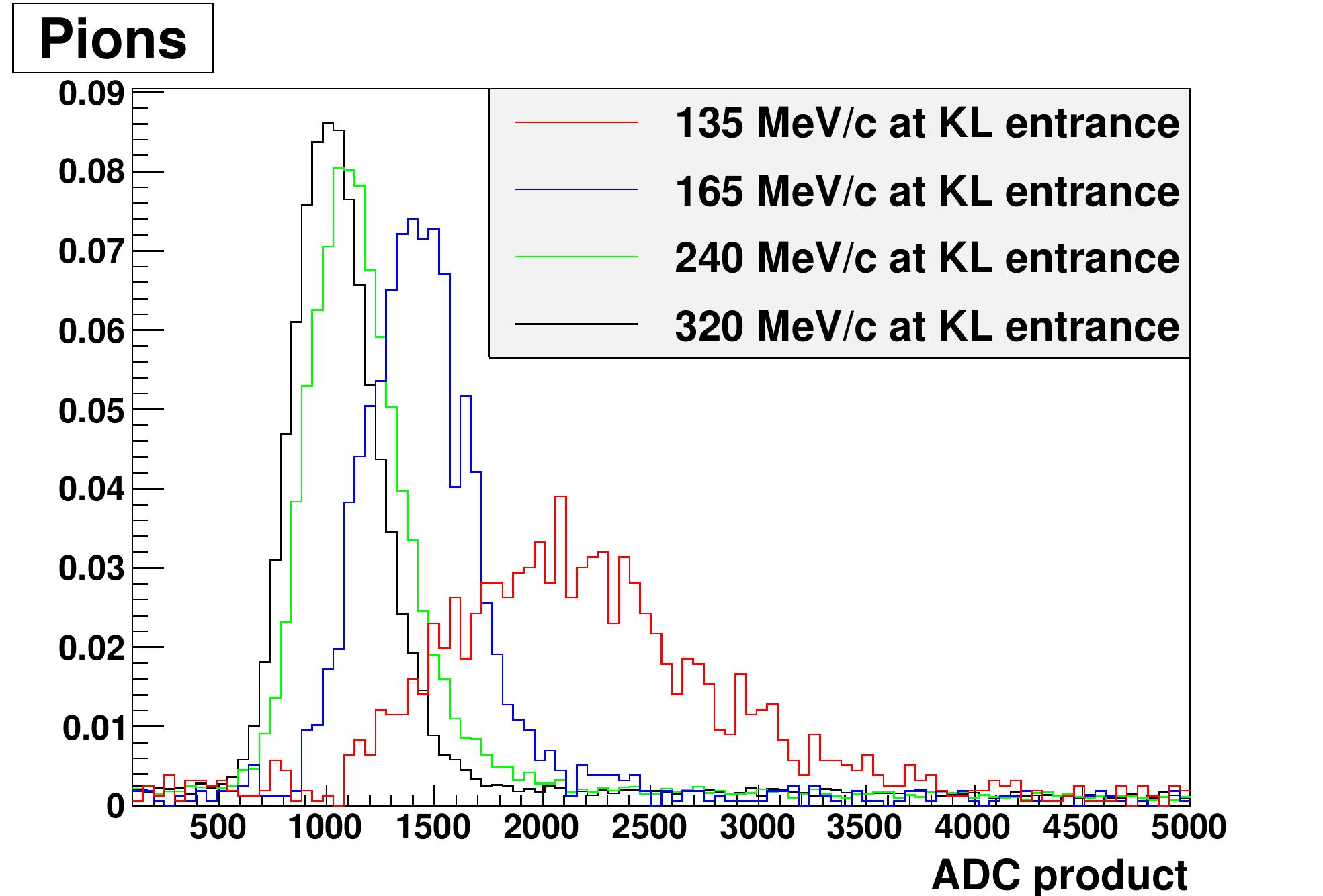}
    \includegraphics[width=0.49\linewidth]%
      {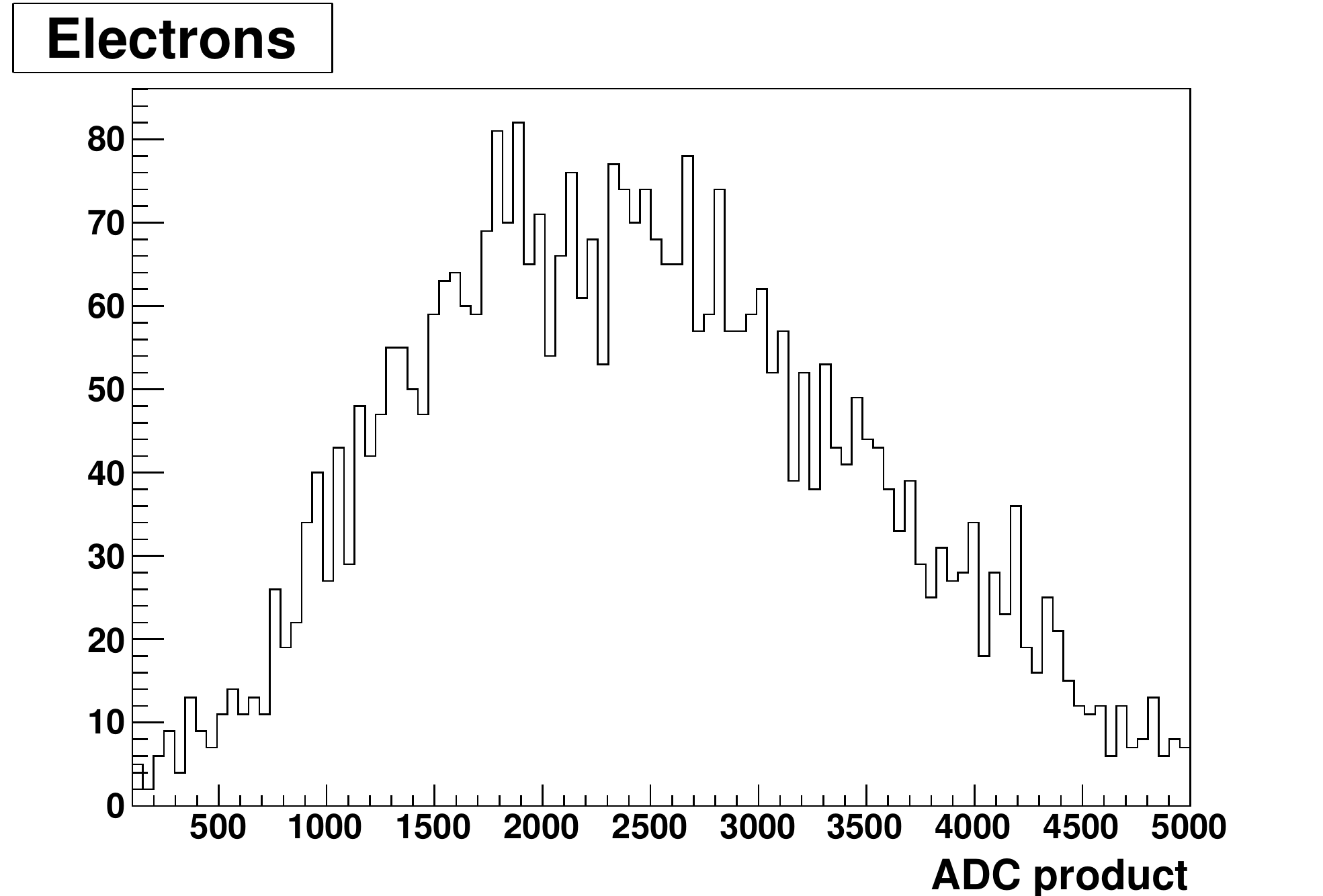}
   \end{center}
  \caption{
    KL response (normalised to the number of events) to muons for
    various incident momenta (top left panel), to pions (top right
    panel), with various momenta, and  to 80 MeV/c electrons at the
    entrance of KL (bottom panel). 
    For an explanation of ``ADC product'' see the text.
  }
  \label{mu}
\end{figure}

The data used to study the response of  KL cover the entire range
of momentum of interest in MICE. 
The ADC product is defined as the product of the digitised signals
from the left and right sides of one slab divided by their sum:
\begin{eqnarray}
  ADC_{prod} = 
    2 \times ADC_{\rm{left}}\times ADC_{\rm{right}} /
    (ADC_{\rm{left}}+ ADC_{\rm{right}}) \, ;
\end{eqnarray}
where the factor of 2 is present for normalisation. 
The product of the two sides compensates for the effect of light
attenuation. 
In the upper left panel of figure \ref{mu}, the KL response to muons
at various momenta is shown.
The deposited energy reaches its minimum ($\sim 30$\,MeV) at a
momentum of 300\,MeV/c (ionization minimum in lead).
The upper-right panel shows  the KL response to pions.
In the plots, the abscissa represents the sum of the ADC product from
all slabs in  KL above a given threshold.
The bottom panel of figure \ref{mu} shows the typical response of 
KL to electrons. 
The fraction of 80\,MeV/c electrons which pass through  KL is 
$\sim 70\%$.

\begin{figure}
  \begin{center}
    \includegraphics[width=0.80\linewidth]%
      {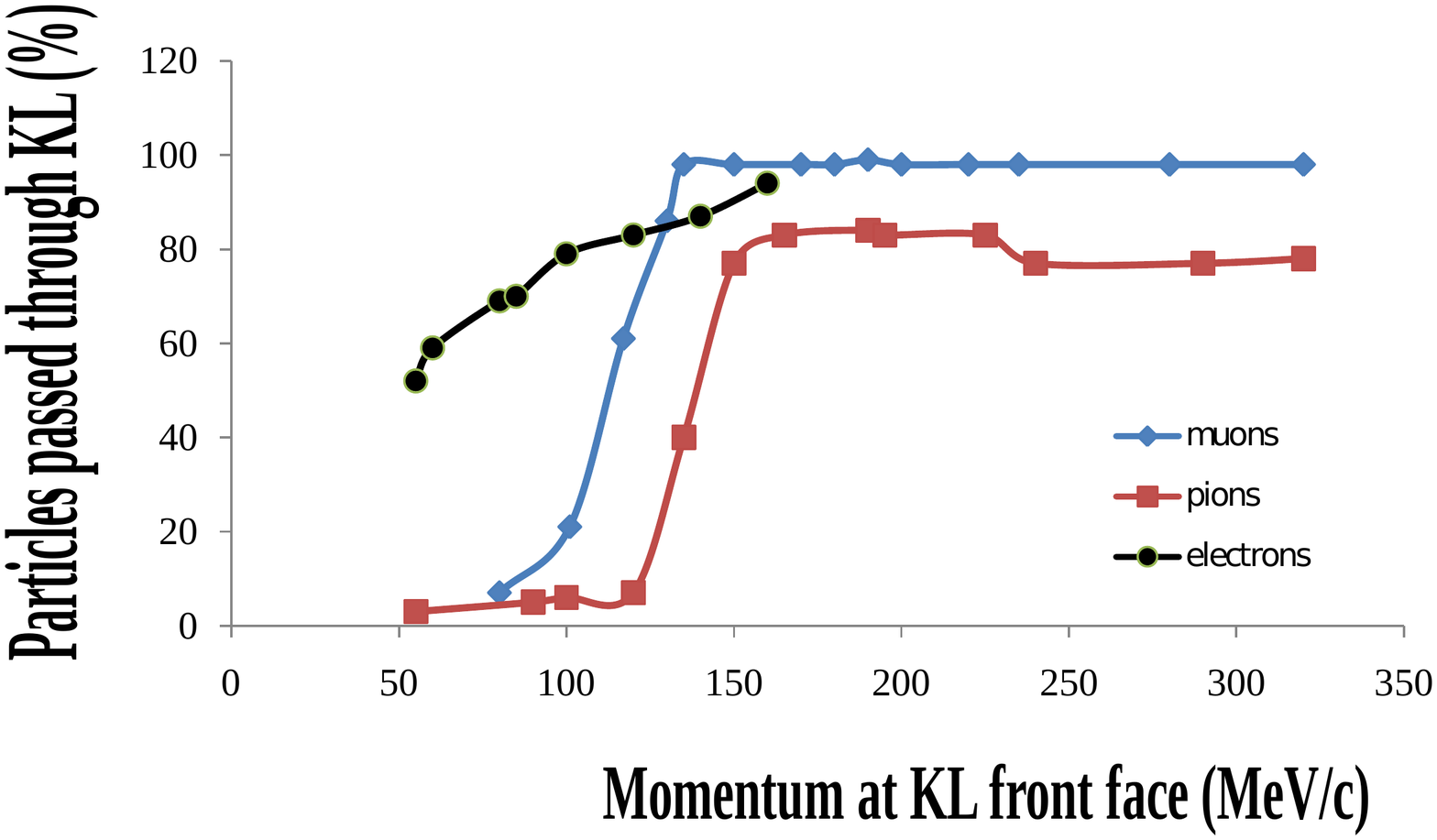}
  \end{center}
  \caption{
    Fractions of electrons, muons and pions passing through KL and
    reaching the Tag counters.
  } 
  \label{penetr}
\end{figure}

\begin{figure}
  \begin{center}
    \includegraphics[width=0.6\linewidth]%
      {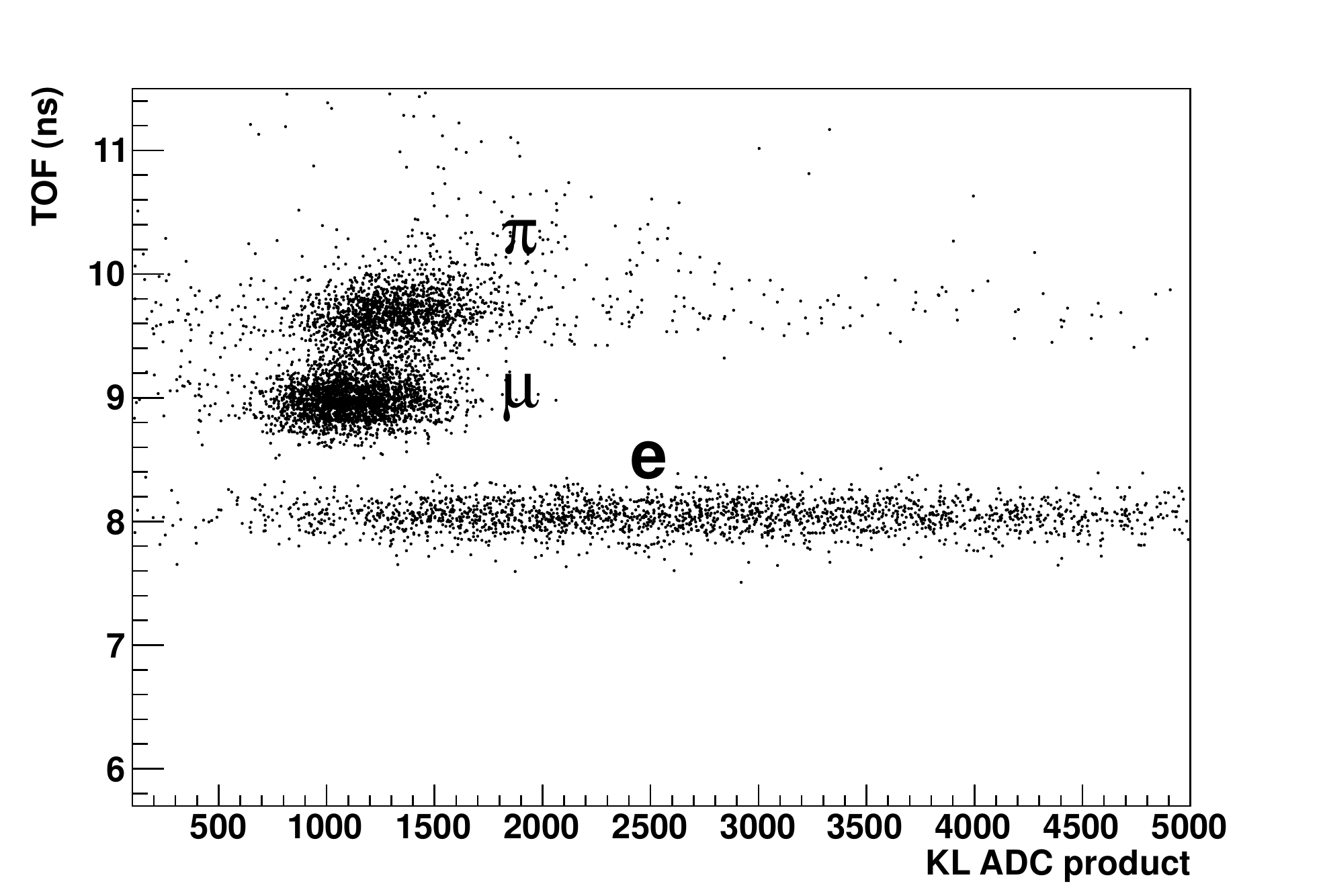}
  \end{center}
  \caption{
    Time of flight, as measured from TOF0 and TOF1, versus KL response
    as ADC product, using a 300\,MeV/c ``calibration'' beam, with
    trigger on TOF1.
  } 
  \label{tof_kl}
\end{figure}

The fractions of electrons, muons and pions passing through KL
(the ``KL transparency'') are shown in figure~\ref{penetr}. 
KL must act as a pre-sampler for the EMR, introducing a minimal 
perturbation to incoming muons and pions. 
%%Figure \ref{penetr} shows that the threshold for this is around
%%140\,MeV/c.
The KL ADC product, plotted as a function of the time-of-flight for 
different particle types, is shown in figure \ref{tof_kl}.  
A clear separation between electrons and  muons/pions is visible.

Preliminary studies of  KL response in different TOF time windows show 
a $\pi$ contamination in the muon beam of the order of $\sim 1 \%$, after
a suitable cut on the KL ADC product. These results will be the subject of 
a forthcoming paper.

A raw time estimate may be made, using a simple linear interpolation
of the rising edge of the sampling FADC signal.
This provides some redundancy in the global time measurement.
Preliminary laboratory measurements have shown that a $\sim 210$ ps preliminary resolution
may be obtained by selecting cosmic muons impinging at the centre of a KL
cell. 

A study of the stability of KL during data-taking showed that
there were no dead or noisy channels.
The pedestals are stable with an rms of $\sim 2.5$ ADC counts,
corresponding to $\sim 0.2 \%$ of a MIP energy deposition.
The KL response to muons and pions is stable in time (over a
several-month period) to within $\sim 1 \%$. 
\section{The MICE Step I Trigger, Data Acquisition and Online/Offline Software}
\label{sec:daq}
\subsection{Trigger and Data acquisition}
The Detector Data Acquisition (DDAQ) system of the MICE experiment
must be able to acquire data at a rate of up to $\sim$600 muons
per spill.
%~\footnote{In all the following, we will use the wording 
%``spill'' for the burst of particles  from one target dip, 
%even if the primary ISIS proton beam is not extracted}.
To fulfil this requirement, the Front-End Electronics (FEE) must 
digitise the analogue signal in less than 500\,ns and store
the digitised data in buffer memory.
The time before the arrival of the next spill ($\sim$1\,s) is used to
read out the buffers and store the data.
Data are transferred off-line to the remote mass storage unit for
subsequent analysis using Grid computing technology
\cite{Forrest:2010zz}.

The acquisition of the data coming from the detectors is based on VME
FEE interfaced to Linux PC processors via  CAEN VME-PCI V2718
interfaces \cite{CAEN}. 
These modules are connected via optical links to PCI cards embedded in
the acquisition PCs and may sustain transfer rates up to 70\,Mbytes/s.

The software framework performing the management of all the
independent readout processes and the event building, as well as the
user interface, has been developed from the DATE package
provided by the ALICE experiment at CERN \cite{Rohrich:2000fa}. 
The event-builder machines receive sub-events from the various VME
crates through the local network, put them together in memory and
store them locally in a 3 TB, RAID6, disk buffer.
These data are subsequently transferred to a second local machine,
from which they can be transferred asynchronously to the RAL Tier 1
centre for archival. 
As this transfer is carried out using Grid protocols, the data files
are also immediately registered in a global file catalogue and
replicated to other remote sites for off-line analysis. 

The Detector DAQ is strongly dependent on the Trigger System, which is
divided into two parts.  
The readout can only be performed at the end of the spill, due to the
time structure of the MICE beam.
Digital information related to the passage of the particles
through the MICE cooling channel during the spill must therefore be buffered
in the FEE.
The first part of the Trigger System is responsible for the generation
of the ``Particle Trigger'' signal.
This signal triggers the digitisation of the analog signals received
from the detectors. A coincidence of the PMTs connected to the same
scintillation counter in one of the TOF stations generates the
Particle Trigger.

Logic units used to generate the Particle Trigger signals are
LeCroy 4516 (CAMAC) logic modules \cite{LeCroy}. 
These use the output signal of the LeCroy 4415 discriminators,
perform the needed logical AND operation between the two PMTs at the
ends of the same counter and issue a logical OR output per group of 8
slabs.
NIM logic is then used to produce a global OR signal for all the slabs
of a TOF  station. 
The resulting signal is distributed to all the front-end boards
involved in the experiment. 
For the moment only TOF0 or TOF1 detectors may be used as a source of
the Particle Trigger signal. 

The second part of the Trigger System generates the so called DAQ
Trigger. 
This signal is generated after the extraction of the ISIS proton beam
and causes the readout and storage of the digital data
corresponding to all the particle triggers received during a spill.
The DAQ Trigger is generated from the same signal as that sent to the
target system to trigger the spill.
\subsection{The control system and online monitoring of MICE}
\label{sec:control}

The EPICS (Experimental Physics and Industrial Control System)
platform was chosen for the MICE controls and monitoring system
because of its reliability, existing drivers for a wide variety of
devices, flexibility and world-wide support network
\cite{Clausen:2008zza}.
The EPICS backbone is a local area network  to which hardware
components are interfaced via EPICS input/output controllers,
while the user has access through EPICS channel access. 

All the equipment required for Step I of MICE, including the MICE Muon
Beam, the TOF and Ckov detectors as well as the systems used for
environmental monitoring of the MICE Hall, was included in the
controls and monitoring system.
The target, beam line, decay solenoid and 
the proton absorber and beam stop (which are manually operated)
are monitored.

The high-voltage system for the PID detectors is based on CAEN SY127 and CAEN
SY527 systems interfaced, via CAENET, with CAEN PCI A1303 cards
\cite{CAEN}.
The control system is used to set-up and monitor the high voltage 
and includes the functionality to turn the detectors on and off and 
set ramp rates, operating voltages and current limits.
The environmental monitoring system includes the measurement of the
temperatures for the TOF detectors and the temperatures and internal
humidity of the two Ckov detectors.

The MICE Hall is monitored using temperature and humidity probes, a
neutron monitor and water-leak detectors.
Selected parameters are used by the EPICS Alarm Handlers which compare
read values to preset limits.  
The EPICS Archiver is used to archive selected parameter values, to be
used later as input to correction algorithms being developed for the
offline analysis.

A MICE Configuration Database (ConfigDB) has been developed and will be
used to store the parameters that determine the operation of the
experiment as well as alarm limits for the various components
\cite{Forrest:2011,PhDForrest}. 
The TOF and Ckov systems read the parameter set-values and limits
from the ConfigDB during initialisation.
Alarms occur when parameter values drift beyond these limits.

In addition to the Control and Monitoring system, the MICE
reconstruction program runs online in parallel to data taking.
This allows a detailed inspection of relevant physical parameters by
the users on shift.
\subsection{The MICE Offline Software System}

The simulation, design and analysis of the experiment are supported
by a variety of software tools.  
Most useful are G4MICE \cite{Rogers:2006zz,Ellis:2007zz} and
G4beamline \cite{G4Beamline}, both of which make use of the GEANT4
toolkit \cite{Agostinelli:2002hh}.
G4MICE provides an extensible framework for offline simulation,
reconstruction and analysis as well as online monitoring and is the
standard software in the experiment for all Step I analysis work. 
G4MICE contains configuration files that describe the geometry of
the experiment (including detectors, magnets, absorbers and cavities),
the cabling of the various detectors and their calibrations. 
Acting as a client to the Configuration Database, G4MICE is able to
reconstruct data taken in any experimental configuration.  

In addition to the GEANT4 toolkit for high-energy physics simulation,
G4MICE also exploits the Fermilab BeamTools package to describe
magnets, absorbers and cavities. 
Inside G4MICE, the Class Library for High Energy Physics (CLHEP)
package~\cite{Lonnblad:1994kt}  provides classes useful for random-number generation and other
applications including mathematical functions (complemented by the GNU
Scientific Library, GSL). 
In addition, the ROOT analysis framework~\cite{Brun:2010zz} is used to develop graphical
user interfaces and to perform data analysis using a wide variety of
libraries including classes for drawing histograms, file persistency
and graph fitting.

G4beamline provides fast simulation of particle behaviour in a variety
of electric and magnetic fields and materials for a given geometry and
set of beam-line optics. 
G4beamline is implemented in C++, although its associated user
interfaces are implemented in Java. 
Users interact with the application exclusively through sets of data
cards describing the parameters and geometry of a run, and are not
required to write their own C++ code. 
G4beamline also includes visualisation features which allow users to
view and explore the geometries generated from their data cards, as
well as to study the particle species and trajectories of a sample of
particles at any point in space.
The QGSP\_BIC hadronic package of GEANT4, which adequately describes
particle production in the MICE target, has been used by both G4MICE
and G4beamline in the optimisation of the beam line.

\section{Characterisation of the MICE Muon Beam}
\label{sec:results}

MICE took data in the Step I configuration in summer 2010.
Data were collected in each of the nine $(\epsilon_N, p_z)$
configurations defined in section 3  for both positive and
negative beams.
With the exception of a single broken PMT in TOF0, all detectors
worked well. 
The MICE Muon Beam was operated at a maximum rate of 34 muons per
target dip (in 1 ms) at a 1\,V$\cdot$ms beam loss.
About 10 million triggers were collected over the various beam
configurations. 
The data were used for detector calibration, to assess the particle
identification capabilities of each of the detectors and to
characterise the beam line.  
The main results on muon rates and comparisons between data and Monte
Carlo are summarised in the next section. 
A first measurement of the beam emittance, using the TOF
detectors only, was also obtained \cite{PAC11_step1} and will be the
topic of a separate publication \cite{ref:mark}. 

\subsection{Particle rates and beam profiles in the MICE beam line}

\begin{figure*}
  \begin{center}
\includegraphics[width=0.49\linewidth]
      {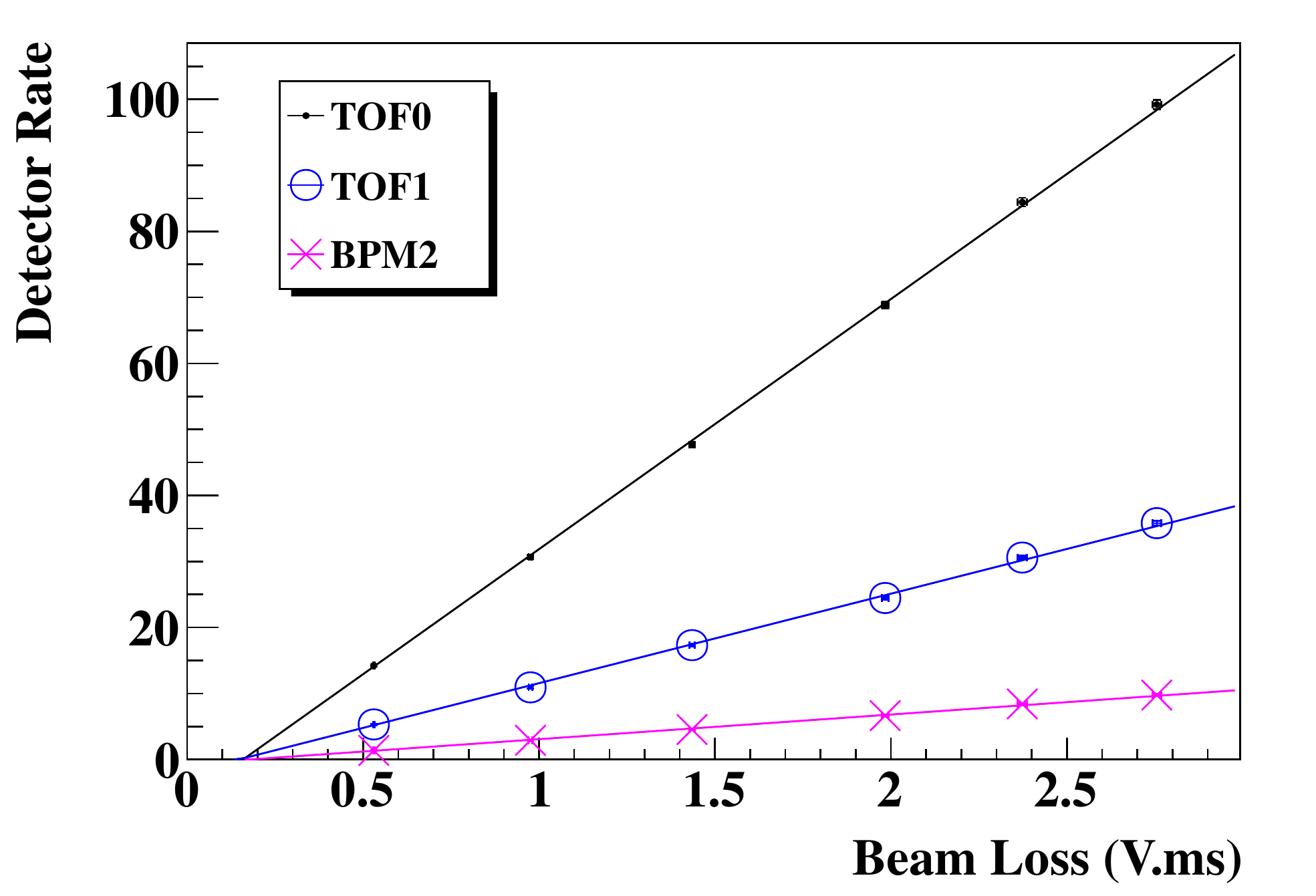}
    \includegraphics[width=0.49\linewidth]
      {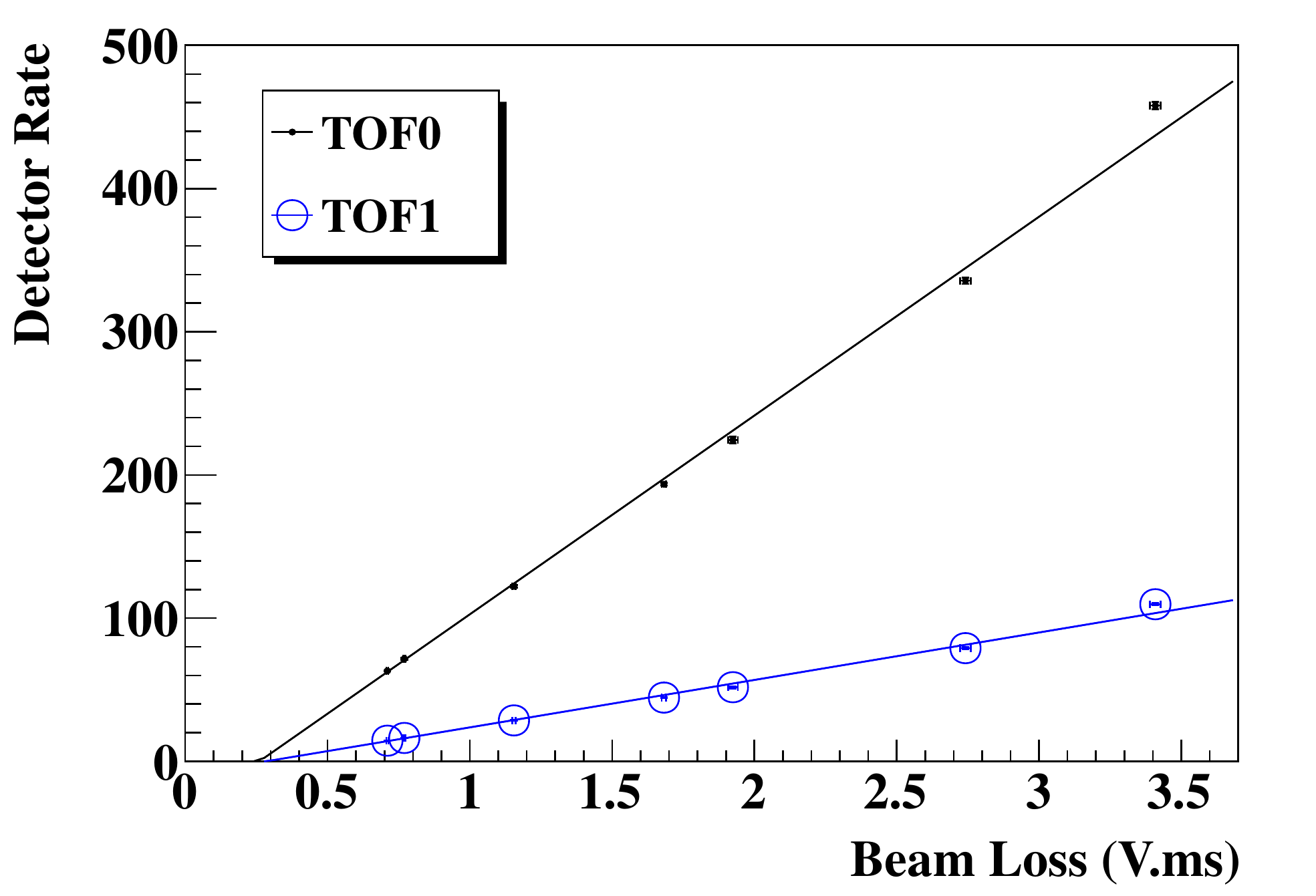}
  \end{center}
  \caption{
    Total particle rates per spill vs induced ISIS beam loss for a
    negative $\pi\rightarrow \mu$ beam, with a 3.2 ms spill gate
    (left), and for a positive $\pi\rightarrow \mu$ beam, with a 1 ms
    spill gate (right).
  } 
  \label{fig:TotalRate}
\end{figure*}

A dedicated study of particle rate in the MICE Muon Beam as a function
of ISIS beam loss was performed. 
The particle rates observed in the GVA1, BPM2, TOF0 and TOF1 
detectors were recorded along with the integrated SP7 beam
loss.
A series of runs, consisting of approximately 400 target dips each,
were taken, each run corresponding to a particular target setting.
The ISIS beam conditions were stable across a run.
The mean beam loss and the mean particle rates were calculated by
averaging the beam loss and particle rates recorded over the 400-dip
sample.
Figure 
\ref{fig:TotalRate} shows the dependence of the BPM2, TOF0 and TOF1 
average particle rates
as a function of the average beam loss. 
Negative $\pi\rightarrow\mu$ beam-line optics with a 3.2\,ms spill
gate are shown in the left-hand panel of Figure \ref{fig:TotalRate},
while positive $\pi\rightarrow\mu$ beam-line optics with a 1\,ms
spill gate are shown in the right-hand
panel.  
Straight-line fits are also shown and give an excellent description of
the data \cite{PAC11_beam},\cite{Dobbs2011}.

Muons were selected by the requirement that the time-of-flight between
TOF0 and TOF1 be in the range $26.2 < \Delta t < 32$\,ns.
Pion contamination of the muon sample was estimated using G4MICE
giving $\sim 1 \%$ for the negative beam
and values between 3 to 5\% for the positive one.
In the future, the $\pi$ contamination will be further reduced using
the two Cherenkov counters (see section 4).

Table \ref{tab:muon_rate} summarises the muon rate determined for each
of the nine beam configurations.  
Errors are a quadratic combination of the statistical error of the
fits and the systematic uncertainties arising from the event
selection. 
The results presented in table \ref{tab:muon_rate} indicate a muon
rate of $\sim 5$\,$\mu$/s per V$\cdot$ms for a 3.2\,ms gate for the
negative beam and between 17 and 34 \,$\mu$/s per V$\cdot$ms for a 1\,ms gate 
for the positive beam.
Studies are underway to determine the maximum beam loss induced by the
MICE target which is acceptable during normal ISIS operations.
In parallel, various proposals are being considered by which the muon
rate per V$\cdot$ms of beam loss can be increased, including adding a
beam ``bump'' to move the ISIS beam towards the MICE target,
increasing the acceleration of the target shaft, and re-optimising the
target geometry or  material. 
\begin{table*}
  \caption{
    Muon track rates for the two  polarities of the MICE beam
    line. Counts are normalised to the measured V.ms beam loss used to
    characterise the target depth. 
  }
  \newcommand{\ty}{\tiny}
  \newcommand{\fn}{\footnotesize}
  \vspace{2mm}
  \begin{tabular}{|c|c|c|c||c|c|c|}
\cline{1-7}
 & \multicolumn{3}{c||}{$\mu^-$ rate (muons/V$\cdot$~ms)} & \multicolumn{3}{c|}{
   $\mu^+$ rate (muons/V$\cdot$~ms)}  \\ \cline{2-7}
 {$\epsilon_N$ ( $\pi$ mm $\cdot$ rad)} & \multicolumn{3}{c||}{p$_z$ (MeV/c)} & \multicolumn{3}{c|}{p$_z$ (MeV/c)} \\ \cline{2-7}
 &140& 200 & 240 &140& 200 & 240  \\ \cline{1-7}
 3 & 4.1$\pm$0.2& 6.3 $\pm$0.2 & 4.9$\pm$0.2 & 16.8$\pm$1.8 & 33.1$\pm$3.2& 33.0$\pm$2.6\\
 6 & 4.1$\pm$0.4 & 4.8$\pm$0.2 & 4.5$\pm$0.2 & 17.8 $\pm$1.8 & 31.0$\pm$2.0 & 31.7$\pm$2.0\\
10 & 4.6 $\pm$0.2 & 5.4$\pm$0.2 & 4.4$\pm$0.1 & 21.6$\pm$2.2 & 34.0$\pm$2.5 & 26.1$\pm$1.5\\
\cline{1-7}
  \end{tabular}
  \label{tab:muon_rate}
\end{table*}

\begin{figure}
  \begin{center}
    \includegraphics[width=0.49\linewidth]%
      {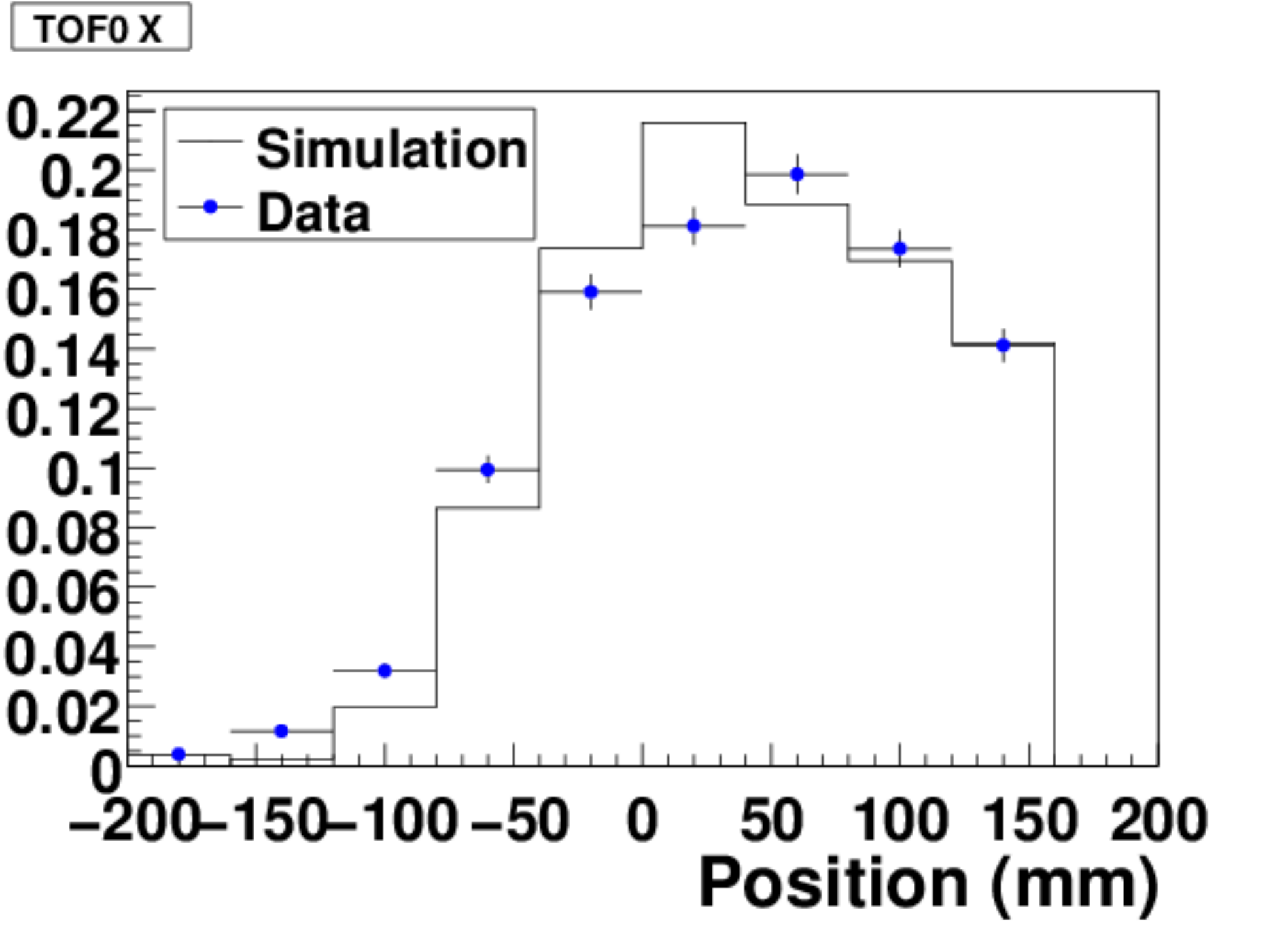}
    \includegraphics[width=0.49\linewidth]%
      {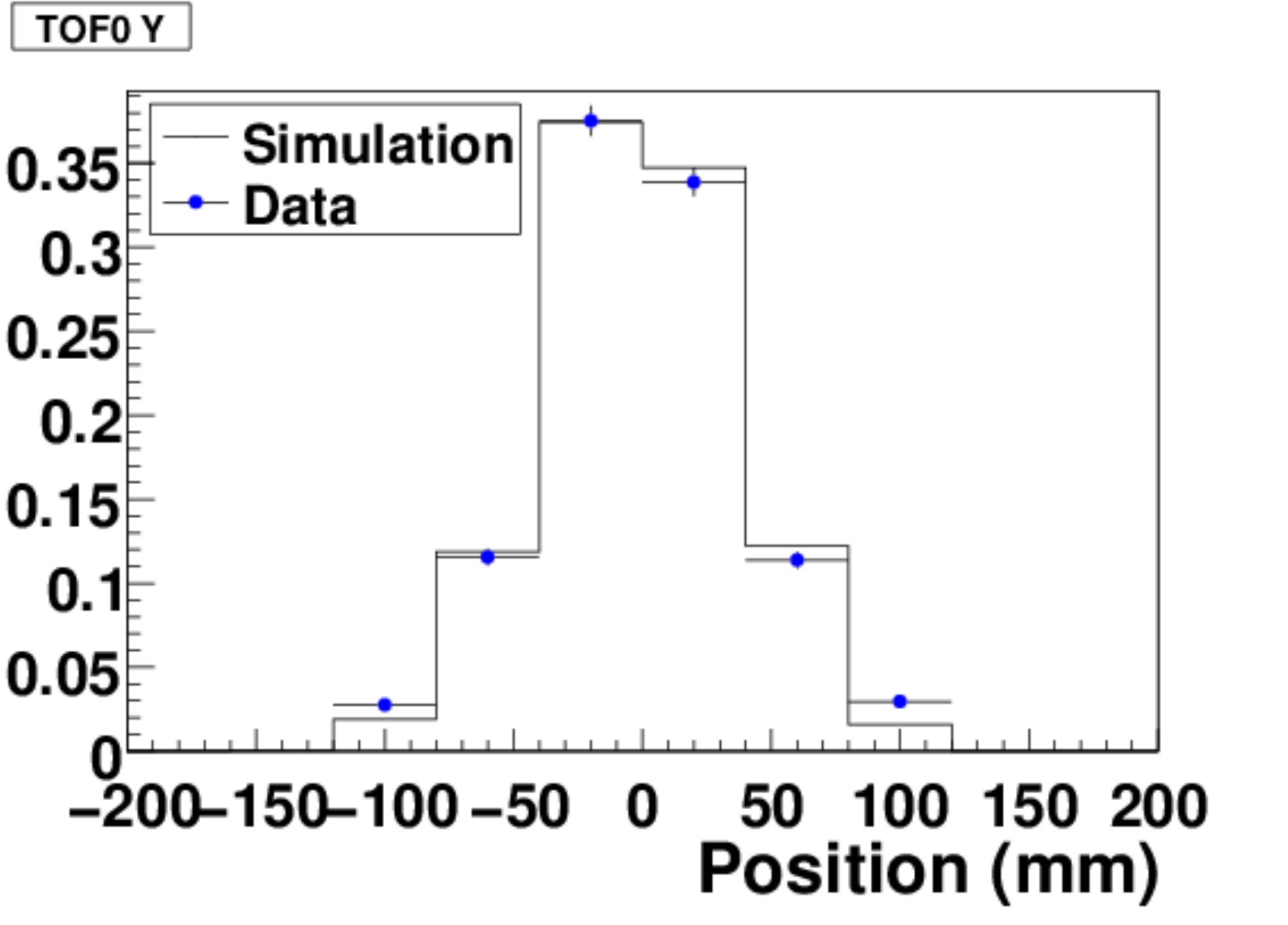}
  \end{center}
  \caption{
    Slab by slab comparison between simulation and data for the $x$
    (left-hand panel) and $y$ (right-hand  panel) planes of TOF0 for
    the (6,200) matrix element, with misalignment corrections applied.
  }
  \label{fig:tErate200t0}
\end{figure}

\begin{figure}
  \begin{center}
    \includegraphics[width=0.49\linewidth]%
      {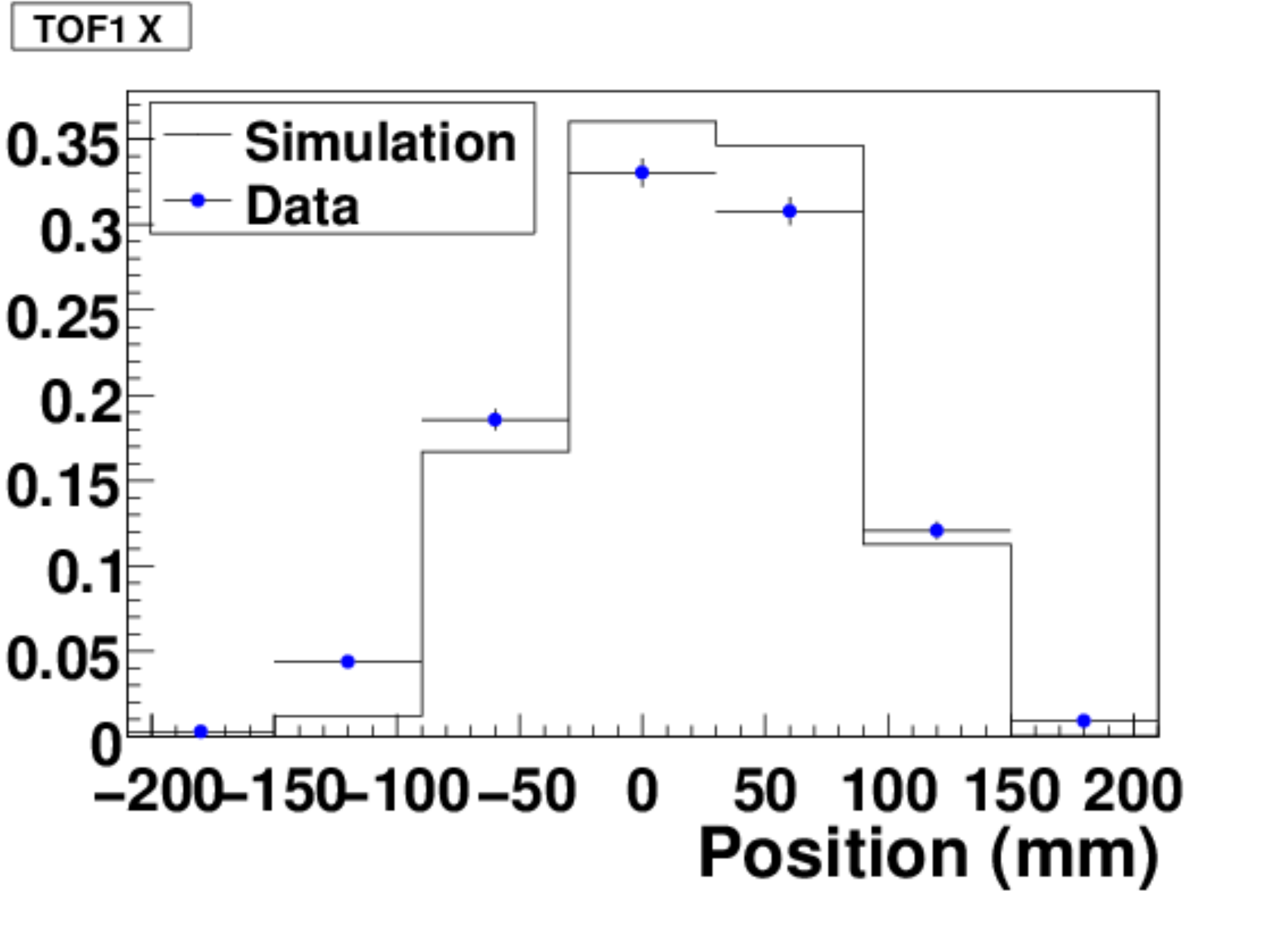}
    \includegraphics[width=0.49\linewidth]%
      {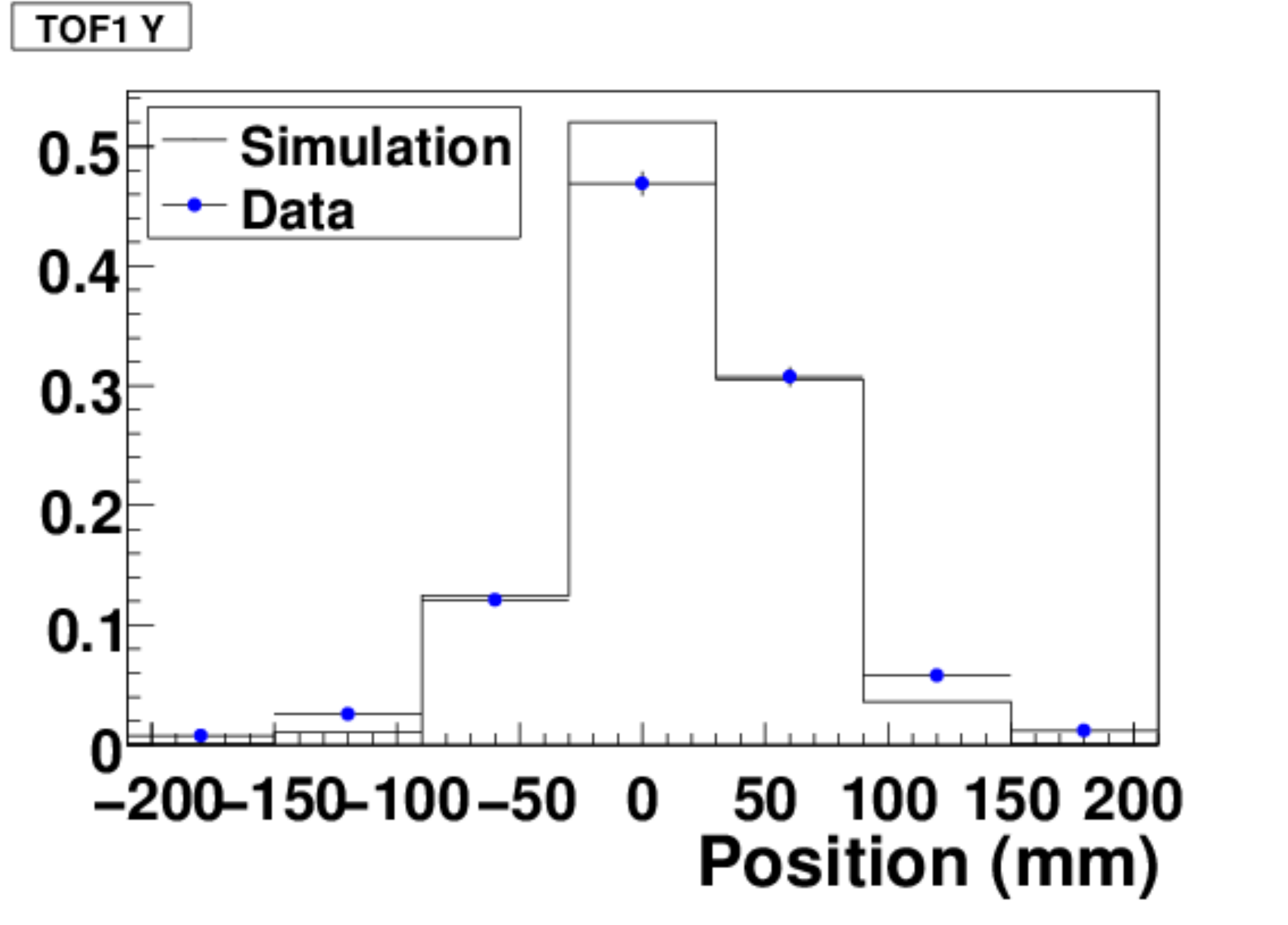}
  \end{center}
  \caption{
    Slab by slab comparison between simulation and data for the $x$
    (left-hand panel) and $y$ (right-hand panel) planes of TOF1 for
    the (6,200) matrix element, with misalignment corrections applied.
  }
  \label{fig:tErate200t1}
\end{figure}

A comparison between the beam profiles observed at TOF0 and TOF1 and
those predicted by G4beamline has been carried out in order to
validate the MICE simulations. 
Small misalignments between the TOF detectors and the magnetic axis formed
by the quadrupole triplets were observed. 
%%For TOF0, average misalignment corrections of $-4.8$\,mm along the $x$
%%axis and $+1.0$\,mm along the $y$ axis were applied. 
%%For TOF1, the misalignments found were $-27.4$\,mm along the $x$ axis
%%and $-15.2$\,mm along the $y$ axis. 
Figures \ref{fig:tErate200t0} and \ref{fig:tErate200t1} show a
comparison of particle rates per TOF slab in the $x$ plane (left) 
and the $y$ plane (right) after applying the misalignment corrections.
The figures show that the beam profile at TOF0 is asymmetric in the
$x$ projection.
Furthermore, the $x$-distributions are  observed to be about 20\%
wider in the data than in the simulation.
These distributions reflect the pion momentum spectrum as produced
in 800 MeV  proton Ti collisions, which is unlikely to be exactly modeled in 
GEANT4.
\section{Conclusions}
\label{sec:conclusions}

The MICE Muon Beam on ISIS at the STFC Rutherford Appleton Laboratory
has been described.
In addition, the design and performance of the MICE beam-line
instrumentation has been presented.
The MICE Muon Beam has been shown to deliver muon beams over the range
of momentum and emittance required to carry out the MICE programme.
Approximately 30 muons
per 1 ms spill (at a rate of $\sim$1~Hz, for a 1 V$\cdot$~ms beam loss) 
for a positive 
muon beam has been achieved, for a negligible activation of the ISIS 
synchrotron. Preliminary results on the pion contamination of the beam 
shows values at the per-cent level. 
The MICE Muon Beam and the instrumentation will serve the experiment
as it is built up in a number of Steps, making it possible to prove
the principle of the ionization-cooling technique that is essential to
the realisation of the Neutrino Factory and the Muon Collider.
\newpage
% ----------------------------------------------------------------
\section*{Acknowledgements}
% ----------------------------------------------------------------

We gratefully acknowledge the help and support of the ISIS staff 
and of the numerous technical collaborators who have contributed to
the design, construction, commissioning and operation of the
experiment. In particular we would like to thank S.~Banfi, F.~Chignoli, 
R.~Gheiger, A.~Gizzi and  V.~Penna.
We wish to acknowledge the essential contributions in the conceptual
development of a muon cooling 
experiment  made by P. Fabbricatore, R. Fernow, D.~Findlay,  
W. Murray, J. Norem, P.R.~Norton, K. Peach, C.~Prior and N. McCubbin. 
We would also wish to acknowledge the work done in the early stages of
the experiment by G. Barr, P. Chimenti, S. Farinon, 
G.~Giannini, E. Radicioni, G. Santin, C. Vaccarezza and S. Terzo.
The experiment was made possible by grants from
National Science Foundation and Department of Energy (USA),
the Istituto Na\-zio\-na\-le di Fisica Nucleare (Italy), 
the Science and Technology 
Facilities Council (UK), the European Community under the European Commission Framework Programe 7,  the Japan Society for the Promotion of Science (Japan)
and the Swiss National Science Foundation (Switzerland), in the framework
of the SCOPES programme.
We gratefully acknowledge their support.

\newpage
\bibliographystyle{JHEP}
\bibliography{STEPI_v9}

\end{document}